\documentclass[12pt,a4paper,openright]{report}

\NewCommandCopy{\oldcleardoublepage}{\cleardoublepage}
\newcommand{\openright}{\pagestyle{empty}\oldcleardoublepage\pagestyle{plain}}
\renewcommand{\cleardoublepage}{\pagestyle{empty}\oldcleardoublepage\pagestyle{plain}}

\usepackage[a-2u,pdf17]{pdfx}

\usepackage[utf8]{inputenc}

\usepackage{lmodern}        
\usepackage{microtype}      
\usepackage{titlesec}       
\usepackage{titletoc}       
\usepackage{parskip}        
\usepackage{lipsum}         
\usepackage{xcolor}         
\usepackage{tocbibind}      

\usepackage{amsmath}        
\usepackage{amsfonts}       
\usepackage{amssymb}        
\usepackage{amsthm}         
\usepackage{accents}        
\usepackage{xparse}         
\usepackage{bm}             
\usepackage{mathtools}      
\usepackage{scalerel}       
\usepackage{esint}          
\usepackage{arydshln}       

\usepackage[sep=3pt, offset=0.82em]{./preamble/packages/simpler-wick/simpler-wick}
\usepackage[italicdiff]{physics} 
\usepackage{tikz-cd}             
\usepackage{ytableau}            

\usepackage{graphicx}       
\graphicspath{{figures/}}
\usepackage{adjustbox}      
\usepackage{booktabs}       
\usepackage[font=footnotesize,labelfont=bf,format=hang]{caption}
\usepackage{subcaption}

\makeatletter
\let\c@figure\c@table
\makeatother

\usepackage{paralist}       
\setdefaultleftmargin{1.87em}{1.7em}{1.5em}{1em}{1em}{1em}
\setdefaultitem{$\bullet$}{\textbullet}{\textasteriskcentered}{\textperiodcentered}
\setdefaultenum{\bfseries (1)}{\bfseries (a)}{\bfseries (i)}{A.}

\usepackage[style=ieee,citestyle=numeric-comp,isbn=false,eprint=true,maxcitenames=4,backref=true]{biblatex}
\DeclareFieldFormat*{date}{\mkbibbold{#1}}
\DeclareFieldFormat*{title}{\textsf{#1}}

\AtEveryBibitem{
    \clearfield{month}
    \clearfield{pages}
    \clearfield{number}
    \ifentrytype{article}{
        \clearfield{volume}
    }{}
}

\DeclareFieldFormat*{eprint:arxiv}{%
    \textsc{arXiv}\addcolon\space 
    \ifhyperref
    {\href{https://arxiv.org/abs/#1}{%
            \nolinkurl{#1}%
            \iffieldundef{eprintclass}
            {}
            {\addspace\texttt{\mkbibbrackets{\thefield{eprintclass}}}}}}
    {\nolinkurl{#1}%
        \iffieldundef{eprintclass}
        {}
        {\addspace\texttt{\mkbibbrackets{\thefield{eprintclass}}}}}}

\DeclareFieldFormat{pagerefformat}{
    {\hypersetup{pdfborder={0 0 0}}\color{gray}
    \nobreak\hfill\penalty50\hskip1em\null\nobreak
    \hfill\scriptsize\mkbibparens{#1}\normalsize
    \parfillskip=0pt \finalhyphendemerits=0 \par}}
\renewbibmacro*{pageref}{%
    \addperiod
    \iflistundef{pageref}
    {}
    {\printtext[pagerefformat]{%
            \printlist[pageref][-\value{listtotal}]{pageref}}}}%

\DeclareSourcemap{
    \maps[datatype=bibtex]{
        \map[overwrite]{
            \step[fieldsource=doi, final]
            \step[fieldset=url, null]
        }
    }
}
\DefineBibliographyStrings{english}{
    url = {\textsc{url}},%
}

\usepackage{hyperref}
\hypersetup{
    unicode,
    breaklinks=true,
    linktoc=all,
    bookmarksnumbered,
}
\usepackage{cleveref}

\makeatletter
\def\@makechapterhead#1{
    {\parindent \z@ \raggedright \normalfont
            \Huge\bfseries \thechapter. #1
            \par\nobreak
            \vskip 20\p@
        }}
\def\@makeschapterhead#1{
    {\parindent \z@ \raggedright \normalfont
            \Huge\bfseries #1
            \par\nobreak
            \vskip 20\p@
        }}
\makeatother

\usepackage{thmtools}
\usepackage[framemethod=TikZ]{mdframed}
\mdfsetup{
    innertopmargin=0.6\topskip,
    innerbottommargin=0.6\topskip,
    skipabove=\baselineskip,
    skipbelow=0,
}

\declaretheoremstyle[headfont=\bfseries\sffamily, bodyfont=\normalfont, mdframed={ nobreak } ]{thmbox}
\declaretheoremstyle[headfont=\bfseries\sffamily, bodyfont=\normalfont]{thmnobox}

\declaretheorem[sibling=table,      style=thmbox,   name=Definition]{definition} 
\declaretheorem[sibling=definition, style=thmbox,   name=Corollary]{corollary}
\declaretheorem[sibling=definition, style=thmbox,   name=Proposition]{prop}
\declaretheorem[sibling=definition, style=thmbox,   name=Theorem]{theorem}

\declaretheorem[sibling=definition, style=thmnobox, name=Example]{example}
\declaretheorem[sibling=definition, style=thmnobox, name=Remark]{remark}
\declaretheorem[sibling=definition, style=thmnobox, name=Derivation]{derivation}
\declaretheorem[sibling=definition, style=thmnobox, name=Calculation]{calculation}

\declaretheoremstyle[headfont=\bfseries\sffamily, bodyfont=\normalfont, mdframed={linewidth=2pt, rightline=false, topline=false, bottomline=false, linecolor=pink, backgroundcolor=pink!20}]{todo}
\declaretheoremstyle[headfont=\bfseries\sffamily, bodyfont=\normalfont, mdframed={linewidth=2pt, rightline=false, topline=false, bottomline=false, linecolor=orange, backgroundcolor=orange!15}]{question}
\declaretheoremstyle[headfont=\bfseries\sffamily, bodyfont=\normalfont, mdframed={linewidth=2pt, rightline=false, topline=false, bottomline=false, linecolor=cyan, backgroundcolor=cyan!10}]{suggestion}

\newenvironment{myproof}[1][\proofname]
{\proof[\normalfont \bfseries\sffamily #1]}
{\endproof}

\newcommand{\iunit}{{\mathring{\imath}}}
\newcommand{\eunit}{\mathinner{\mathrm{e}}}

\newcommand{\R}{\mathbb{R}}

\newcommand{\C}{\mathbb{C}}
\newcommand{\F}{\mathbb{F}}
\newcommand{\N}{\mathbb{N}}
\newcommand{\Id}{\bm{1}}

\newcommand{\J}{\mathcal{J}}
\newcommand{\K}{\mathcal{K}}
\newcommand{\B}{\mathcal{B}}
\newcommand{\D}{\mathcal{D}}
\newcommand{\opers}[2]{d_{\bm{#1}\mkern 1mu #2}}

\DeclareMathOperator{\Sect}{Sect}
\DeclareMathOperator{\End}{End}

\DeclareMathOperator{\GL}{\mathsf{GL}}

\DeclareMathOperator{\OO}{\mathsf{O}}
\DeclareMathOperator{\SO}{\mathsf{SO}}
\DeclareMathOperator{\Spin}{\mathsf{Spin}}

\DeclareMathOperator{\SU}{\mathsf{SU}}
\DeclareMathOperator{\Sp}{\mathsf{Sp}}
\DeclareMathOperator{\g}{\mathfrak{g}}
\renewcommand{\d}{\mathsf{d}}
\renewcommand{\r}{\mathsf{r}}

\DeclareMathOperator{\rt}{R}

\newcommand{\Sym}{S}

\DeclareMathOperator{\PE}{PE}
\DeclareMathOperator{\Span}{Span}
\DeclareMathOperator{\Ker}{Ker}
\DeclareMathOperator{\Hom}{Hom}

\newcommand{\AD}{{\normalfont\text{AD}}\mkern-1mu}
\newcommand{\Ad}{{\normalfont\text{Ad}}}
\newcommand{\ad}{{\normalfont\text{ad}}}

\DeclareMathSymbol{\shortminus}{\mathbin}{AMSa}{"39}

\makeatletter
\newcommand{\Ext}{\@ifnextchar^\@Ext{\@Ext^{\,}}}
\def\@Ext^#1{\mathop{\bigwedge\nolimits^{\mspace{-5mu}\raisebox{-0.13em}{\scriptsize$#1$}}}}
\makeatother

\NewDocumentCommand{\lowerindex}{o e{_^}}{%
    _{\IfValueT{#2}{#2}\vphantom{\smash[b]{\scalebox{#1}{\big|}}}}
        ^{\IfValueT{#3}{#3}\vphantom{\smash[b]{\scalebox{#1}{\big|}}}}
}
\NewDocumentCommand{\lowerindexnew}{o e{_^}}{%
    _{\IfValueT{#2}{\raisebox{#1}{$\scriptstyle #2$}}}
        ^{\IfValueT{#3}{#3}}
}

\NewCommandCopy{\oldepsilon}{\epsilon}
\renewcommand{\epsilon}{\varepsilon}
\NewCommandCopy{\oldchi}{\chi}
\renewcommand{\chi}{\oldchi\lowerindexnew[-1.5pt]}

\NewCommandCopy{\olddet}{\det}
\renewcommand{\det}{\olddet\nolimits}

\NewCommandCopy{\originalmiddle}{\middle}
\def\middle#1{\mathrel{}\originalmiddle#1\mathrel{}}

\newcommand{\smallerrel}[1]{\mathrel{\mathpalette\smallerrelaux{#1}}}
\newcommand{\smallerrelaux}[2]{\raisebox{.1ex}{\scalebox{.75}{$#1#2$}}}

\newcommand{\smallin}{\smallerrel{\in}}

\NewCommandCopy{\originalleft}{\left}
\NewCommandCopy{\originalright}{\right}
\renewcommand{\left}{\mathopen{}\mathclose\bgroup\originalleft}
\renewcommand{\right}{\aftergroup\egroup\originalright}

\NewCommandCopy{\oldotimes}{\otimes}
\NewCommandCopy{\oldbullet}{\bullet}
\newcommand\sbullet[1][.5]{\mathbin{\ThisStyle{\vcenter{\hbox{%
                    \scalebox{#1}{$\SavedStyle\bullet$}}}}}}

\newcommand{\bigslant}[2]{{\left.\raisebox{.2em}{$#1$} \mkern -5mu \middle/\mkern -6mu \raisebox{-.2em}{$#2$}\right.}}
\newcommand{\gt}{\bigslant{\g\mkern-1mu}{\mkern-1mu\mathfrak{t}\mkern-1mu}}
\newcommand{\gts}{\left.\raisebox{0.15em}{$\scriptstyle\g$}\mkern-1.7mu\middle/\mkern-2.7mu\raisebox{-0.15em}{$\scriptstyle\mathfrak{t}$}\right.}

\definecolor{mygray}{gray}{0.85}
\tikzset{
    main node/.style={inner sep=0,outer sep=0},
    label node/.style={inner sep=0.3em,font=\tiny,overlay},
    strike out/.style={shorten <=-.2em,shorten >=-.2em,overlay,thick,double distance = 0em,line cap=round}
}
\newcommand{\cancel}[3][mygray]{\tikz[baseline=(N.base)]{
        \node[main node](N){$#2$};
        \node[label node,gray, anchor=south] at (N.north){#3};
        \draw[strike out,#1]  (N.south west) -- (N.north east);
    }}

\ytableausetup{centertableaux}
\NewCommandCopy{\oldsquare}{\square}
\RenewDocumentCommand{\square}{O{0.55em}}{\ytableausetup{boxsize=#1}\operatorname{\ydiagram{1}}}
\NewDocumentCommand{\smallsquare}{O{0.35em}}{\raisebox{-0.07em}{$\square[#1]$}}
\NewDocumentCommand{\smallotimes}{}{\mspace{+1mu}\raisebox{.275ex}{\scalebox{.5}{$\oldotimes$}}\mspace{+1mu}}

\begin{document}

\pagenumbering{roman}
\def\ThesisTitle{Counting operators in Effective Field Theories}
\def\ThesisTitleFront{Counting operators\\[0.2em] in Effective Field Theories}

\def\ThesisAuthor{Jonáš Dujava}

\def\YearSubmitted{2022}
\def\YearRevision{2024}

\def\Department{Institute of Particle and Nuclear Physics}

\def\DeptType{Institute}

\def\Supervisor{Mgr. Petr Vaško, Ph.D.}

\def\SupervisorsDepartment{Institute of Particle and Nuclear Physics}

\def\StudyProgramme{Physics}
\def\StudyBranch{FP Physics}

\def\Abstract{%
We systematically develop the Hilbert series technique for counting independent operators in Effective Field Theories. In the hope of providing more approachable entry point to the subject we include a detailed introduction of all necessary group theoretic tools (in a rather mathematical definition--theorem--proof style). Finally, we apply the formalism in the case of a single scalar field and also electromagnetic field, partly reproducing and partly extending known results.
}

\def\Keywords{%
    {Effective Field Theory}, {Operator counting}, {Hilbert series}, {Group theory}, {Lie groups}, {Representation theory}, {Graded representations}
}

\pagestyle{empty} 
\begin{center}
    \graphicspath{{preliminaries/img/}}
    \centerline{\mbox{\includegraphics[width=166mm]{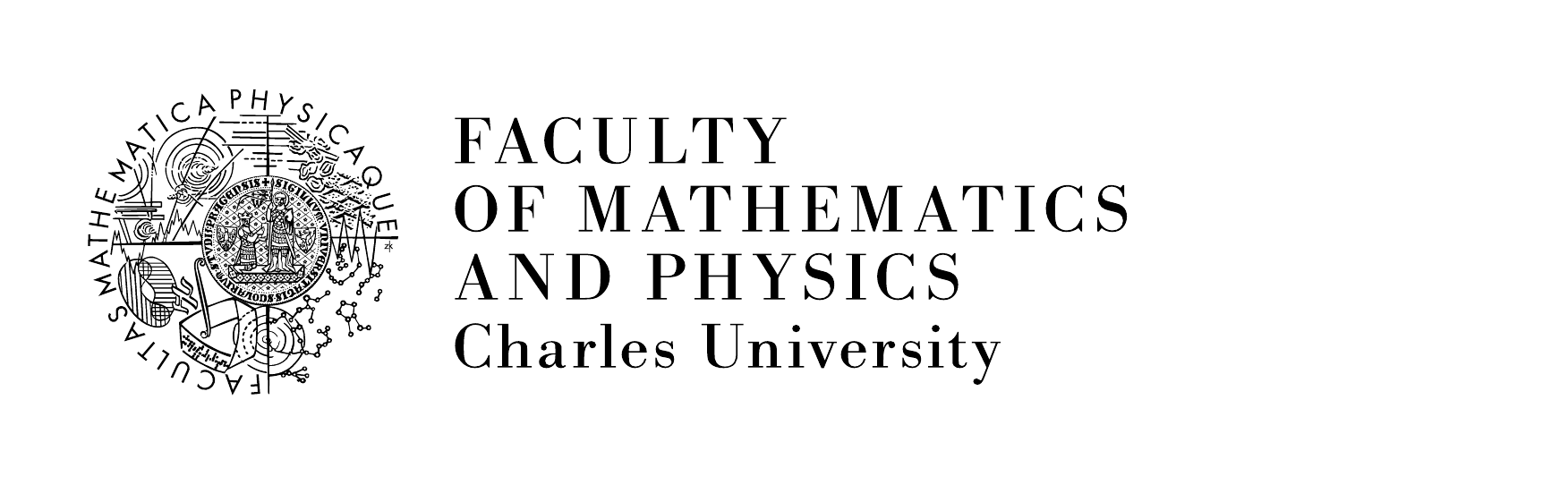}}}

    \vspace{-8mm}
    \vfill

    {\bfseries\Large BACHELOR THESIS}

    \vfill

    {\LARGE\ThesisAuthor}

    \vspace{15mm}

    {\Huge \bfseries \ThesisTitleFront}

    \vfill

    \Department

    \vfill

    {\centerline{\vbox{\halign{
                    \hbox to 0.45\hsize{\hfil #}&\hskip 0.5em\parbox[t]{0.45\hsize}{\raggedright #}\cr
                    Supervisor of the bachelor thesis:&\Supervisor \cr
                    \noalign{\vspace{2mm}}
                    Study programme:&\StudyProgramme \cr
                    \noalign{\vspace{2mm}}
                    Study branch:&\StudyBranch \cr
                }}}}

    \vfill

    Prague \YearSubmitted \\
    \emph{Revised version \YearRevision}

\end{center}

\newpage

\openright

{ \setlength\parskip{5mm}

    \phantomsection
    \section*{Abstract}
    \addcontentsline{toc}{chapter}{Abstract}
    \Abstract

    \section*{Keywords}
    \Keywords

    \section*{Quick summary}

    In \textbf{\Cref{ch:effective}} we briefly introduce the framework of Effective Field Theories, giving us background and physical motivation to study the structure of operators.

    In \textbf{\Cref{ch:operatorbasis}} we recognize possible relations between operators and define the fundamental object of our interest --- \emph{the Hilbert series} --- which effectively encodes information about the independent operators of a given type.

    In \textbf{\Cref{ch:grouptheory}} we thoroughly build parts of the \emph{representation theory} of compact Lie groups, which shall prove useful for computing the Hilbert series.

    In \textbf{\Cref{ch:masterformuladerivation}} we finally derive \emph{the Master Formula} for the Hilbert series, utilizing everything we learned in previous chapters.

    In \textbf{\Cref{ch:applications}} we apply the developed formalism to count operators for a single \emph{scalar field}, and also in a more complex case of the \emph{electromagnetic field}. We highlight numerous intricacies, which are present even in these basic theories.
}

{
    \hypersetup{hidelinks}
    \tableofcontents
}

\chapter*{Introduction}
\addcontentsline{toc}{chapter}{Introduction}

Every physical calculation applied to the real world is only an approximation. This is mainly a consequence of two factors --- we can not measure and prepare initial states with infinite precision, and we do not possess the \emph{exact} theory. Even if we did, it is almost impossible to calculate anything without making some approximations and mathematical shortcuts along the way.

Take for example particle physics, where it is hard to complete the calculations in full generality using the present ``full-fledged'' theory, which furthermore still might miss even more particles. However, in a given energy range it is enough to consider only a couple of particles to obtain relevant quantitative predictions, and the existence of additional particles only leads to certain corrections. Adressing them systematically would be desirable.

This is the main idea of \emph{Effective Field Theories} (EFTs). Being perturbative in their nature, EFT Lagrangians are essentially infinite series, every next order giving additional corrections. Even if one wishes to calculate up to a fixed order, there may be many operators to consider. It is therefore of interest to find the \emph{operator basis}, which parametrizes all physical phenomena in the realm of the given effective theory, but already does not contain any redundancies.

In general, construction of such a basis is immensely difficult because of various non-trivial relations between them. Nonetheless, it is possible to count the number of independent operators of a given type, which certainly helps. This thesis aims to derive the formula for these multiplicities, and to develop all the necessary tools along the way.

The group-theoretical approach we will develop was pioneered by \textcite{Operator2017}. The research paper is lengthy, spanning over 100 pages and presents multiple approaches, with the main one presupposing knowledge of conformal representations. As a result, inexperienced readers may struggle to understand both the details and the main ideas. We will try to complement this.

We assume the reader is well versed in the subjects of linear algebra and differential geometry, as well as familiar with some elementary constructions in group theory such as cosets and quotient groups. Otherwise, we have attempted to provide a mostly self-contained exposition, building up essentially all necessary theory about Lie groups and their representations from the beginning.

\pagenumbering{arabic}
\chapter{Effective Field Theories} \label{ch:effective}

Wondering about physics one may stumble upon a fundamental question:
\begin{center}
    \emph{How can we calculate anything without knowing everything?}
\end{center}
We do not have the \emph{Theory of Everything}, yet we are able to perform vast amount of precise calculations in practically every domain of the Universe.

The predictive power of physical theories is rooted in the fact that only a couple of relevant degrees of freedom play an important role in a given problem. The act of neglecting the rest is the first step in the progress of understanding any physical phenomena. One could claim that in a certain sense \cite{Penco2020}
\begin{center}
    \textbf{all physical theories are \emph{effective theories}},
\end{center}
where the word \emph{effective} reflects the restriction of a given theory to only a certain range of effects while being oblivious to the rest.

Just take for example \emph{Newtonian mechanics}, which often works only with point masses as an idealization of particles with possibly complicated internal structure. Yet, it offers brilliant tools for our everyday life and explanations of many effects, ranging from the usage of a lever to the motion of celestial bodies.

Another example can be found in well-known theories of \emph{electrostatics}, \emph{magnetism} and \emph{optics}. Historically, these were separate subjects without many references between each other, and it was only later realized that they were manifestations of the unified \emph{Maxwell theory of electromagnetism}.

Our current ``fundamental'' theories of electroweak and strong interactions are formulated in the language of \emph{quantum field theory} and \emph{gauge theory}. Together with the classification of all known elementary particles they are contained in what is called the \emph{Standard Model} of particle physics. Still, we know that the Standard Model can not be complete, as it (among other things) lacks account of General Relativity. Currently, the Standard Model is often viewed as a low-energy \emph{effective field theory} (EFT) of some more complete theory \cite{Weinberg2021}.

\section{Advantages of EFTs}

In the rest of the thesis, we restrict ourselves to \emph{field} theories, that is theories fully described by the action as an integral of the Lagrangian density over the whole space-time manifold. Moreover, we will assume the context of particle physics, where together with \emph{Lorentz transformations} we often encounter additional \emph{gauge transformations}. Naturally, the action must be invariant with respect to both.

To be more precise, the term \emph{effective field theory} in the context of \emph{particle physics} is used in a more strict sense. We will not need most of the following, but to have some background and motivation for our problem we mention some important features, examples, and advantages of EFTs.

\begin{definition}[Effective field theory] \label{def:EFT}
    A low-energy \emph{effective field theory} (EFT) $\mathcal{T}$ is described by \parencite{Penco2020,Jones2020}:
    \begin{enumerate}[\bfseries (a)]
        \item The \emph{degrees of freedom}/\emph{particle fields}, whose behavior $\mathcal{T}$ tries to explain.
        \item The \emph{domain of validity}, usually characterized by an energy scale $\Lambda$. For energies well below the characteristic energy scale ($E\ll\Lambda$), $\mathcal{T}$ properly explains physical phenomena of interest. For energies approaching the characteristic energy scale ($E \approx \Lambda$) a transition to a more complete theory or a different and more appropriate EFT may be necessary.
        \item The \emph{perturbative expansion} of the EFT action $S_{\mathcal{T}}$ in a suitable parameter (prime example being $E/\Lambda$), correcting for effects of initially ignored ``irrelevant'' degrees of freedom by introducing additional interaction terms.
    \end{enumerate}
\end{definition}

\begin{remark}
    Previous statement about all theories being effective theories can be viewed as talking about leading-order effective theories.
\end{remark}

\begin{example} Here we just mention some established EFTs \cite{manohar2018}:
    \begin{itemize}
        \item \emph{Fermi theory of weak interactions} --- effective theory for the \emph{weak interactions} at energies below the $W^{\pm}$ and $Z^{0}$ masses.
        \item \emph{Heavy Quark Effective Theory} (HQET) --- low-energy dynamics of \emph{hadrons} containing a \emph{heavy quark}.
        \item \emph{Chiral Perturbation Theory} ($\chi$PT) --- dynamics of \emph{pions} in a strongly coupled low-energy limit of \emph{Quantum ChromoDynamics} (QCD).
        \item \emph{Standard Model Effective Field Theory} (SMEFT) --- used to analyze deviations from the \emph{Standard Model} and search for \emph{Beyond Standard Model} physics.
    \end{itemize}
\end{example}

\begin{remark}
    There are numerous advantages of using EFTs, many being hard to appreciate without a deeper dive. We only list some of broad relevance \cite{manohar2018,Penco2020}:
    \begin{itemize}
        \item Even if we are equipped with a more complete theory, calculations in a certain energy range can be drastically simplified by using an appropriate EFT.
              \begin{itemize}
                  \item From the beginning we are neglecting aspects unimportant to the problem and remain focused on the relevant degrees of freedom.
                  \item Isolation of relevant degrees of freedom may reveal new symmetries that otherwise would have remained obscured.
              \end{itemize}
        \item When dealing with yet unknown physics, EFTs provide us with a systematic perturbative parametrization of physics in a certain energy domain.
              \begin{itemize}
                  \item The \emph{main idea} --- if we believe our problem can be understood by means and tools of quantum field theory, we can not go wrong by starting with \emph{``the most general Lagrangian''} consistent with our assumptions of locality, unitarity, Lorentz and gauge invariance, \ldots
                  \item A unified framework utilizing a perturbative expansion containing only a few coefficients can be useful for placing model-independent constraints.
              \end{itemize}
    \end{itemize}

\end{remark}

\section{Construction of EFTs}

The construction of some particular low-energy EFT can be outlined by the following prescription \cite{Jones2020}:
\begin{enumerate}
    \item Identify all relevant degrees of freedom $\{\Phi_{i}\}$ and symmetries they satisfy.
    \item Construct the most general effective action consistent with locality, Lorentz invariance and additional (gauge, global, etc.) symmetries,
          \begin{equation} \label{eq:mostgeneralaction}
              S_{\text{eff}}[\{\Phi_{i}\}]= \int_{\mathcal{M}} \dd[\d]{x} \left[ \mathcal{L}_{\text{kin}}(\{\Phi_{i},\partial_{a}\Phi_{i}\}) + \sum_{j} \frac{c_{j}}{\Lambda^{\Delta_{j}-\d}} \mathcal{O}_{j} \right], \tag{$\bm{\infty}$}
          \end{equation}
          where $\d$ is the dimension of the space-time manifold $\mathcal{M}$, $\mathcal{L}_{\text{kin}}$ is the kinetic part of the Lagrangian density, $\Lambda$ is the characteristic energetic scale of the EFT and $c_{j}$ is a dimensionless ``Wilson coefficient'' corresponding to the local operator $\mathcal{O}_{j}$ with the mass dimension $\Delta_{j}\equiv \Delta[\mathcal{O}_{j}]$.
    \item Calculate physical observables (to the required order of accuracy).
    \item Determine Wilson coefficients $c_{j}$ by matching the EFT to the already existent underlying model or to the experimental measurements.
\end{enumerate}

In this thesis we will focus on a part of the step \textbf{(2)}, which has certain freedom in the choice of the set of operators $\{\mathcal{O}_{j}\}$. Assuming someone already did step \textbf{(1)} for us, our goal will be to count all independent operators figuring in the effective action \eqref{eq:mostgeneralaction}. Precise meaning of this will be clarified in \Cref{ch:operatorbasis}.

Steps \textbf{(3)} and \textbf{(4)} are mentioned here only for completeness, nonetheless they are of paramount importance and can be immensely deep.

For simplicity, we restrict our attention only to \emph{linearly realized} symmetries, and we will not worry about discrete global symmetries, such as parity, charge conjugation or time inversion.
Throughout the thesis we also assume an implicit \emph{Wick rotation} from Minkowski to Euclidean space, thus for our (Euclidean) \emph{Lorentz group} we can take $\SO(\d)$, since we will ignore inversions.

Of course, explicitly constructing the basis of $\{\mathcal{O}_{j}\}$ would be even better, but it is in general a formidable task. The next best thing is to at least gather some information about the number of operators of different types, which fortunately turns out to be a feasible goal.

\chapter{The Operator Basis} \label{ch:operatorbasis}

The notion of operators is probably clear, but before advancing further, it is better to fix the terminology to avoid possible ambiguities.
\begin{definition}[Operator, invariant/scalar operator] \label{def:operator}
    An \emph{operator} $\mathcal{O}$ is a local object composed only from the particle fields and their derivatives at a given point of space-time. An operator is called \emph{invariant} or \emph{scalar} if it is Lorentz and gauge invariant (and also invariant under any other symmetry group). Only scalar operators can appear in the Lagrangian density, since the action can not depend on the choice of an observer.
\end{definition}
\begin{remark}
    Operators naturally form a linear space, since we are free to multiply them by a constant and add them together. We can thus specify the EFT by the choice of a single scalar \emph{Lagrangian} operator $\mathcal{L}$ from the big space of all operators (actually, it is usually a series in a suitable parameter, see \Cref{def:EFT}).
\end{remark}

As we suggested before, operators with a different appearance can lead to the same physical effect. Equivalently, a mapping from the space of all Lagrangians to the space of $S$-matrix operators has a nonzero kernel. Unsurprisingly, it is advantageous to work with a minimal set of operators covering all possible physical phenomena in a given EFT, thus we come to the following definition.
\begin{definition}[Operator basis] \label{def:operatorbasis}
    The \emph{operator basis} $\B$ of the EFT is a minimal set of operators leading to all physical phenomena in the realm of the EFT. We will denote the space of all operators generated by $\B$ as $\K\equiv\Span\B$.
\end{definition}

\begin{remark}
    As in any vector space, there is a bunch of equivalent operator bases for a given EFT. We will loosely refer to any chosen one as the \emph{operator basis}.
\end{remark}

In this chapter we will recognize possible \emph{relations}/redundancies/constraints between operators, which effectively lower the number of linearly independent operators, thus reduce the size of $\B$ and the dimension of $\K$.

Next, we will define a \emph{weighting scheme} for categorizing different types of operators, enabling us to compactly encode reduced information about the operator basis into a generating function called the \emph{Hilbert series}, our primary object of interest. More specifically, it will contain the numbers of independent operators for all distinct types.

\section{Equivalence relations between operators} \label{sec:relations}

From classical theoretical mechanics we know very well that the form of the Lagrangian is not entirely fixed.
By performing a transformation to a different set of \emph{generalized coordinates}, we visually change the Lagrangian, but the physics is the same (we have just changed the parametrization of the configuration space).

Note that terms proportional to the \emph{equations of motion} are in fact just trivial classically (along the classical trajectory).
Additionally, \emph{total derivative terms} do not change the dynamics, as they reduce to boundary terms in the actions, which do not play a role in a variational principle with fixed boundary conditions.

But how does the story play out for an effective quantum field theory? Quite similarly, the total derivative (divergence) terms can be straightforwardly excluded from the effective action \eqref{eq:mostgeneralaction}. However, in quantum theory we can not eliminate the operators proportional to the equations of motion simply by substitution as in the classical case.

Nonetheless, we are still able to perform \emph{field redefinitions}, which in general do not change the \(S\)-matrix. This allows us to systematically eliminate operators proportional to the lowest order equations of motion (obtained from $\mathcal{L}_{\text{kin}}$) in exchange for a higher order terms in the perturbation expansion. A more in-depth explanation is provided by \textcite[Chapter 6]{manohar2018}. Thus, for the purpose of counting independent operators, we can simply ignore such operators.

Last, but by no means least, we have relations induced by the finite-dimensionality of the space-time manifold. To give a simple example, assume we are in $\d=2$ and we have one scalar field $\Phi$. Since in $\d=2$ the antisymmetrization of $3$ indices is automatically zero, we obtain the following relation between scalar operators (we first make the antisymmetrization, and after that we contract the indices)
\begin{align*}
    0 & \overset{!}{=} 3!\,\partial^{\bm{[}a}\partial_{a}\Phi\partial^{b}\partial_{b}\Phi\partial^{c\bm{]}}\partial_{c}\Phi \\[-1.5em]
      & \equiv\phantom{3!\,} \partial^{a}\partial_{a}\Phi\partial^{b}\partial_{b}\Phi\partial^{c}\partial_{c}\Phi + \overbrace{\partial^{c}\partial_{a}\Phi\partial^{a}\partial_{b}\Phi\partial^{b}\partial_{c}\Phi +\partial^{b}\partial_{a}\Phi\partial^{c}\partial_{b}\Phi\partial^{a}\partial_{c}\Phi}^{2\partial^{b}\partial_{a}\Phi\partial^{c}\partial_{b}\Phi\partial^{a}\partial_{c}\Phi} \\
      & \phantom{\equiv 3!\,}\ \underbrace{- \partial^{b}\partial_{a}\Phi\partial^{a}\partial_{b}\Phi\partial^{c}\partial_{c}\Phi - \partial^{c}\partial_{a}\Phi\partial^{b}\partial_{b}\Phi\partial^{a}\partial_{c}\Phi - \partial^{a}\partial_{a}\Phi\partial^{c}\partial_{b}\Phi\partial^{b}\partial_{c}\Phi}_{-3\partial_{a}\partial_{b}\Phi\partial^{a}\partial^{b}\Phi\partial^{c}\partial_{c}\Phi} \\[0.7em]
      & \Longrightarrow \quad 0 \overset{!}{=} \wick{\c1\partial\c1\partial\Phi\c1\partial\c1\partial\Phi\c1\partial\c1\partial\Phi} + 2\wick{\c2\partial\c1\partial\Phi\c1\partial\c1\partial\Phi\c1\partial\c2\partial\Phi} -3\wick{\c2\partial\c1\partial\Phi\c1\partial\c2\partial\Phi\c1\partial\c1\partial\Phi},
\end{align*}
where the derivatives $\partial$ always act only on the fields $\Phi$ standing directly after them, and in the last line we just opted for an alternative notation for contractions.

\begin{remark}[Gram determinant conditions]
    These finite-dimensionality conditions are analogous to the statement that $r>\d$ vectors $\{\bm{v}_{i}\}_{i=1}^{r>\d}$ can not be linearly independent, thus invariants created from them obey some relations. Assume we are in Euclidean space and organize the vectors in the matrix $A\equiv (\bm{v}_1| \ldots | \bm{v}_{r})$. We can easily compute the corresponding \emph{Gram matrix}, which is essentially a matrix of all scalar products $\bm{v}_{i}\sbullet \bm{v}_{j}$ ($A^{^{\scriptstyle\intercal}}$ denotes the transpose of $A$),
    \begin{equation*}
        G\equiv A^{^{\scriptstyle\intercal}}\!A =
        \begin{pmatrix}
            \bm{v}_{1} \sbullet \bm{v}_{1} & \bm{v}_{1} \sbullet \bm{v}_{2} & \hdots & \bm{v}_{1} \sbullet \bm{v}_{r} \\
            \bm{v}_{2} \sbullet \bm{v}_{1} & \bm{v}_{2} \sbullet \bm{v}_{2} & \hdots & \bm{v}_{2} \sbullet \bm{v}_{r} \\
            \vdots                         & \vdots                         & \ddots & \vdots \\
            \bm{v}_{r} \sbullet \bm{v}_{1} & \bm{v}_{r} \sbullet \bm{v}_{2} & \hdots & \bm{v}_{r} \sbullet \bm{v}_{r}
        \end{pmatrix}.
    \end{equation*}
    It is a basic fact from linear algebra that $\Ker A = \Ker A^{^{\scriptstyle\intercal}}\!A$, from which follows that $G$ is singular if and only if $\{\bm{v}_{i}\}$ are linearly dependent. We can apply the same argument for every $(\d+1)\!\times\!(\d+1)$ submatrix of $G$, thus obtaining the so-called \emph{Gram determinant conditions}. They assert that every $(\d+1)\!\times\!(\d+1)$ minor of $G$ is zero, giving us relations between scalar products of vectors $\{\bm{v}_{i}\}$.
\end{remark}


Bringing it all together, we conclude our discussion in the following definition. \vspace{-0.4em}
\begin{definition}[Operator relations]
    Operators $\mathcal{O}_{m}$ and $\mathcal{O}_{n}$ are considered equivalent (denoted by $\mathcal{O}_{m}\sim\mathcal{O}_{n}$), if they satisfy either one of the following relations or possibly their combination \cite{Operator2015}:
    \begin{enumerate}[\bfseries (a)]
        \item \emph{Equations of motion} --- \textbf{EOM} --- the usage of field redefinition argument generates relations of the form \vspace{-0.5em}
              \[ \mathcal{O}_{m} \overset{\text{\bfseries\scalebox{.7}{EOM}}}{\sim} \mathcal{O}_{n} \quad \overset{\text{def}}{\Longleftrightarrow}  \quad \exists\,\mathcal{O}',\Phi_{j}\colon \ \mathcal{O}_{m} = \mathcal{O}_{n} + \fdv{S_{\text{kin}}}{\Phi_{j}} \mathcal{O}'. \vspace{-1em} \]
        \item \emph{Integration by parts} --- \textbf{IBP} --- we also have relations of the type
              \[ \mathcal{O}_{m} \overset{\text{\bfseries\scalebox{.7}{IBP}}}{\sim} \mathcal{O}_{n} \quad \overset{\text{def}}{\Longleftrightarrow} \quad \exists\,\mathcal{O}'\colon \ \mathcal{O}_{m} = \mathcal{O}_{n} + \partial\cdot\mathcal{O}', \]
              since by the generalized Stokes theorem $\int_{\mathcal{M}} \partial\cdot\mathcal{O} = \int_{\partial\mathcal{M}} \mathcal{O}$ is a boundary term, which leaves the dynamics unchanged.
        \item \emph{Gram determinant conditions} --- \textbf{GDC} --- some operators are trivially zero by the finite-dimensionality of $\mathcal{M}$, inducing relations
              \[ \mathcal{O}_{m} \overset{\text{\bfseries\scalebox{.7}{GDC}}}{\sim} \mathcal{O}_{n} \quad \overset{\text{def}}{\Longleftrightarrow} \quad \exists\,\mathcal{O}'
                  \textcolor{gray}{
                  \big|_{d}
                  \left\{
                      \begin{subarray}{c}
                          =\,0  \text{ for } d\;=\;\dim \mathcal{M} \\[0.2em]
                          \neq\,0   \text{ for a general } d
                      \end{subarray}
                  \right\} }
                  \colon \ \mathcal{O}_{m} = \mathcal{O}_{n} + \mathcal{O}'\textcolor{gray}{\big|_{\dim \mathcal{M}}}. \]
    \end{enumerate}
\end{definition}
\vspace{-0.2em}
\begin{remark}[Kinematic polynomial rings]
    Another useful approach to counting and even building the operator basis passes to the momentum space by the Fourier transform, thereby replacing operators with derivatives by \emph{polynomials} in momenta. Operator relations are then represented by relations between polynomials, namely EOM act as \emph{on-shell} conditions, IBP manifest as the statement of \emph{momentum conservation}, and GDC are precisely Gram determinant conditions between scalar products of particles momenta or \emph{Mandelstam variables}. The operator basis is then formed as a \emph{quotient ring} with the ideal generated by these relations. Reader is recommended to take a look at the beginning of Section 5 in \cite{Operator2017}. However, GDC relations are highly non-trivial in this approach, whereas they will be efficiently addressed within the group-theoretical formalism we are going to develop.
\end{remark}

\section{Generating functions, Hilbert series} \label{sec:generatingfunctions}

Consider an EFT with $N$ degrees of freedom (particle fields) $\{\Phi_{i}\}_{i=1}^{N}$. As it is desirable to reduce the complexity of full information about the operator basis, we introduce the concept of \emph{operator weight}, which defines the operator ``type''. \vspace{-0.3em}
\begin{definition}[Weight of an operator] \label{def:weight}
    Let $\mathcal{O}$ be an operator of the type $\partial^{n}\bm{\Phi}^{\bm{r}}$, meaning that it is composed of $\bm{r}\equiv (r_{1}, \ldots, r_{N})$ powers of $\bm{\Phi}\equiv (\Phi_{1}, \ldots, \Phi_{N})$ and $n$ derivatives distributed amongst them. Then we define its \emph{weight} as
    \[ w[\mathcal{O}] \equiv \bm{\phi}^{\bm{r}}\D^{n}, \]
    where we use notation $\bm{\phi}^{\bm{r}} \equiv \phi_{1}^{r_{1}}\ldots \phi_{N}^{r_{N}}$.
\end{definition}
\begin{remark}
    The weighting scheme in \Cref{def:weight} retains most information about the basis while still having the advantage of reducing the full information. Other weightings can be obtained by a change of parametrization, for example $\D \mapsto t$ and $\phi_{i} \mapsto t^{\Delta[\Phi_{i}]}$ gives us the \emph{mass dimension weighting}, $w[\mathcal{O}] \mapsto t^{\Delta[\mathcal{O}]}$.
\end{remark}

Now, we would like to have an object encoding the information about the basis, which in our case is the number of independent operators with a given weight. The notion of \emph{generating functions} turns out to be exactly what we need.
\begin{definition}[Hilbert series] \label{def:hilbertseries}
    The Hilbert series of a given EFT is a formal series in complex parameters $\D$ and $\bm{\phi}\equiv (\phi_{i})_{i=1}^{N}$ defined by
    \[ H(\bm{\phi},\D) = \sum_{\mathcal{O} \in \B} w[\mathcal{O}] = \sum_{\bm{r}} \sum_{n=0}^{\infty} \opers{r}{n} \bm{\phi}^{\bm{r}}\D^{n}, \]
    where $\B$ is the operator basis and $\opers{r}{n} \equiv d_{r_{1}\ldots r_{N}\mkern 1mu n} \in \N_{0}$ is the number of independent operators in $\B$ with the weight $\bm{\phi}^{\bm{r}}\D^{n}$.

    In a certain domain, usually for $\abs{\D}, \abs{\phi}<1$, the formal series in the definition can be interpreted as a converging series defining a holomorphic function.

    \textcolor{gray}{To obtain simpler expressions, we also include the trivial constant operator with weight $1$, which is the only operator with no field content.}
\end{definition}

\begin{remark}
    Hilbert series are usually defined in the context of graded algebras or their generalizations \cite{Eisenbud1995}. Nonetheless, we will always refer to Hilbert series in the spirit of \Cref{def:hilbertseries}.
\end{remark}

\begin{remark}
    The power of Hilbert series (or generating functions in general) arises in cases where it can be expressed in a closed form. This can be understood either in the context of a formal power series, or (in the domain of convergence) as really the function obtained by summation of the convergent series.
\end{remark}

\begin{remark} \label{rem:hilbertseriesform}
    As will be shown several times, a typical summed up Hilbert series takes the form of
    \[ H=\frac{N}{D}, \quad \text{with }\ D\equiv\prod_{\mathcal{O}\in\mathcal{G}} (1-w[\mathcal{O}]), \]
    where $\mathcal{G}$ is the set of \emph{generators}, which can be multiplied repeatedly to form operators of higher weight. The form of the denominator $D$ reflects this fact, which is seen by expanding it as a geometric series. The numerator $N$ encodes additional operators (which upon repetition decompose to operators generated from $\mathcal{G}$) and possible relations between operators. In the case of a \emph{freely generated} basis, the numerator is unity and all operators can be generated from $\mathcal{G}$.
\end{remark}

\begin{remark} \label{rem:partialhilbertseries}
    Sometimes it is useful to write the Hilbert series as
    \[ H(\bm{\phi},\D) = \sum_{\bm{r}} \bm{\phi}^{\bm{r}} H_{\bm{r}}(\D), \quad\text{where}\quad H_{\bm{r}}(\D) \equiv \sum_{n=0}^{\infty} \opers{r}{n} \D^{n}. \]
    We will refer to $H_{\bm{r}}(\D)$ also by the term Hilbert series, where we are just considering operators with a fixed field content, that is of the type $\partial^{n}\bm{\Phi}^{\bm{r}}$ for a given $\bm{r}$.
\end{remark}

\begin{example} \label{ex:dim1}
    Some examples of the Hilbert series are obtained by considering the theory of $N$ scalar fields in $\d=1$ with the kinetic term
    \[ \mathcal{L}_{\text{kin}}(\{\Phi_{i},\partial\Phi_{i}\}) \equiv \sum_{i=1}^{N} \frac{1}{2} (\partial \Phi_{i})^{2}. \]
    It is particularly simple, because in the absence of the indices there are no Gram determinant conditions, and every operator is automatically invariant. Nevertheless, it will be very instructive.
    For a comprehensive analysis take a look at the work of \textcite{Operator2015}. We will go through some of the edge cases:

    \begin{enumerate}
        \item \emph{No relations.} This case is rather trivial, because the operator basis can be easily guessed, namely it is freely generated by the set $\{\partial^{n}\Phi_{i}\}$ with $i=1,\ldots,N$ and~$n \in \N_{0}$. In other words, every operator is obtained precisely once in the expansion of
              \begin{multline*}
                  \prod_{i=1}^{N} (1+\Phi_{i}+\Phi_{i}^2 + \ldots)(1+ \partial\Phi_{i}+(\partial\Phi_{i})^2 + \ldots)(1+ \partial^2\Phi_{i}+(\partial^2\Phi_{i})^2 + \ldots) \cdots \\[-0.3em]
                  = \prod_{i=1}^{N} \frac{1}{(1-\Phi_{i})(1-\partial\Phi_{i})(1-\partial^2\Phi_{i}) \cdots } = \prod_{i=1}^{N} \prod_{n=0}^{\infty} \frac{1}{1-\partial^{n}\Phi_{i}}.
              \end{multline*}

              The corresponding \emph{free Hilbert series} is obtained by substituting $(\Phi_{i},\partial)$ for their corresponding labels $(\phi_{i},\D)$,
              \[ H_{N}^{\text{free}}(\bm{\phi},\D) = \prod_{i=1}^{N} \prod_{n=0}^{\infty}\frac{1}{1-\D^{n}\phi_{i}}. \]

        \item \emph{Only EOM relations.} Considering the kinetic Lagrangian density $\mathcal{L}_{\text{kin}}$, the relations generated by free equations of motion are $\partial^{2}\Phi_{i}=0$.
              Even simpler than before, because we have $\partial^{2}\Phi_{i}=0 \Rightarrow \partial^{n}\Phi_{i}=0$ for $n\ge 2$, the operator basis is now freely and finitely generated by the set $\{\Phi_{i},\partial\Phi_{i}\}_{i=1}^{N}$, giving us the \emph{EOM Hilbert series} of the form
              \[ H_{N}^{\text{EOM}}(\bm{\phi},\D) = \prod_{i=1}^{N} \frac{1}{(1-\phi_{i})(1-\D\phi_{i})}. \]

        \item \emph{Only IBP relations.} Relations between $\opers{r}{n,\text{free}}$ operators of the type $\partial^{n}\bm{\Phi}^{\bm{r}}$ are formed from $\opers{r}{n-1,\text{free}}$ operators of the type $\partial^{n-1}\bm{\Phi}^{\bm{r}}$ (with one less derivative) by performing a total derivative. One particular example in the case of only one field flavor would be
              \[ 0 \sim \partial\left( (\partial^{n-1}\Phi)\Phi^{r-1} \right) = (\partial^{n}\Phi)\Phi^{r-1} + (r-1)(\partial^{n-1}\Phi)(\partial \Phi) \Phi^{r-2}. \]

              Due to linearity of the derivative, we obtain a linearly independent set of relations from a linearly independent set of operators, with only exception being the trivial constant operator, which does not generate any relation. Therefore, we have ($\delta_{ij}$ denotes the Kronecker delta and $\opers{r}{n}^{\text{free}} \equiv  0$ for $n<0$)
              \begin{equation*}
                  \opers{r}{n}^{\text{IBP}} =
                  \begin{dcases}
                      \opers{r}{n}^{\text{free}} - \opers{r}{n-1}^{\text{free}}         & \bm{r} \neq \bm{0}, \\
                      \delta_{0n} \textcolor{gray}{\ \equiv \opers{0}{n}^{\text{free}}} & \bm{r} = \bm{0}.    \\
                  \end{dcases}
              \end{equation*}
              A simple reordering of the summation gives us
              \begin{equation*}
                  \begin{multlined}
                      H_{N}^{\text{IBP}}(\bm{\phi},\D) = \sum_{\bm{r}} \sum_{n=0}^{\infty} \opers{r}{n}^{\text{IBP}} \bm{\phi}^{\bm{r}}\D^{n}  = 1 + \sum_{\bm{r \neq \bm{0}}} \sum_{n=0}^{\infty} (\opers{r}{n}^{\text{free}} - \opers{r}{n-1}^{\text{free}} ) \bm{\phi}^{\bm{r}}\D^{n} \\
                      = 1+(1-\D)\sum_{\bm{r\neq \bm{0}}} \sum_{n=0}^{\infty} \opers{r}{n}^{\text{free}} \bm{\phi}^{\bm{r}} \D^{n} = \D+(1-\D) H_{N}^{\text{free}}(\D,\{\phi_{i}\}).
                  \end{multlined}
              \end{equation*}
              Again, the \emph{IBP Hilbert series} is given by a straightforward modification of the \emph{free Hilbert series} as
              \[ H_{N}^{\text{IBP}}(\bm{\phi},\D) = \D + (1-\D)\prod_{i=1}^{N} \prod_{n=0}^{\infty}\frac{1}{1-\D^{n}\phi_{i}} . \]

        \item \emph{Both EOM and IBP relations.} Interplay of both types of relations makes this problem harder (and therefore interesting), but it is still possible to find analytic expression using techniques similar to those we will present in the following chapters \autocite{Operator2015}. We do not have the space to cover it here, but it is advised for the reader to take a look at its solution (mainly Section 4 of \cite{Operator2015}), probably best as a complement to our \Cref{ch:masterformuladerivation}.

              In fact, this was really meant as a training toy model for exploring the simplest possible but already non-trivial computation of an operator basis including both EOM and IBP relations.
    \end{enumerate}

\end{example}

A natural question arises:
\begin{center}
    \emph{How do we obtain the Hilbert series in a more general setting, for an arbitrary space-time dimension and field content with possible internal symmetries?}
\end{center}

So far we have only encountered the case with $\d=1$ and no internal symmetries. Only in this dimension the Lorentz group is trivial and the application of the derivatives is always unambiguous. From now on, we always assume $\d\ge 2$, where each derivative carries an index with non-trivial transformation properties.

To construct a Lorentz invariant operator, we are forced to somehow contract all the indices, and usually there are multiple possibilities. A similar thing goes for internal symmetries, both additions bringing \emph{substantial complexity}. It suffices to say that it is far from obvious how to generalize the discussion in \ref{ex:dim1}.

Nonetheless, the toy model in \ref{ex:dim1} has proved to be full of inspiring lessons, mainly case \textbf{(4)}, paving the way for vast generalization developed by the same authors --- \textcite{Operator2017}. In the next chapter, we will start the preparatory work essential for a systematic treatment of our problem.

\chapter{Representations of Lie Groups} \label{ch:grouptheory}

\emph{Groups} are omnipresent in the realm of mathematics and physics, and so are \emph{manifolds} in the context of field theories. It is impossible to go even past the proper formulation of the physical theory without specifying \emph{transformation properties} of the present particle fields. Since everything plays out on a smooth manifold, the corresponding symmetry groups are themselves often smooth manifolds.

Groups and their \emph{representations} will precisely turn out to be the tools we need to systematically count the Hilbert series. The idea is that by continuously being aware of the transformation properties --- representations --- of different operators, we can automatically rule out operators that generate GDC conditions, and also project out the number of only invariant operators. Furthermore, we will learn to project out the number of operators corresponding to any representation, which will be crucial for addressing IBP relations.

In \textbf{\Cref{sec:liegroup}} we will introduce the notion of \emph{Lie groups} as well as present particular examples we are interested in. Moreover, by studying their differential structure, we will be able to show unique existence of the \emph{invariant Haar integral} on \emph{compact} Lie groups, on which much of the following results build upon.

In \textbf{\Cref{sec:representations}} we will start the study of group \emph{representations}. By borrowing concepts from linear algebra we will be able to obtain a wide range of representations describing various transformation properties of different objects.

In \textbf{\Cref{sec:irreducibilityschurlemma}} we will dive deeper into representation theory with the goal to systematically examine the structure of representations. We will introduce the notion of \emph{irreducible} representations and show that any finite-dimensional representation of a compact Lie group \emph{decomposes} to such atomic pieces.

In \textbf{\Cref{sec:charactersprojectionformula}} we will finish our endeavor to find the \emph{multiplicities} of irreducible representations in any given representation by deriving the \emph{projection formula}. It works by assigning \emph{characters} to corresponding representations and integrating over the group, utilizing the fact that \emph{irreducible} characters turn out to be orthonormal.

In \textbf{\Cref{sec:gradedrepresentations}} we will extend our knowledge to \emph{graded representations}, which enable us to work with a countable number of representations in parallel, since they are in direct relation with generating functions.

In \textbf{\Cref{sec:weylintegrationformula}} we will once again return to a direct study of compact Lie groups. We will introduce numerous important concepts, such as \emph{maximal torus} and \emph{roots} of a group, ultimately leading to the \emph{Weyl integration formula}, which provides us with explicit means of performing the integral in the projection formula.

We only introduce important notions from group and representation theory, which shall prove useful for our problem. This chapter is mainly based on the beautiful works of \textcite{BrockerDieck1985, Fulton1999, Sepanski2007}.

\section{Compact Lie groups} \label{sec:liegroup}

Lie groups are essentially a harmonious combination of the algebraic structure of a group and the analytic and geometric structure of a differentiable manifold. They underlie the fundamental theory describing continuous transformations.

\begin{definition}[Lie group]
    A \emph{Lie group} $G$ is a \emph{group} that is also a \emph{differentiable manifold}, for which the group multiplication $m\colon G\times G\to G$ is a \emph{smooth} map.
\end{definition}
\begin{remark}
    It can be shown that the definition of a Lie group already implies also the smoothness of the inverse map $\iota:G\to  G, g\mapsto g^{-1}$ \cite{BrockerDieck1985}.
\end{remark}
\begin{remark}
    We will use $e$ or $\bm{1}$ to denote the identity of a group $G$. For ease of notation we write just $gh$ instead of $m(g;h)$ for any $g,h \in G$.
\end{remark}

Some Lie groups can have \emph{similar structures}. To get a grasp of what it means, we introduce the elementary notion of \emph{homomorphisms}.
\begin{definition}[Homomorphism of Lie groups] \label{def:homomorphism}
    A \emph{homomorphism of Lie groups} is a smooth map $\phi \colon G \to H$ between two Lie groups $G$ and $H$, such that the \emph{group operation is preserved}, that is
    \[ \phi(g h) = \phi(g)\phi(h) \quad\text{for all } g,h \in G, \]
    where the multiplication on the left side is in $G$ and on the right side in $H$.
\end{definition}

\begin{remark} \label{rem:isomoporhism}
    There always exists a \emph{trivial homomorphism}, which maps every element of $G$ to the identity element of $H$. If homomorphism between groups is non-trivial, it is an indication they share certain similarities. A bijective homomorphism is called an \emph{isomorphism}, the corresponding groups are then called \emph{isomorphic} ($G \cong H$), which means they are identical up to differences in the notation of their elements.
\end{remark}
\begin{example}
    Every finite-dimensional \emph{vector space} $V$ with its additive group structure is a Lie group. By any choice of basis it is isomorphic to $\R^{\dim V}$.
\end{example}

We can construct new groups by a simple combination of already known ones. \vspace{-0.5em}
\begin{definition}[Direct product of Lie groups] \label{def:groupsdirectproduct}
    Let $G$ and $H$ be Lie groups. Then their \emph{direct product} $\mathsf{G} \equiv G\times H$ is also a Lie group, where we understand it both as a direct product of their manifold structures and their group structures. Namely, the group multiplication of $\mathsf{g}_1,\mathsf{g}_{2} \in \mathsf{G}$ is defined component-wise via
    \[ \mathtt{g}_{1}\mathtt{g}_{2}\equiv (g_1;h_1)(g_2;h_2)\equiv (g_1 g_2;h_1 h_2), \quad\text{where } g_1,g_2 \in G \text{ and } h_1,h_2 \in H. \]
\end{definition}

\begin{example} \label{ex:torus}
    The \emph{torus} $\mathbb{T}^{n}$ ($n\in\N$) defined by  \vspace{-0.7em}
    \[ \mathbb{T}^{n} \equiv \R^{n}/\Z^{n}=(\R/\Z)^{n}\cong (S^{1})^{n}\equiv \overbrace{S^{1} \times \cdots \times S^{1}}^{n} \]
    is a Lie group, where the circle group $S^{1}\equiv \left\{z \in \C \middle| \abs{z}=1 \right\}$ is the unit circle in $\C$ viewed as a multiplicative subgroup of $\C^{\times}\equiv \C \setminus \{0\}$. The isomorphism $\R/\Z \cong S^{1}$ is given by the map $\R/\Z \to S^{1}\colon t\mapsto \eunit^{2 \pi \iunit t}$.
\end{example}

\begin{figure}[!ht]
    \centering
    \includegraphics[width=0.8\linewidth]{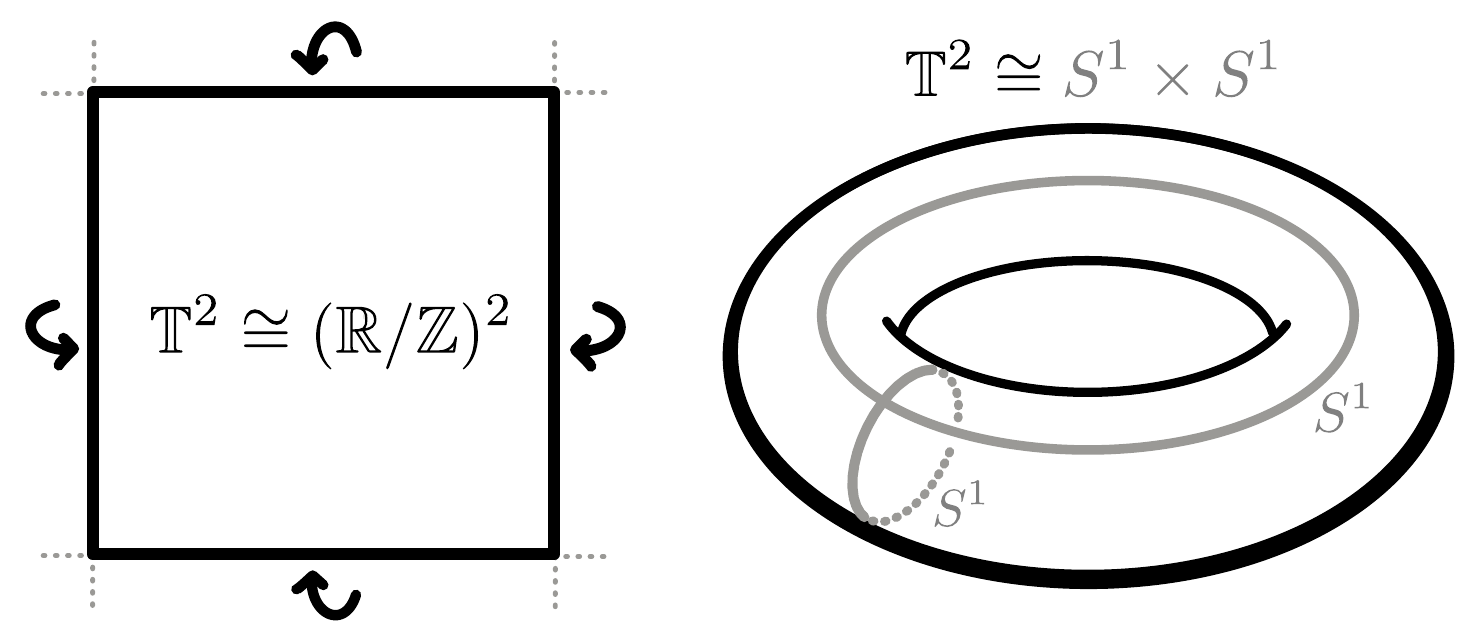}
    \caption{Illustration of alternative isomorphic definitions of the torus $\mathbb{T}^{2}.$}
    \label{fig:torus}
\end{figure}

\begin{example}[General linear groups] \label{ex:generallinear}
    Let $V$ be a finite-dimensional vector space over $\R$ or $\C$. By $\End(V)$ we denote the space of \emph{endomorphisms} of $V$, that is the vector space of linear maps from $V$ to itself. Very important is the set $\GL(V) \subset \End(V)$ of \emph{automorphisms} (invertible endomorphisms) of $V$. Equivalently, we have
    \[ \GL(V) = \left\{ A \in \End(V) \middle| \det A \neq 0 \right\}. \]
    The determinant is a continuous function, thus $\GL(V)$ is an open subset of $\End(V)$. It is easy to see that $\GL(V)$ has the structure of a differentiable manifold and in coordinates the group operation is the matrix multiplication, which is smooth. $\GL(V)$ is therefore a Lie group. In particular, we have the \emph{General Linear} groups
    \[ \GL(n,\R)\equiv \GL_{\R}(\R^{n}) \qq{and} \GL(n,\C)\equiv \GL_{\C}(\C^{n}), \]
    which are canonically isomorphic to the groups of invertible $n\times n$ matrices over $\R$ and $\C$, respectively. Thus, we can think of $\GL(n,\R)$, $\GL(n,\C)$, and their subgroups (see \Cref{ex:compactliegroups}) as \emph{matrix groups} with the operation of matrix multiplication.
\end{example}
%

\begin{example}[\textcolor{gray}{Special} Orthogonal and Unitary groups] \label{ex:compactliegroups}
    We will be mostly interested in compact Lie groups, which will represent our (Euclidean) Lorentz and gauge groups. Since a closed subgroup of a Lie group inherits the Lie group structure, we obtain the following \emph{compact classical groups} \cite{BrockerDieck1985}:
    \begin{enumerate}[\bfseries (a)]
        \item The \emph{Orthogonal} $\OO(n)$ and the \emph{Special Orthogonal} $\SO(n)$ groups defined by
              \[ \OO(n) \equiv \left\{ g \in \GL(n,\R) \middle| g^{\intercal}g=\Id \right\}, \quad \SO(n) \equiv \left\{ g \in \OO(n) \middle| \det g = 1 \right\}, \]
              where $g^{\intercal}$ denotes the transpose of $g$. Group $\OO(n)$ splits into two connected components with the values $\pm 1$ of the determinant, one being $\SO(n)$.
        \item The \emph{Unitary} $\U(n)$ and the \emph{Special Unitary} $\SU(n)$ groups defined by
              \[ \U(n) \equiv \left\{ g \in \GL(n,\C) \middle| g^{+}g=\Id \right\}, \quad \SU(n) \equiv \left\{ g \in \U(n) \middle| \det g = 1 \right\}, \]
              where $g^{+}\equiv \overline{g^{\intercal}}$ denotes the conjugate transpose of $g$. It can be shown that both $\U(n)$ and $\SU(n)$ are connected.
    \end{enumerate}
    These groups preserve the standard inner products on $\R^{n}$ and $\C^{n}$, particularly the norms. Thus, it can be seen they are compact, since they are closed and bounded in the finite-dimensional vector spaces $\End(\R^{n})$ and $\End(\C^{n})$.


\end{example}
\begin{remark}[Symplectic groups]
    There is one other compact classical Lie group family, namely the \emph{symplectic} groups
    \[ \Sp(n) \equiv \left\{ g \in \GL(n,\mathbb{H}) \middle| g^{*}g=\Id \right\}, \]
    where $\mathbb{H}$ denotes the \emph{quaternions} and $g^{*}$ denotes the quaternionic conjugate transpose of $g$.
    Since understanding a noncommutative division algebra $\mathbb{H}$ takes some work and we will not need the symplectic groups, we refer the interested reader to practically any book on the subject of Lie groups \cite{BrockerDieck1985, Fulton1999, Sepanski2007}.
\end{remark}
\begin{remark} \label{rem:bilinearformonSOgroup}
    More generally, classical groups are defined as automorphism groups that preserve a \emph{bilinear} or \emph{sesquilinear} form on finite-dimensional vector spaces over $\R$, $\C$, or $\mathbb{H}$. For example, $\OO(n)$ is defined as a subgroup of $\GL(n,\R)$ preserving some symmetric positive-definite bilinear form $Q\colon \R^{n}\times \R^{n} \to \R$, that is
    \[ Q(g \bm{v}, g \bm{w}) = Q(\bm{v},\bm{w}) \quad\text{for any } g \in \OO(n) \text{ and } \bm{v},\bm{w} \in \R^{n}.\]
    Since we can always bring such form $Q$ by a change of basis to the standard scalar product on $\R^{n}$, we precisely obtain the definition in \textbf{(a)} of \Cref{ex:compactliegroups}.
\end{remark}

What can we reveal about the Lie groups (namely the compact ones) by studying the interplay of their algebraic and analytic structures? Perhaps not so surprisingly, many strong results are just waiting to be uncovered. We will start by introducing the tensor fields which are adjusted to the group structure of a Lie group.

\begin{definition}[Left translations, left-invariant fields] \label{def:invariantfields}
    Let $G$ be a Lie group. The \emph{left translation} by $g \in G$ is a diffeomorphism $L_{g}\colon G\to G$ defined by $L_{g}h \equiv gh$ for any $h \in G$. The inverse is given by $L_{g}^{-1} = L_{g^{-1}}$.
    %

    A tensor field (section of the tensor bundle) on $G$ is called \emph{left-invariant} if it is invariant under the induced action of left translations. That is, a left invariant tensor field $\bm{A} \in \Sect\bm{T}^{p}_{\,q}G$ satisfies
    \[ L_{g*}\bm{A}=\bm{A} \quad\text{for every } g \in G, \]
    where $L_{g*}$ denotes the pushforward by $L_{g}$.
\end{definition}
\begin{remark}
    We implicitly used the fact that $L_{g}$ is a diffeomorphism, thus we can pushforward the whole vector field to obtain another vector field. We also extend the definition of pushforward to forms by $L_{g*}\equiv (L^{*}_{g})^{-1}= L^{*}_{g^{-1}}$.
\end{remark}
\begin{remark}
    Analogously we can define the \emph{right translations} by $R_{g}h \equiv hg$ for any $g,h\in G$, and the corresponding right-invariant tensor fields.
\end{remark}

It follows that left-invariant vector (and more generally tensor) fields are uniquely given by their value at the identity.
\begin{prop}[Left-invariant lift] \label{prop:lefinvariantlift}
    The tensor space ${\bm{T}_{e}}^{p}_{q}G$ at the identity of a Lie group $G$ is \emph{canonically isomorphic} to the space of the left-invariant tensor fields $\Sect_{L}{\bm{T}}^{p}_{\,q}G \subset \Sect{\bm{T}}^{p}_{\,q}G$ on $G$. We denote this isomorphism by
    \begin{align*}
        \bm{\ell}_{\bullet}\colon {\bm{T}_{e}}^{p}_{q}G & \to \Sect_{L}{\bm{T}}^{p}_{\,q}G \\
        \bm{a}                                          & \mapsto \bm{A}\equiv \bm{\ell}_{\bm{a}},
    \end{align*}
    and it is defined by $\bm{\ell}_{\bm{a}}|_{g}\equiv L_{g*}\bm{a}$.
\end{prop} \vspace{-0.5em}
\begin{myproof}
    To any left-invariant tensor field $\bm{A}$ we can assign uniquely a tensor $\bm{A}|_{e}$ at the identity, which follows from (we use the left-invariance of $\bm{A}$)
    \[ \bm{A}|_{g}=(L_{g*}\bm{A})|_{g}\equiv L_{g*}(\bm{A}|_{e}) \equiv \bm{\ell}_{\bm{A}|_{e}}|_{g} \quad\Longrightarrow\quad \bm{A}=\bm{\ell}_{\bm{A}|_{e}}.\]
    The map $\bm{A}|_{e} \xmapsto{\bm{\ell}} \bm{A} $ is evidently one-to-one and linear, thus an isomorphism.
\end{myproof}
\begin{remark} \label{ref:volumeformorientability}
    Since the space of top-dimensional forms $\bm{\Lambda}^{\!\d}_{e}G$ at the identity of $G$ is one-dimensional, and the pushforward of a nonzero form by a diffeomorphism $L_{g}$ leads to a nonzero form, we obtain a unique (up to multiplication by a constant) left-invariant \emph{volume form} on $G$. Specially, every Lie group is orientable.
\end{remark}

Now we restrict ourselves to the case of a compact Lie group $G$, where it is possible to perform integration over the whole $G$ without any problem. We will show there actually exists a unique normalized \emph{invariant Haar integral}.

\begin{theorem}[Invariant Haar integral] \label{th:haarmeasure}
    Let $G$ be a compact Lie group and $h\in G$ be any element of $G$. Denote by $C(G)$ the vector space of continuous real-valued functions on $G$. Then there exists the \emph{invariant Haar integral}
    \begin{align*}
        \int\colon C(G) & \to \R \\
        f               & \mapsto \int f \equiv \int_{G}f(g)\dd{g},
    \end{align*}
    and is uniquely determined by the following properties:
    \begin{enumerate}[\bfseries (a)]
        \item It is linear, monotone, and \emph{normalized}, that is $\displaystyle\int_{G}\dd{g} = 1$.
        \item It is \emph{left invariant}, that is $\displaystyle\int_{G} f(hg)\dd{g} \equiv \int_{G} f \circ L_{h}(g)\dd{g} = \int_{G} f(g) \dd{g}$.
    \end{enumerate}
\end{theorem}
\begin{myproof}
    Existence follows from \Cref{ref:volumeformorientability}, giving us a volume form $\bm{\omega}$ on $G$, with respect to which we define the integral. Property \textbf{(a)} is then satisfied by properly rescaling the $\bm{\omega}$. Property $\textbf{(b)}$ follows from the left-invariance of $\bm{\omega}$ by
    \[\int_{G}\!f\circ L_{h} \dd{g} \equiv \int_{G} (f \circ L_{h}) \bm{\omega} = \int_{G} (f \circ L_{h}) L_{h}^{*}\bm{\omega} = \int_{G} \!L_{h}^{*}(f\bm{\omega}) = \int_{L_{h}G\equiv G} \!f\bm{\omega}\equiv \int_{G} \!f \dd{g}, \]
    where we used that $L_{h}$ is orientation preserving (since $L_{h}^{*}$ preserves $\bm{\omega}$).

    We will prove uniqueness by comparing all left-invariant integrals with a single right-invariant integral (can be constructed similarly as the left-invariant one).
    Let $f \mapsto \int_{G} f(g) \dd{g}$ be any integral with the properties from the definition, and $f\mapsto \int_{G} f(g) \delta g$ is a given right-invariant normalized integral, that is
    \[ \int_{G} \delta g=1 \qq{and} \int_{G} f(gh) \delta g =  \int_{G} f(g) \delta g. \]
    Then for any $f \in C(G)$ by use of left-invariance/right-invariance, normalization, and \emph{Fubini's theorem} to exchange the order of integration, we obtain
    \begin{alignat*}{2}
        \int_{G} f(g) \dd{g} & = \int_{G} \left( \int_{G} f(\phantom{h}g) \dd{g} \right)  \delta h &  & = \int_{G} \left( \int_{G} f(hg) \dd{g} \right)  \delta h \\
                             & = \int_{G} \left( \int_{G} f(hg) \delta h \right) \dd{g}            &  & = \int_{G} \left( \int_{G} f(h\phantom{g}) \delta h \right) \dd{g} = \int_{G} f(h) \delta h.
    \end{alignat*}
    This is sufficient to establish uniqueness.
\end{myproof}
\vspace{-0.5em}
\begin{corollary}[Properies of invariant Haar integral] \label{cor:propertiesHaar}
    Let $G$ be a compact Lie group, $h\in G$ be any element of $G$, and $\phi$ be any automorphism $\phi:G\to G$ (isomorphism from $G$ to itself). Then the \emph{invariant Haar integral} satisfies
    \[ \int_{G}f(g)\dd{g} \overset{\textbf{(1)}}{=} \int_{G}f(hg)\dd{g} \overset{\textbf{(2)}}{=}  \int_{G}f(gh)\dd{g} \overset{\textbf{(3)}}{=} \int_{G}f(g^{-1})\dd{g} \overset{\textbf{(4)}}{=} \int_{G}f\circ \phi(g)\dd{g}.\]
\end{corollary}\vspace{-0.7em}
\begin{myproof}
    Suppose $\varphi:G\to G$ is a diffeomorphism, then the map $f\mapsto \int_{G}f\circ \varphi(g) \dd{g}$ is clearly a normalized integral. If it happens to also be left-invariant, that is $\int_{G} f \circ L_{h} \circ \varphi(g) \dd{g}=\int_{G} f\circ \varphi(g) \dd{g}$, from uniqueness (see \Cref{th:haarmeasure}) it must be that $\int_{G} f \circ \varphi(g) \dd{g}=\int_{G} f(g) \dd{g}$. We apply this idea $\bm{(\star)}$ to prove:
    \begin{enumerate}
        \item Follows directly from the definition of the invariant Haar integral.
        \item Choose $\varphi(g) \equiv R_{h}(g)\equiv gh$ and any $k\in G$. Left-invariance follows from
              \[ \int_{G} f\circ L_{k} \circ R_{h}(g) \dd{g} = \int_{G} f(kgh) \dd{g} = \int_{G} f \circ R_{h}(kg) \dd{g} \overset{\textbf{(1)}}{=} \int_{G} f \circ R_{h}(g) \dd{g}. \]
        \item Choose $\varphi(g) \equiv \iota(g) \equiv  g^{-1}$. We can use the already proved \textbf{(2)} to show
              \[ \int_{G} f\circ L_{h} \circ \iota(g) \dd{g} = \int_{G} f(h g^{-1}) \dd{g} = \int_{G} f \circ \iota(g h^{-1})  \dd{g} \overset{\textbf{(2)}}{=} \int_{G} f \circ \iota(g) \dd{g}. \]

        \item Choose $\varphi(g) \equiv \phi(g)$. By homomorphism of $\phi$ (see \Cref{def:homomorphism}) we have
              \[ \int_{G} f\circ L_{h} \circ \phi(g) \dd{g} = \int_{G} f(h \phi(g)) \dd{g} = \int_{G} f \circ \phi(\phi^{-1}(h) g) \dd{g} \overset{\textbf{(1)}}{=} \int_{G} f \circ \phi(g) \dd{g}. \]
    \end{enumerate}
    Application of the idea $\bm{(\star)}$ finishes the proof.
\end{myproof}

\begin{remark}[Haar measure]
    We sometimes refer to $\dd{g}$ figuring in the invariant Haar integral as the (invariant) \emph{Haar measure}.
\end{remark}
\begin{remark}
    Every \emph{finite group} is actually a zero-dimensional compact Lie group (with discrete topology). Everything what we will present about representations applies to finite groups with the appropriate transition from integrals to sums. Historically, it was knowledge about finite groups that was extended to the special case of compact Lie groups by existence of the Haar measure.
\end{remark}

The importance of the invariant Haar integral on compact Lie groups can not be understated. We will fundamentally use it in \Cref{sec:irreducibilityschurlemma,sec:charactersprojectionformula} to obtain invariant objects by averaging over the group.

For now, we leave the direct analysis of the group structure, as we will approach them somewhat indirectly through their \emph{representations}. We will return once more to their differential geometry in \Cref{sec:weylintegrationformula}.

\section{Representations} \label{sec:representations}

Groups themselves can be pretty abstract entities. However, their properties can be probed by examining their action on certain simple structures we understand very well, in our case \emph{vector spaces}. This is the subject of \emph{representation theory}, which in a \emph{group structure preserving} way assigns to every group element a linear transformation of a vector space. In essence, it is able to reduce some problems in abstract algebra to problems in linear algebra.

\begin{definition}[Representation] \label{def:representation}
    A \emph{representation} of a Lie group $G$ on a finite-dimensional complex vector space $V$ is a \emph{Lie group homomorphism} $\rho\colon G\to\GL(V)$. The \emph{dimension of the representation} $V$ is $\dim V \equiv  \dim_{\C}V$.
\end{definition}

\begin{figure}[!ht]
    \centering
    \includegraphics[width=1\linewidth]{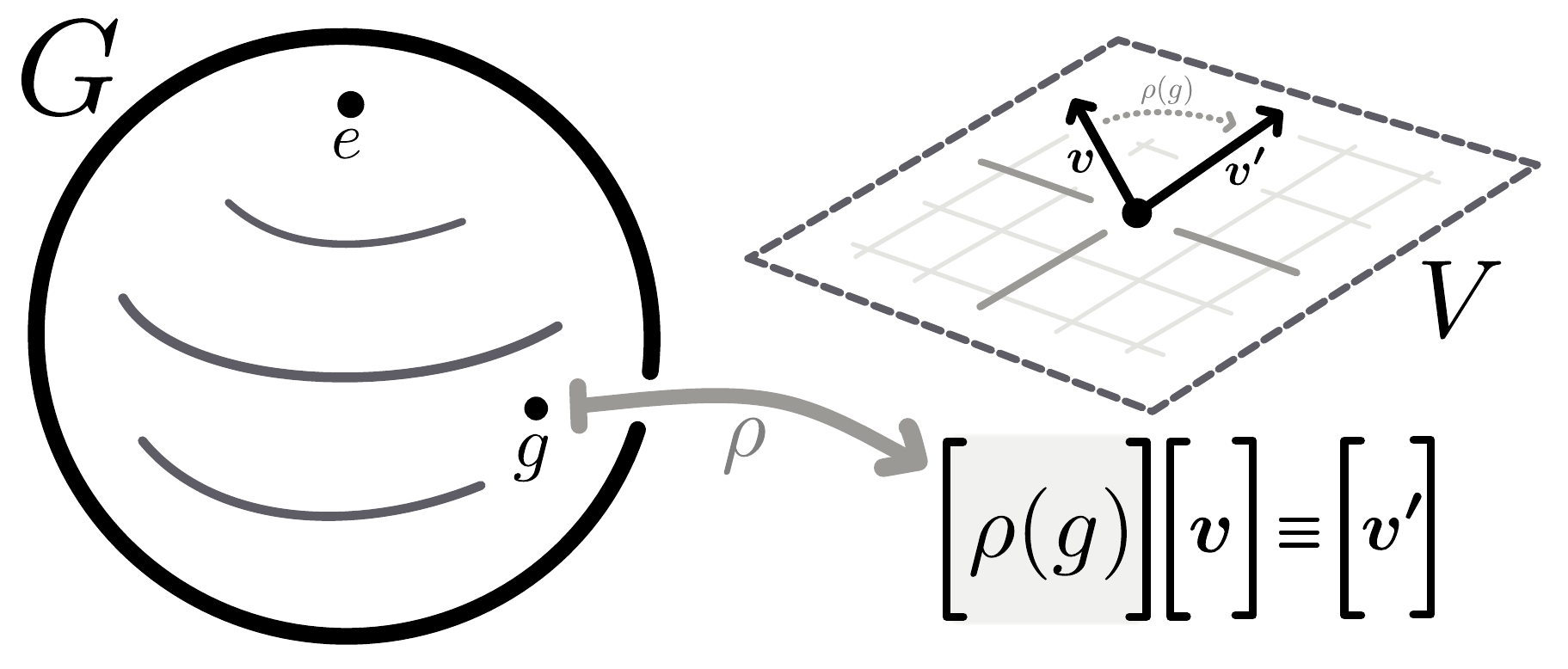}
    \caption{Illustration of a representation $\rho$ of group $G$ on vector space $V$. A group element $g \in G$ is represented by a linear operator $\rho(g) \in \mathsf{GL}(V)$, in particular transforming a vector $\bm{v}\in V$ to the vector $\bm{v}'\equiv \rho(g)(\bm{v}) \in V$.}
    \label{fig:grouprepresentationfinal}
\end{figure}

\begin{remark}
    When there is little ambiguity about the map $\rho$, we call $V$ itself a representation of $G$. In that case we say that $g$ has an \emph{action} on $V$. We often write $g|_{V}$ for $\rho(g)$ and directly $g\bm{v}$ for $\rho(g)(\bm{v})$, where $\bm{v} \in V$. For a visual representation of the notion of group representations see \Cref{fig:grouprepresentationfinal}.
\end{remark}

\begin{remark}
    We could extend the definition to infinite-dimensional vector spaces or to vector spaces over different fields $\F$, for example $\R$. Since we will not find much use for such generalizations, and sometimes we will distinctly use finite-dimensionality and properties of complex numbers, we assume finite-dimensional representations over $\C$ unless stated differently.
\end{remark}

\begin{example}
    Some elementary representations appear automatically:
    \begin{enumerate}[\bfseries (a)]
        \item \emph{Trivial representation} --- every group element is represented by an identity operator on any given vector space $V$. Although this does not seem exciting at first sight, we are frequently interested in objects which are invariant under the action of the group.
        \item \emph{Standard representation} --- the \emph{matrix} groups $\GL(n,\R)$, $\GL(n,\C)$, and their subgroups (see \Cref{ex:generallinear,ex:compactliegroups}) are canonically represented by matrices on the corresponding complex vector space $\C^{n}$ (we understand $\R \subset \C$).
    \end{enumerate}
\end{example}

To enlarge our collection of representations, we will borrow some concepts from linear algebra, which allow us to build more elaborate representations out of known simpler ones. Typical and the most basic constructions are the direct sum and the tensor product of vector spaces.
\begin{definition}[Direct sum and tensor product of representations] \label{def:dirsumtensprodrepre}
    Let $V$ and $W$ be representations of a group $G$. Then the \emph{direct sum} $V\oplus W$ and the \emph{tensor product} $V\otimes W$ are representations with the actions of any $g \in G$ defined through relations
    \[ g(\bm{v}\oplus \bm{w})\equiv g\bm{v} \oplus g\bm{w}, \quad g(\bm{v} \otimes \bm{w})\equiv g\bm{v} \otimes g\bm{w}, \]
    for any $\bm{v} \in V$ and $\bm{w} \in W$.
\end{definition}

We can also induce representation of $V$ on the dual space $V^{*}\equiv \Hom(V,\C)$ of linear functionals on $V$ by respecting the duality $\langle \bm{\alpha},\bm{v}\rangle \equiv \bm{\alpha}(\bm{v})$, that is by requiring $\langle g \bm{\alpha},g \bm{v}\rangle \equiv \langle \rho^{*}(g)(\bm{\alpha}),\rho(g)(\bm{v})\rangle=\langle \bm{\alpha},\bm{v}\rangle $ for all $g\in G,\bm{\alpha}\in V^{*}$ and $\bm{v} \in V$.
\begin{definition}[Dual representation] \label{def:dualrepre}
    Let $V$ be a representation of a group $G$. Then the \emph{dual representation} $\rho^{*}\colon G\to \GL(V^{*})$ is defined by
    \[ \rho^{*}(g)=\rho(g^{-1})^{\intercal} \quad\text{for any } g \in G. \]
\end{definition}

Now that we have already defined the dual representation and the tensor product of representations, we can similarly realize a representation on $\Hom(V,W)$, the space of linear maps between vector spaces $V$ and $W$ (which themselves are representations), motivated by the canonical isomorphism $\Hom(V,W) \cong V^{*} \otimes W$ for finite-dimensional $V$ and $W$.
\begin{definition}[$\Hom(V,W)$ representation] \label{def:homrepre}
    Let $V$ and $W$ be representations of a group $G$. Then $\Hom(V,W) \cong V^{*} \otimes W$ is a representation with the action of any $g \in G$ defined through the relation
    \[ (g \varphi)(\bm{v}) \equiv g\varphi(g^{-1}\bm{v}) \quad\text{ for any } \varphi \in \Hom(V,W), \bm{v} \in V. \]
\end{definition}

It is often interesting to study maps between representations, which preserve their structure (in a similar sense as homomorphisms preserve group structure).

\begin{definition}[$G$-map, $\Hom_G(V,W)$, equivalent representations] \label{def:g-map}
    A \emph{$G$-map} (\emph{$G$-linear} map) is a linear map $\varphi\colon V\to W$ between representations $V$ and $W$ of a group $G$, such that it preserves the group representation structure, that is
    \[ \varphi(g \bm{v}) = g \varphi(\bm{v}) \quad\text{for all $g\in G, \bm{v} \in V$}. \]
    We denote by $\Hom_{G}(V,W)$ the space of all $G$-linear maps between $V$ and $W$.

    Representations $V$ and $W$ are \emph{equivalent} ($V\cong W$) if there exists a bijective $G$-map between them, which means they are identical up to a change of basis.
\end{definition}

\begin{remark} \label{rem:homG}
    Comparing \Cref{def:homrepre,def:g-map} we observe that $\Hom_{G}(V,W)$ is just a subspace of $\Hom(V,W)$ that transforms trivially under the action of $G$.
\end{remark}

\begin{remark}
    Previous definitions are equivalent to the statement, that the following diagrams are commutative for every $g \in G$: \vspace{-0.5em}
    \begin{figure}[!ht]
        \centering
        \begin{subfigure}[c]{0.25\linewidth}
            \centering
            \begin{tikzcd}
                V \arrow[r, "\varphi"] \arrow[d, swap, "g"]& \C \arrow[d, "\Id"] \\
                V \arrow[r, swap, "\varphi"]         & \C
            \end{tikzcd}
            \caption{$V^{*}\equiv \Hom(V,\C)$}
        \end{subfigure}
        \begin{subfigure}[c]{0.25\linewidth}
            \centering
            \begin{tikzcd}
                V \arrow[r, "\varphi"] \arrow[d, swap, "g"]& W \arrow[d, "g"] \\
                V \arrow[r, swap, "g\varphi"]        & W
            \end{tikzcd}
            \caption{$\Hom(V,W)$}
        \end{subfigure}
        \begin{subfigure}[c]{0.25\linewidth}
            \centering
            \begin{tikzcd}
                V \arrow[r, "\varphi"] \arrow[d, swap, "g"]& W \arrow[d, "g"] \\
                V \arrow[r, swap, "\varphi"]         & W
            \end{tikzcd}
            \caption{$\Hom_{G}(V,W)$}
        \end{subfigure}
        \caption{Commutative diagrams corresponding to \Cref{def:dualrepre,,def:homrepre,def:g-map}.}
    \end{figure}
\end{remark} \vspace{-2em}

\section{Complete reducibility and \texorpdfstring{Schur's}{Schur’s} Lemma} \label{sec:irreducibilityschurlemma} \vspace{-0.5em}
We have already encountered numerous possibilities, how we can construct other representations from already known ones, most simply by taking the direct sum. Steps in the other direction would be of similar if not greater interest. If we were able to systematically decompose any representation to its smallest pieces, understanding just them would be enough, leading to a major simplification.

In order to decompose a given representation $V$, there necessarily must exist some proper subspace of $V$ that is somehow closed under the action of a group. We can create another representation by restricting the group on such a subspace. \vspace{-0.5em}
\begin{definition}[Subrepresentation]
    A vector subspace $W$ of a representation $V$ of a group $G$ is called \emph{$G$-invariant} or a \emph{subrepresentation} if $g \bm{w} \in W$ for every $g \in G$ and $\bm{w} \in W$. Thus, $W \le  V$ itself is a representation of $G$.
\end{definition}

Now we define ``atomic'' representations, in which the action of the group already mixes up the whole space, therefore they can not be decomposed any further.
\begin{definition}[Irreducible representation]
    A nonzero representation $V$ is called \emph{irreducible} if the only subrepresentations are the zero space $\{\bm{0}\}$ and the whole $V$.
    Otherwise, $V$ is called \emph{reducible}.
\end{definition}

It would be ideal if any representation could be decomposed into irreducible representations. In the case of general Lie groups, this is not the case. However, as we will show later, representations of compact Lie groups satisfy this property.
\begin{definition}[Complete reducibility]
    A representation $V$ of a group $G$ is called \emph{completely reducible} if it is a direct sum of irreducible representations.
\end{definition}

But first, to have a glimpse of how useful is to work with irreducible representations, we present an essential tool in Representation Theory as a whole.
\begin{theorem}[Schur's lemma] \label{th:schur}
    Let $V$ and $W$ be irreducible representations of a group $G$, and $\varphi\colon V \to W$ be a $G$-map between them. Then the following results hold:
    \begin{enumerate}
        \item Either $\varphi$ is an isomorphism, or $\varphi=\bm{0}$.
        \item If $V=W$, then $\varphi = \lambda\Id$ for some $\lambda \in \C$.
    \end{enumerate}
\end{theorem}
\begin{myproof}
    Subspaces $\Ker \varphi$ and $\Im \varphi$ of a $G$-map $\varphi$ are $G$-invariant, which follows directly from \Cref{def:g-map}. Claim \textbf{(1)} then follows from irreducibility of $V$ and $W$, because only possibilities are that either $\Ker \varphi = \{\bm{0}\}$ and $\Im \varphi = W$, or that $\Ker \varphi = V$ and $\Im \varphi = \{\bm{0}\}$.

    Since $V$ is a finite-dimensional vector space over the algebraically closed field $\C$, there exists a solution $\lambda \in \C$ of the equation $\det( \varphi - \lambda \Id ) = 0$. Claim \textbf{(2)} then follows from \textbf{(1)} applied to a $G$-map $\varphi - \lambda \Id$, giving us $\varphi - \lambda \Id = 0 \Longrightarrow \varphi = \lambda \textbf{1}$.
\end{myproof}

\begin{corollary}[Dimension of $\Hom_G$-space between irreducible representations] \label{cor:schur}
    Let $V$ and $W$ be two irreducible representations of a group $G$. Then
    \[ \dim \Hom_{G}(V,W) =
        \begin{dcases}
            \,1 \quad\text{ if } V\cong W, \\
            \,0 \quad\text{ if } V\ncong W.
        \end{dcases} \]
\end{corollary}
\begin{myproof}
    From the claim \textbf{(1)} of \Cref{th:schur} follows, that there exists a nonzero $\varphi \in \Hom_{G}(V,W)$ if and only if $V\cong W$.

    In the case $V\cong W$ we fix a bijective $\varphi_0 \in \Hom_{G}(V,W)$. If also $\varphi \in \Hom_{G}(V,W)$, then $\varphi \circ \varphi_0^{-1} \in \Hom_{G}(V,V)$, and from the claim \textbf{(2)} of \Cref{th:schur} follows
    \[ \varphi \circ \varphi_0^{-1}= \lambda \Id \quad\text{for some } \lambda\in\C  \quad\Longrightarrow\quad \Hom_{G}(V,W)=\C \varphi_{0}.\qedhere  \]
\end{myproof}

Now we embark on the journey of proving the complete reducibility of compact Lie group representations. After that, \Cref{th:schur} and mainly \Cref{cor:schur} will result in some far-reaching consequences.

We will exploit the existence of the invariant Haar integral to construct an inner product compatible with the structure of the group. This will enable us to decompose any reducible representation to its proper subrepresentation and the corresponding orthogonal complement, which will also be its subrepresentation.
\begin{definition}[Unitary representation]
    A representation $V$ of a group $G$ is called \emph{unitary} if there exists a $G$-invariant sesquilinear inner product $\left( \bullet,\bullet \right) \colon V \times V \to \C$, that is
    \[ \left( g \bm{v},g \bm{w} \right) = \left( \bm{v},\bm{w} \right) \quad\text{for any } g\in G \text{ and } \bm{v},\bm{w} \in V. \]
\end{definition}
\begin{remark} \label{rem:diagonalizable}
    Any $g \in G$ is thus represented on any unitary representation $V$ by a \emph{unitary} operator $g|_{V}$. So $g|_{V}$ is also a normal operator, hence \emph{diagonalizable}.

\end{remark}

\begin{theorem}[Unitarity of compact Lie group representations] \label{th:compactisunitary}
    \hfill

    Let $V$ be a representation of a compact Lie group $G$. Then $V$ is unitary.
\end{theorem}
\begin{myproof}
    Let $\left\langle \bullet,\bullet \right\rangle$ be any inner product. We will average it over the whole $G$ using the invariant Haar measure $\dd{g}$ (see \Cref{th:haarmeasure}). Since integration preserves sesquilinearity and positive-definitness, we can define the inner product
    \[ \left( \bullet,\bullet \right) \equiv \int_{G} \left< g\bullet,g\bullet \right> \dd{g}. \]
    $G$-invariance of $(\bullet,\bullet)$ follows from \Cref{cor:propertiesHaar} applied to the function defined by $f(g) \equiv \langle g \bm{v} ,g \bm{w} \rangle$ for any fixed $\bm{v},\bm{w} \in V$, simply by the calculation
    \[\left(h \bm{v} ,h \bm{w} \right) = \int_{G} \langle gh \bm{v} ,gh \bm{w} \rangle \dd{g} = \int_{G} \langle g \bm{v} ,g \bm{w} \rangle \dd{g} \equiv \left(\bm{v},\bm{w}\right) \]
    for any $h \in G$.
\end{myproof}

\begin{corollary}[Complete reducibility of compact Lie group representations] \label{cor:completereducibility}
    Every representation $V$ of a compact Lie group $G$ is completely reducible.
\end{corollary}
\begin{myproof}
    Suppose $V$ is reducible with $W \le V$ being its proper subrepresentation. From \Cref{th:compactisunitary} we obtain a $G$-invariant inner product $( \bullet, \bullet)$, thus we have the decomposition $V = W \oplus W^{\perp}$, where the orthogonal complement $W^{\perp}$ of $W$ is also a proper subrepresentation of $V$, since
    \[ ( g \bm{v},\bm{w} ) = ( \bm{v},g^{-1}\bm{w} ) = 0 \quad\text{for any } g\in G, \bm{v}\in W^{\perp}, \bm{w}\in W. \]
    We finish the proof by induction and finite dimensionality.
\end{myproof}
Therefore, to study any finite-dimensional representation of a compact Lie group it is enough to understand the set of irreducible representations, and to know how to compute multiplicities in its decomposition.

\begin{corollary}[Decomposition of compact Lie group representations] \label{cor:decomposition}
    Let $V$ be a representation of a compact Lie group $G$. Then there exists a decomposition
    \[ V = \bigoplus_{i=1}^{k} V_{i}^{\oplus a_{i}} \equiv V_{1}^{\oplus a_{1}} \oplus \cdots \oplus V_{k}^{\oplus a_{k}}, \]
    where $V_{i}$ are distinct irreducible representations with multiplicities $a_{i}$ given uniquely by
    \[ a_{i} = \dim \Hom_{G}(V_{i},V) .\]
\end{corollary}
\begin{myproof}
    Existence follows from \Cref{cor:completereducibility}. Multiplicities are calculated as
    \begin{multline*}
        \dim \Hom_{G}(V_{i},V) = \dim \Hom_{G}\left(V_{i},\bigoplus_{j=1}^{k} V_{j}^{\oplus a_{j}}\right) = \dim\left[ \bigoplus_{j=1}^{k} \Hom_{G}\left(V_{i},V_{j}^{\oplus a_{j}}\right) \right]  \\
        =\sum_{j=1}^{k} \dim \Hom_{G}\left(V_{i},V_{j}^{\oplus a_{j}}\right) = \sum_{j=1}^{k} a_{j} \underbrace{\dim \Hom_{G}\left(V_{i},V_{j}\right)}_{\delta_{ij} \text{ from \Cref{cor:schur}}} = a_{i},
    \end{multline*}
    where we have used the isomorphism (valid for arbitrary representations $U,V,W$)
    \[ \Hom_{G}(U,V \oplus W) \cong \Hom_{G}(U,V)\oplus \Hom_{G}(U,W) \]
    following from the basic linear algebra equivalence
    \[ U^{*}\otimes( V \oplus W) \cong (U^{*}\otimes V) \oplus ( U^{*} \otimes W).\qedhere \]
\end{myproof}

We will show one last consequence of \Cref{th:compactisunitary}, namely that dual representations of unitary representations are actually isomorphic to so-called \emph{conjugate} representations, where multiplication by scalars from $\C$ is always conjugated.

\begin{corollary}[Equivalence of $\overline{V}$ and $V^{*}$] \label{cor:equivconjanddual}
    Let $V$ be a representation of a compact Lie group $G$. Then we have an equivalence of representations $\overline{V}\cong V^{*}$, where the \emph{conjugate} representation $\overline{V}$ has the same underlying additive structure as $V$, but is equipped with a new scalar multiplication structure $\cdot'\colon \C \times V \to V$ given by $z \cdot' \bm{v} = \overline{z} \bm{v}$.
\end{corollary}
\begin{myproof}
    Define a bijective linear map \(\varphi: \overline{V}\to V^{*}$ by $\varphi \bm{v} \equiv \left( \bm{v},\bullet  \right) \) for any $\bm{v} \in V$, where $(\bullet,\bullet)$ is the $G$-invariant inner product from \Cref{th:compactisunitary}. To see that $\varphi$ is a $G$-map, and therefore realizes the equivalence $\overline{V} \cong V^{*}$, we calculate
    \[ g(\varphi \bm{v}) = ( \bm{v},g^{-1}\bullet ) = ( g \bm{v}, \bullet ) \equiv \varphi(g \bm{v}),  \]
    where the first equality follows from \Cref{def:dualrepre} of the action of $G$ on $V^{*}$, and the second from $G$-invariance of the inner product.
\end{myproof}
\begin{remark}
    From the definition of $\overline{V}$ we have $g|_{\overline{V}}\equiv \overline{g|_{V}}$ for any $g \in G$.
\end{remark}

\section{Characters and the Projection formula} \label{sec:charactersprojectionformula}

We would like to explicitly compute the factors in the decomposition of a representation into irreducible ones. The first step will be finding an explicit formula for the projection onto the direct sum of trivial factors in the decomposition. To start, we introduce a notation for this trivial factor.

\begin{definition}[Fixed point set]
    Let $V$ be a representation of a group $G$. Then the \emph{fixed point set} is a subrepresentation of $V$ defined by
    \[ V^{G} \equiv \left\{ \bm{v} \in V \middle| g\bm{v}=\bm{v} \text{ for all } g \in G \right\}. \]
\end{definition}

We again exploit the idea of averaging over $G$ already used in the proof of \Cref{th:compactisunitary} to obtain a $G$-invariant projection.

\begin{prop}[Projection onto $V^{G}$] \label{prop:projectionontoVG}
    Let $V$ be a representation of a compact Lie group $G$.
    Then the map $\mathsf{p} \in \End(V)$ defined using the vector integration for any $\bm{v} \in V$ by
    \[ \mathsf{p}(\bm{v}) \equiv \int_{G} g \bm{v} \dd{g} \]
    is a projection of $V$ onto $V^{G}$.
\end{prop}

\begin{myproof}
    Using invariance of $\dd{g}$ for any $h\in G$ we have
    \[ h \mathsf{p}(\bm{v}) \equiv \int_{G} hg\bm{v} \dd{g} = \int_{G} g\bm{v} \dd{g} \equiv \mathsf{p}(\bm{v}), \]
    so $\Im \mathsf{p} \subset V^{G}$. Conversely, if $\bm{v}\in V^{G}$, then
    \[ \mathsf{p}(\bm{v}) \equiv \int_{G} g\bm{v} \dd{g} = \int_{G} \bm{v} \dd{g}= \bm{v}\int_{G} \dd{g} \equiv \bm{v}, \]
    so $V^{G} \subset \Im\mathsf{p}$ and $\mathsf{p} = \mathsf{p} \circ \mathsf{p}$ is a projection onto $\Im \mathsf{p} = V^{G}$.
\end{myproof}

\begin{figure}[!ht]
    \centering
    \includegraphics[width=1\linewidth]{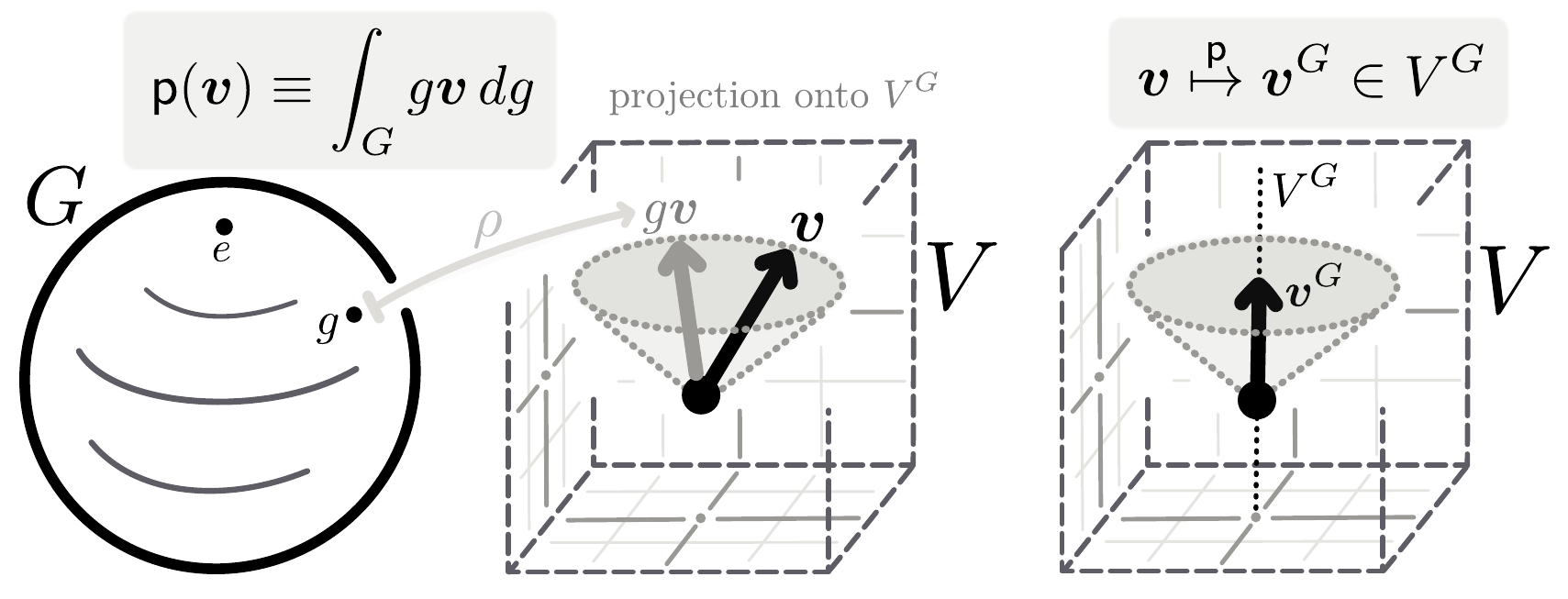}
    \caption{Illustration of \Cref{prop:projectionontoVG}. We imagine group $G=\SO(2)\cong S^{1}$ and its real $3$-dimensional representation \(V\), which decomposes as $V = S \oplus \R$, $S$ being the standard representation in the $x$-$y$ plane, and $\R$ being the trivial representation along the $z$-axis. As the projection map $\mathsf{p}$ averages over the whole group $G$, the action of $g \in G$ rotates the components of the vector in the $x$-$y$ plane, thereby only leaving the component along the $z$-direction, which is precisely $\bm{v}^{G} \in V^{G}$.}
    \label{fig:projection}
\end{figure}

\begin{remark} \label{rem:dimVG}
    We can calculate the dimension of $V^{G}$ using a standard property of projections, that is (we can commute $\Tr$ and integration by linearity)
    \[ \dim V^{G} = \Tr(\mathsf{p}) \equiv \Tr\left( \int_{G} g|_{V} \dd{g} \right) = \int_{G} \Tr(g|_{V}) \dd{g} \]
    which follows from $\mathsf{p}|_{V^{G}}=\Id_{V^{G}}$ and a convenient choice of basis adjusted to the decomposition $V = \Ker \mathsf{p} \oplus \Im \mathsf{p} = \Ker \mathsf{p} \oplus V^{G}$. For a visual intuition about the projection map $\mathsf{p}$ see \Cref{fig:projection}.
\end{remark}

The previous remark is a straightforward demonstration, that traces of $g|_{V}$ appear naturally when one is exploring the structure of representations. This suggests the fundamental notion of \emph{characters}, the basis of \emph{Character Theory}.
\begin{definition}[Character] \label{def:character}
    Let $V$ be a representation of a group $G$. Then its \emph{character} $\chi_V\colon G\to \C$ is a function on $G$ defined by
    \[ \chi_{V}(g) = \Tr(g|_{V}) \equiv \Tr_{V}(g) \quad\text{ for any } g \in G. \]
\end{definition} \vspace{-0.5em}
\begin{remark} \label{rem:charatidentityisdimension}
    Note that $\chi_{V}(e) \equiv \Tr(\Id_{V}) = \dim V$.
\end{remark}

Characters are a condensed form of information about the representation. For finite groups and compact Lie groups they give us a unique (up to isomorphism) characterization of their finite-dimensional representations, while for general Lie groups they still encode very important properties. Before we state the fundamental result concerning the calculation of multiplicities in decompositions, we first find out some useful properties of characters. \vspace{-0.3em}
\begin{prop}[Properties of characters] \label{prop:charactersproperies}
    Let $V$ and $W$ be representations of a compact Lie group $G$. Then the following properties hold:
    \begin{enumerate}
        \item Characters are \emph{class functions}, meaning they are invariant under the conjugation map, that is $\chi_{V}(g)=\chi_{V}(h g h^{-1})$ for any $g,h \in G$.
        \item If $V \cong W$, then $\chi_{V}=\chi_{W}$. Specifically, $\chi_{V^{*}} = \chi_{\overline{V}} = \overline{\chi_{V}}$.
        \item The character of a direct sum $V \oplus W$ is given by $\chi_{V\oplus W} = \chi_{V} + \chi_{W}$.
        \item The character of a tensor product $V \otimes W$ is given by $\chi_{V \otimes W} = \chi_{V}\mkern+1mu\chi_{W}$.
        \item The character of the dual space $V^{*}$ satisfies the relation $\chi_{V^{*}}(g) = \chi_{V}(g^{-1})$.
    \end{enumerate}
\end{prop}\vspace{-0.7em}
\begin{myproof}
    We fix any $g,h \in G$. Claim \textbf{(1)} follows from the cyclic property of $\Tr$.

    Action on equivalent representations (see \Cref{def:g-map}) differs just by a change of basis, with respect to which the trace is invariant. The second part of claim \textbf{(2)} then follows from \Cref{cor:equivconjanddual}.

    Actions of $g$ are automatically diagonalizable on $V$ and $W$ (see remark \ref{rem:diagonalizable}).
    Denote by $\{\alpha_{i}\}$ and $\{\beta_{j}\}$ the eigenvalues of the action $g$ on $V$ and $W$, respectively. Using \Cref{def:dirsumtensprodrepre} we obtain that $\{\alpha_{i}\} \cup \{\beta_{j}\}$ and $\{\alpha_{i}\beta_{j}\}$ are eigenvalues on $V \oplus W$ and $V \otimes W$, from which the claims \textbf{(3)} and \textbf{(4)} follow.

    Claim \textbf{(5)} follows from \Cref{def:dualrepre} and the fact that $\Tr(A)=\Tr(A^{\intercal})$.
\end{myproof}


%
%

We are now fully prepared to state the main result of this section.

\begin{corollary}[Projection formula for the multiplicity] \label{cor:projectionformula}
    The multiplicity $a_{i}$ of the irreducible representation $V_{i}$ in a decomposition of a representation $V$ of a group $G$ is given by
    \[ a_{i} \overset{\bm{(1)}}{=} \dim \Hom_{G}(V_{i},V) \overset{\bm{(2)}}{=} \int_{G} \chi_{V_{i}}(g^{-1})\,\chi_{V}(g) \dd{g} \overset{\bm{(3)}}{=} \int_{G} \overline{\chi_{V_{i}}(g)}\,\chi_{V}(g) \dd{g}. \]
\end{corollary}
\begin{myproof}
    Equality $\bm{(1)}$ was already proved in \Cref{cor:decomposition}. Equalities $\bm{(2)}$ and $\bm{(3)}$ follow from \Cref{rem:dimVG} applied to the representation $\Hom(V_{i},V)$ by noticing $\Hom_{G}(V_{i},V) \equiv \Hom(V_{i},V)^{G}$ (see \Cref{rem:homG}), and applying \Cref{prop:charactersproperies} to the equivalence $\Hom(V,W) \cong V^{*} \otimes W \cong \overline{V} \otimes W$, that is
    \[ \chi_{\Hom(V,W)}(g) = \chi_{V^{*}\otimes W}(g) = \chi_{V}(g^{-1})\,\chi_{W}(g) =  \chi_{\overline{V}\otimes W}(g) = \overline{\chi_{V}(g)}\,\chi_{W}(g).\qedhere \]
\end{myproof}

\begin{remark}
    \Cref{rem:dimVG} is a special case of \Cref{cor:projectionformula}, where we take the projection onto the trivial representation $V_{i} \equiv \C$, so $g|_{\C} \equiv \Id \Longrightarrow \chi_{V_{i}}(g)=1$.
\end{remark}
\begin{remark}
    \emph{Irreducible characters} (characters of the irreducible representations) turn out to be \emph{orthonormal}, which follows from \Cref{cor:projectionformula} and \Cref{cor:schur}.
\end{remark}

\begin{remark}[Quick summary]
    In the last two sections we learned that every finite-dimensional representation of a compact Lie group is completely reducible (\Cref{cor:completereducibility}). The decomposition is given uniquely (up to isomorphism) by the multiplicities of corresponding irreducible representations (\Cref{cor:decomposition}), which can be calculated using just the irreducible characters (\Cref{cor:projectionformula}). Thus, characters give us a unique characterization up to isomorphism.
\end{remark}

\section{Graded representations and characters} \label{sec:gradedrepresentations}
We learned a fair amount about the finite-dimensional representations, enough to be capable of extracting interesting information about the operators by assigning them appropriate representations and characters. Our goal is to find the Hilbert series $H(\bm{\phi},\D)$ encoding information about the operator basis containing an infinite number of operators.

It is hopeless to work with one operator representation at a time, not only efficiency-wise, but also because we have non-trivial IBP relations between them. Thankfully, we can organize them to so-called graded representations and calculate the corresponding characters all at once.

We start by realizing that the tensor product of representations $V \otimes W$ corresponds to the representation of two distinguishable particle fields, one transforming under  $V$, the other under $W$. Thus, if we want $n$ distinguishable copies of the same particle field, we just work with the $n$th tensor power.

\begin{definition}[Tensor power of representation] \label{def:tensorpower}
    Let $V$ be a representation of a group $G$. Then  the $n$th tensor power $V^{\otimes n}$ is (by convention $V^{\otimes 0}\equiv \C$)
    \[ V^{\otimes n} \equiv\mkern+1mu \underbrace{V \otimes \cdots \otimes V}_{n} \quad\text{for } n\in\N. \]
\end{definition}

But all known physical particles are in fact indistinguishable. Exchange of two particles cannot change any physical observable, so the wave-function can only change by a phase. Two subsequent exchanges of the same pair is an identity operation (ignoring peculiarities such as \emph{anyons}), so the change of phase can only be by a factor of $\pm$, corresponding to the bosonic/fermionic particles, respectively.

Since the action of the group on $V^{\otimes n}$ commutes with the permutation of the factors, a fully symmetric subspace $\displaystyle\Sym^{n}(V)$ and a fully antisymmetric subspace $\displaystyle\Ext^{n}(V)$ are actually subrepresentations of $V^{\otimes n}$. In the following, the symmetric product symbol $\otimes_{\text{sym}}$ is just omitted, and the exterior product is denoted as usual by the wedge $\wedge$.

\begin{definition}[Symmetric and exterior power of representation] \label{def:symmetricexterior}
    Let $V$ be a representation of a group $G$. Then the $n$th \emph{symmetric power} $\displaystyle\Sym^{n}(V)$ and the $n$th \emph{exterior power} $\displaystyle\Ext^{n}(V)$ are subrepresentations of the $n$th tensor power $V^{\otimes n}$, with actions of any $g \in G$ given by
    \[ g\left( \bm{v}_{1} \cdots \bm{v}_{n} \right) \equiv (g \bm{v}_{1})\cdots(g \bm{v}_{n}), \quad g\left( \bm{v}_{1}\wedge \cdots \wedge\bm{v}_{n} \right) \equiv (g \bm{v}_{1})\wedge\cdots\wedge(g \bm{v}_{n}), \]
    for any $\{\bm{v}_{i}\}_{i=1}^{n} \subset V$. By convention let $\displaystyle\Sym^{0}(V)=\Ext^{0}(V)\equiv \C$.

\end{definition}
\vspace{-0.5em}
\begin{remark}
    This construction is further generalized by \emph{Schur functors} $\mathbb{S}^{\lambda}$ \cite{GoodmanWallach2009}.
\end{remark}

We can extend the notions of tensor, symmetric, and exterior power to construct corresponding algebras. But first, we define the notion of graded representations.
\begin{definition}[Graded representation] \label{def:gradedrep}
    A representation \(V\) of a group \(G\) is called a \emph{graded representation}  \textcolor{gray}{(possibly infinite-dimensional)} if it has the extra structure of a \emph{grading}, meaning it has a decomposition of the form
    \[ V \equiv \bigoplus_{n=0}^{\infty} t^{n} V_{n}, \]
    where each $V_{n}$ is a finite-dimensional representation of $G$. For now, the formal factors $t^{n}$ are just labels helping us to distinguish different graded pieces (useful when combining different graded representations together).
\end{definition}
\begin{remark} \label{rem:gradedofgraded}
    We defined $V_{n}$ to be finite-dimensional, but we would often like them to be themselves graded representations. Standard reordering of summation makes this requirement reasonable in the sense of \Cref{def:gradedrep} \vspace{-1em}
    \[ V \equiv \bigoplus_{n=0}^{\infty} t^{n} V_{n} \equiv \bigoplus_{n=0}^{\infty} t^{n} \left( \bigoplus_{r=0}^{\infty} q^{r} W_{n,r} \right) = \bigoplus_{n=0}^{\infty} \bigoplus_{r=0}^{\infty} t^{n}q^{r} W_{n,r} = \bigoplus_{k=0}^{\infty} \overbrace{ \left( \bigoplus_{n+r=k} t^{n}q^{r} W_{n,r} \right)  }^{U_{k}},\]
    where the representations $W_{n,r}$ are already finite-dimensional, thus also $U_{k}$. \\
    Alternatively, we could extend the definition of graded representations to utilize grading with multiple labels right from the start.
\end{remark}

Natural examples of graded representations arise from algebra by considering the tensor algebra and its quotients, the symmetric and the exterior algebra.
\begin{definition}[Tensor, symmetric, and exterior graded representations] \label{def:examplesgradedreps}
    Let $V$ be a \textcolor{gray}{graded} representation of a group $G$. We define the \emph{tensor}, \emph{symmetric}, and \emph{exterior graded representations} of $V$, respectively, as
    \[ T(V) \equiv \bigoplus_{n=0}^{\infty} t^{n} V^{\otimes n}, \quad \Sym(V) \equiv \bigoplus_{n=0}^{\infty} t^{n}\Sym^{n}(V), \quad \Ext(V) \equiv \bigoplus_{n=0}^{\;\dim V} t^{n}\!\Ext^n(V). \]
\end{definition}
\vspace{-0.2em}
\begin{remark}
    We naturally extend the definitions of tensor, symmetric, and exterior powers (\Cref{def:tensorpower,def:symmetricexterior}) to the case of graded representations. Similarly as in \Cref{rem:gradedofgraded}, the labels keep track of the appropriate gradings.
\end{remark}
\begin{remark}
    Note that for a finite-dimensional vector space $V$ the direct sum in the definition of the exterior graded representation terminates, because there are no non-trivial $n$-forms with $n>\dim V$. We still understand $\displaystyle\Ext(V)$ as a graded representation, just for $n>\dim V$ we have only trivial \emph{zero representations}.
\end{remark}
\begin{remark}
    We say that $\Sym(V)$ and $\displaystyle\Ext(V)$ are \emph{graded subrepresentations} of $T(V)$, since they are subrepresentations for every graded piece.
\end{remark}

If we want to calculate some property of a graded representation, it would be often useful to know which graded piece contributed by what amount. This is where the labels will play a fundamental role. Take for example the dimension.

\begin{definition}[Graded dimension] \label{def:gradeddim}
    Let $V$ be a graded representation of a group $G$. Then its \emph{graded dimension} $\dim_{t}V$ is a formal series in a complex parameter $t$ defined by
    \[ \dim_{t}V \equiv \sum_{n=0}^{\infty} t^{n} \dim V_{n}. \]
\end{definition}
\vspace{-0.2em}
\begin{remark}
    The use of labels enables us to extract useful information even in cases, when the ordinary dimension of $V$ would simply be just infinity.
\end{remark}
\begin{remark}
    The graded dimension of $V$ in the context of \Cref{rem:gradedofgraded} is a series with two parameters, greatly helping us to identify the former grading.
\end{remark}
\begin{remark}[$\K$ as a graded representation and $H(\bm{\phi},\D)$ as its graded dimension] \label{rem:Kasgradedrephilbertseries}
    The space $\K\equiv \Span \B$ in \Cref{def:operatorbasis} can be viewed as a graded representation of the Lorentz (and possibly gauge or some other) group, with the grading given by \Cref{def:weight}. Every graded piece is composed of trivial representations, one for each independent operator of a given weight, since all operators in $\B$ must be invariant. By comparing \Cref{def:gradeddim} with \Cref{def:hilbertseries} we recognize that
    \[ \dim_{\left( \bm{\phi},\D \right) }\K = H(\bm{\phi},\D), \]
    thus the Hilbert series $H$ is the graded dimension of the representation $\K\equiv \Span \B$. That is actually why the coefficients of $H(\bm{\phi},\D)$ are denoted as $\opers{r}{n}$, because they are the dimensions of the corresponding graded pieces.
\end{remark}

For efficient manipulations it is essential to extend the definition of characters to graded representations. Graded characters will play a fundamental role in the computation of the Hilbert series.
\begin{definition}[Graded character] \label{def:gradedchar}
    Let $V$ be a graded representation of a group $G$. Then its \emph{graded character} $\chi_{V}$ is a formal series on $G$ in a complex parameter $t$ defined by
    \[ \chi_{V}(t;g) \equiv  \sum_{n = 0}^{\infty} t^n \chi_{V_{n}}(g), \quad \text{for any } g \in G. \]
\end{definition}

\begin{remark}
    In particular, we have $\chi_{V}(t;e) = \dim_{t}V$, see \Cref{rem:charatidentityisdimension}.
\end{remark}
\begin{remark} \label{rem:gradedmultiplicity}
    Similarly as in \Cref{cor:projectionformula}, we can use the graded character $\chi_{V}$ to obtain \emph{graded multiplicities} of any irreducible representation $W$ in $V$ as
    \[ \dim_{t}a_{W} \equiv \dim_{t} \Hom_{G}(W,V) \equiv \sum_{n=0}^{\infty} t^{n} \dim \Hom_{G}(W,V_{n}) = \int_{G} \overline{\chi_{W}(g)}\,\chi_{V}(t;g) \dd{g}. \]
\end{remark}

\begin{remark}[Alternative formulation of graded representations] \label{rem:reformulationgradedrep}
    In the light of \Cref{def:gradedchar} it appears that the $n$th graded piece $V_{n}$ is also being scaled by the factor of $t^{n}$ (in addition to the ordinary action of $g$ on $V_{n}$ represented by $g|_{V_{n}}$). Indeed, we can make an alternative formulation of \Cref{def:gradedrep} which gives a precise meaning to the labels $t^n$.

    First, we extend our group by a \emph{scaling group} of non-zero complex numbers with a multiplication operation, that is (see \Cref{def:groupsdirectproduct})
    \[ \mathsf{G} \equiv \\ \GL(\C) \times G = \left( \C \setminus \{0\} \right) \times G \equiv \C^{\times} \times G. \]
    We can now define a set of representations \(\mathsf{C}^{n}\) of \(\C^{\times}\), which still act on \(\C\), but with a modified action of $t \in \C^{\times}$ to $t|_{\mathsf{C}^{n}}\equiv t^{n}$. This enables us to distinguish different graded pieces $V_{n}$ by constructing the graded representation
    \[ \mathsf{V} \equiv \bigoplus_{n=0}^{\infty} \mathsf{V}_{n} \equiv \bigoplus_{n=0}^{\infty} \mathsf{C}^{n} \otimes V_{n}, \]
    with the action of $\mathsf{g}=(t;g) \in \mathsf{G}$ on $\bm{\mathsf{v}} \in \mathsf{V}$ defined by
    \[ \mathsf{g}\bm{\mathsf{v}} \equiv (t;g)\left[ \bigoplus_{n=0}^{\infty} \bm{v}_{n} \right] \equiv \bigoplus_{n=0}^{\infty} t^{n} g|_{V_{n}} \bm{v}_{n}, \quad\text{where } \bm{v}_{n} \in \mathsf{C}^{n} \otimes V_{n}. \]
    Note that $\mathsf{V}_{n}\equiv \mathsf{C}^{n}\otimes V_{n}$ is isomorphic to \(V_{n}\) as a complex vector space, but not as a representation of $\mathsf{G}$. Factors of $\mathsf{C}^{n}$ just \emph{charge} the graded pieces $V_{n}$, granting them an additional scaling action. Now we can easily see, that we reproduced
    \[ \chi_{\mathsf{V}}(\mathsf{g})\equiv \Tr(g|_\mathsf{V})\equiv \Tr\left( \bigoplus_{n=0}^{\infty} t^{n} g|_{V_{n}} \right) = \sum_{n = 0}^{\infty} t^n \chi_{V_{n}}(g), \quad\text{for any } \mathsf{g}=(t;g) \in \mathsf{G}. \]
    Since most of the time we work only with a single group element $\mathsf{g} = (\bm{t};g)$ with possibly multiple scaling parameters $\bm{t}\equiv (t_1, \ldots, t_{l})$, we can think of them as just being labels, just like in \Cref{def:gradedrep}. Nonetheless, in \Cref{cor:plethysticexponentialgradedchars} we will really need to use the interpretation of scaling introduced in this reformulation.
\end{remark}

Graded characters corresponding to graded representations in \Cref{def:examplesgradedreps} can be (formally) summed up to quite pretty formulas. It is important, that we are able to find such closed forms, otherwise the calculations would be intractable.

\begin{prop}[Selected graded characters] \label{prop:gradedchars}
    Let $V$ be a \textcolor{gray}{graded} representation of a group $G$. The following graded characters are given by: \vspace{-0.5em}
    \[
        \renewcommand\arraystretch{2.2} 
        \everymath={\displaystyle}
        \begin{array}{c @{{}={}} c @{{}={}} c @{{}\equiv{}} c}
            \chi_{T(V)}(t;g)    & \sum_{n = 0}^{\infty} t^n \chi_{V^{\otimes n}}(g)                                 & \frac{1}{1- t \Tr(g|_{V})}  & \frac{1}{1 - t\mkern+1mu\chi_{V}(g)}, \\
            \chi_{\Sym(V)}(t;g) & \sum_{n = 0}^{\infty} t^n \chi_{\Sym^{n}(V)}(g)                                   & \frac{1}{\det(\Id-t\mkern+1mu g|_{V})} & \frac{1}{\det_{V}(1-tg)}, \\
            \chi_{\Ext(V)}(t;g) & \sum_{n = 0}^{\infty} t^n \chi_{\Ext^{\raisebox{0.10em}{$\scriptstyle n$}}(V)}(g) & \det(\Id+t\mkern+1mu g|_{V})           & \det_{V}(1+tg).
        \end{array} \vspace{-0.2em}
    \]
    \textcolor{gray}{We suppressed the label corresponding to the grading of \(V\).}
\end{prop}
\begin{myproof}
    We perform the proof only for a finite-dimensional $V$ with $m \equiv \dim V$. The proof would go along the same lines for a graded representation, only $\chi_{V}$ itself would be a series, and $V$ would have a countable basis.

    The first relation follows from \Cref{prop:charactersproperies} applied to $V^{\otimes n}$, thus
    \[ \sum_{n = 0}^{\infty} t^n \chi_{V^{\otimes n}}(g) = \sum_{n = 0}^{\infty} \left[ t\,\chi_{V}(g) \right]^{n}   = \frac{1}{1 - t\mkern+1mu\chi_{V}(g)}, \]
    where the last equality is just evaluating the sum of a geometric series.

    Fix any $g$ and let $\{\bm{e}_{i}\}_{i=1}^{m}$ be a basis of $V$ created from eigenvectors of $g|_{V}$ with corresponding eigenvalues $\{\lambda_{i}\}_{i=1}^{m}$.

    For any given $n \in \N_{0}$, it is easy to see that a basis of $\displaystyle\Sym^{n}(V)$ is
    \[ \left\{ \bm{e}_{1}^{r_1} \cdots \bm{e}_{m}^{r_{m}} \middle| r_{i} \in \N_{0}, \sum_{i=0}^{m} r_{i} = n \right\}, \]
    where $r_{i}$ determines what power of $\bm{e}_{i}$ is present in a given basis element. Since we have (see \Cref{def:examplesgradedreps})
    \[ g\left( \bm{e}_{1}^{r_1} \cdots \bm{e}_{m}^{r_{m}} \right) = \lambda_{1}^{r_1} \cdots \lambda_{m}^{r_{m}} \left( \bm{e}_{1}^{r_1} \cdots \bm{e}_{m}^{r_{m}} \right),  \]
    it follows that (using the multi-index notation $\bm{\lambda}^{\bm{r}}\equiv \prod_{i} \lambda_{i}^{r_{i}}$, $|\bm{r}|\equiv \sum_{i} r_{i}$)
    \begin{align*}
        \sum_{n = 0}^{\infty} t^n \chi_{\Sym^{n}(V)}(g)
         & =
        \sum_{n = 0}^{\infty} t^n \sum_{|\bm{r}|=n} \bm{\lambda}^{\bm{r}} =
        \sum_{n = 0}^{\infty} \sum_{|\bm{r}|=n} (t\bm{\lambda})^{\bm{r}} =
        \sum_{\bm{r}} (t\bm{\lambda})^{\bm{r}} \\
         & =
        \prod_{i=1}^{m} \sum_{r_{i}=0}^{\infty} (t\lambda_{i})^{r_{i}} =
        \prod_{i=1}^{m} \frac{1}{1-t\lambda_{i}} =
        \frac{1}{\det(\Id-t\mkern+1mu g|_{V})}.
    \end{align*}

    For any given $n \in \N_{0} \cap [0,m]$, a basis of $\displaystyle\Ext^{n}(V)$ is
    \[ \left\{ \bm{e}_{1}^{r_1} \wedge \cdots \wedge \bm{e}_{m}^{r_{m}} \middle| r_{i} \in \{0,1\}, \sum_{i=0}^{m} r_{i} = n \right\}, \]
    where $r_{i}$ determines whether $\bm{e}_{i}$ is present or not.
    Similarly, since we have
    \[ g\left( \bm{e}_{1}^{r_1} \wedge \cdots \wedge \bm{e}_{m}^{r_{m}} \right) = \lambda_{1}^{r_1} \cdots \lambda_{m}^{r_{m}} \left( \bm{e}_{1}^{r_1} \wedge \cdots \wedge \bm{e}_{m}^{r_{m}} \right),  \]
    it follows that
    \begin{align*}
        \sum_{n = 0}^{m} t^n \chi_{\Ext^{\raisebox{0.10em}{$\scriptstyle n$}}(V)}(g)
         & =
        \sum_{n = 0}^{m} t^n \sum_{\substack{|\bm{r}|=n \\ \mathclap{r_{i} \in \{0,1\}}}} \bm{\lambda}^{\bm{r}} =
        \sum_{n = 0}^{m} \;\sum_{\substack{|\bm{r}|=n \\ \mathclap{r_{i} \in \{0,1\}}}} (t\bm{\lambda})^{\bm{r}} =
        \;\sum_{\substack{\bm{r} \\ \mathclap{r_{i} \in \{0,1\}}}}\; (t\bm{\lambda})^{\bm{r}} \\
         & =
        \prod_{i=1}^{m} \sum_{r_{i}=0}^{1} (t\lambda_{i})^{r_{i}} =
        \prod_{i=1}^{m} (1+t\lambda_{i}) =
        \det(\Id+t\mkern+1mu g|_{V}).
        \qedhere
    \end{align*}
\end{myproof}

\begin{remark}
    Summed up graded characters are reminiscent of \emph{partition functions} with corresponding statistics found in statistical mechanics, namely the grand-canonical partition functions for Bose-Einstein/Fermi-Dirac ideal quantum gases
    \[ \mathcal{Z}_{g} = \prod_{n} \left[ 1 \mp \eunit^{-\beta (E_{n}-\mu)} \right]^{\mp} \equiv \prod_{n} \left[ 1 \mp z \eunit^{-\beta E_{n}} \right]^{\mp}, \]
    where $n$ indexes over all energy states with the corresponding energies $E_{n}$, and as usual $\beta\equiv \frac{1}{k_{B}T}$ is the \emph{thermodynamic beta}. The \emph{fugacity} $z\equiv \eunit^{\beta\mu}$ plays the role of the grading label $t$, and $\eunit^{-\beta E_{n}}$ plays the role of $\lambda_{i}$.
\end{remark}

There is another way to express graded characters of $\Sym(V)$ and $\displaystyle\Ext(V)$, mainly useful in the case when $V$ itself is a graded representation. It enables us to sidestep the calculation of the determinant just by knowing its graded character.

\begin{corollary}[Graded characters as a Plethystic Exponential] \label{cor:plethysticexponentialgradedchars}
    Let $V$ be a graded representation of a group $G$ with label $q$.
    The graded characters in \Cref{prop:gradedchars} can be also calculated as \vspace{-0.5em}
    \[
        \renewcommand\arraystretch{2.2} 
        \everymath={\displaystyle}
        \begin{array}{c @{{}={}} c @{{}\equiv{}} c}
            \chi_{\Sym(V)}(t,q;g) & \exp\left(\mspace{2mu}\sum_{r=1}^{\infty} \textcolor{gray}{(+1)^{r+1}} \frac{t^{r}}{r} \chi_{V}(q^{r}; g^{r}) \right) & \PE\left[t\mkern+1mu\chi_{V}(q; g)\right], \\
            \chi_{\Ext(V)}(t,q;g) & \exp\left(\mspace{2mu}\sum_{r=1}^{\infty} (-1)^{r+1} \frac{t^{r}}{r} \chi_{V}(q^{r}; g^{r}) \right)                   & \PE_{f}\left[t\mkern+1mu\chi_{V}(q; g)\right].
        \end{array}
    \]
\end{corollary}
\begin{myproof}
    First, we perform manipulations with an arbitrary matrix of the form $\bm{1}\mp t A$, where we use the well-known formula $\det\exp = \exp\Tr \Rightarrow \det = \exp\Tr\ln$, and expansion of $\ln$ around $\bm{1}$, to obtain
    \begin{multline*}
        [\det\left( \bm{1} \mp tA \right)]^{\mp} = [\exp\Tr\ln\left( \bm{1}\mp tA \right)]^{\mp} \\
        = \exp\left[ \mp \Tr \left(\sum_{r=1}^{\infty} (-1)^{r+1} \frac{(\mp tA)^{r}}{r}  \right) \right] = \exp\left(\sum_{r=1}^{\infty} (\pm 1)^{r+1} \frac{t^{r}}{r} \Tr(A^{r}) \right).
    \end{multline*}
    Since the action of $\mathsf{g}$ on $\mathsf{V}_{n}$ is given by $q^{n}g|_{V_{n}}$ (see \Cref{rem:reformulationgradedrep}), the result follows from \Cref{prop:gradedchars} and the calculation
    \begin{align*}
         & [\det_{\mathsf{V}}(1 \mp tg)]^{\mp } = \left[ \prod_{n=0}^{\infty} \det_{\mathsf{V}_{n}}(1 \mp tg) \right]^{\mp } \equiv \prod_{n=0}^{\infty} \left[ \det(\bm{1}\mp t q^{n}g|_{V_{n}}) \right]^{\mp } \overset{\text{\scriptsize\textcolor{gray}{[from before]}}}{=\mkern-2.7mu=\mkern-2.7mu=\mkern-2.7mu=\mkern-2.7mu=\mkern-2.7mu=\mkern-2.7mu=} \\
         & \mspace{10mu}=\prod_{n=0}^{\infty}\exp \left(\sum_{r=1}^{\infty} (\pm 1)^{r+1} \frac{t^{r}}{r} \Tr[(q^{n}g|_{V_{n}})^{r}] \right) = \exp \left(\sum_{n=0}^{\infty}\sum_{r=1}^{\infty} (\pm 1)^{r+1} \frac{t^{r}}{r} \Tr[q^{nr}g^{r}|_{V_{n}}] \right) \\
         & \mspace{10mu}= \exp \left(\sum_{r=1}^{\infty} (\pm 1)^{r+1} \frac{t^{r}}{r} \sum_{n=0}^{\infty}(q^{r})^{n}\Tr[g^{r}|_{V_{n}}] \right) \equiv  \exp \left(\sum_{r=1}^{\infty} (\pm 1)^{r+1} \frac{t^{r}}{r} \chi_{V}(q^{r}; g^{r}) \right),
    \end{align*}
    or more simply using $\mathsf{g}^{r}=(q;g)^{r}=(q^{r};g^{r})$ as
    \[ [\det_{\mathsf{V}}(1 \mp tg)]^{\mp } = \exp \left(\sum_{r=1}^{\infty} (\pm 1)^{r+1} \frac{t^{r}}{r} \chi_{\mathsf{V}}(\mathsf{g}^{r}) \right) \equiv  \exp \left(\sum_{r=1}^{\infty} (\pm 1)^{r+1} \frac{t^{r}}{r} \chi_{V}(q^{r}; g^{r}) \right).\qedhere \]
\end{myproof}
\begin{remark}[Plethystic Exponential] \label{rem:plethysticexponential}
    For a function $\alpha(t_1,\ldots,t_{k})$ satisfying $\alpha(0,\ldots,0) = 0$, the \textcolor{gray}{(fermionic)} Plethystic Exponential is defined by \cite{LehmanMartin}
    \[ \PE_{\textcolor{gray}{f}}\left[\alpha(t_1, \ldots, t_{k})\right] \equiv \exp\left(\sum_{r=1}^{\infty} \textcolor{gray}{(-1)^{r+1}} \frac{1}{r} \alpha(t_1^{r}, \ldots, t_{k}^{r}) \right). \]
    It is customary to leave out $f$ in $\PE_{f}$, and just automatically include the factor $(-1)^{r+1}$ for ``fermionic'' functions. Already working with this convention, for two functions $\alpha,\beta$ (either one can be bosonic or fermionic) of some common variables, the Plethystic Exponential satisfies the \emph{sum-to-product} property
    \[ \PE[\alpha+\beta]=\PE[\alpha]\PE[\beta]. \]
    One simple example is $\alpha(t,q)\equiv t$ and $\beta(t,q)\equiv q$, both bosonic or both \textcolor{gray}{fermionic}, for which it generates all \textcolor{gray}{anti}symmetric combinations of the variables, that is
    \begin{equation*}
        \PE[t+q]= \PE[t]\PE[q]=\begin{cases}
            \frac{1}{(1-t)(1-q)} = 1+t+q+t^2 + tq + q^2 + \cdots, \\
            \textcolor{gray}{(1+t)(1+q) = 1+t+q+tq,}
        \end{cases}
    \end{equation*}
    thus reflecting the corresponding statistics.
\end{remark}

\section{Weyl integration formula} \label{sec:weylintegrationformula}

A particularly hard question is, how do we actually compute integrals appearing in \Cref{cor:projectionformula}. Normally, we would need to have a parametrization of a compact Lie group $G$, construct the invariant Haar measure $dg$ by (for example) left-translating the volume form in the identity of $G$ to the whole $G$, express characters in a given parametrization, and perform the $\mathsf{D}$-dimensional integral, where $\mathsf{D}$ is the dimension of $G$. More often than not, already the first part of finding the parametrization can be extremely tough.

But there exists a remedy, which by utilizing the fact that characters are \emph{class functions} and a deep analysis of the structure of compact Lie groups enables us to restrict the integration to a simpler subgroup of $G$, moreover setting us up just right to use the powerful method of complex integration.

But first, we must expand our toolbox. Mostly we will just quickly recall some elementary definitions, since one can find more details in any book on Lie groups. Already in \Cref{prop:lefinvariantlift} we found the isomorphism between left-invariant vector fields and vectors at the identity of a Lie group. Furthermore, this enables us to induce a Lie algebra structure on the tangent space at identity.
\begin{definition}[Lie algebra] \label{def:liealgebra}
    A \emph{Lie algebra} $\g\equiv\bm{T}_{e}G$ of a Lie group $G$ is the tangent vector space at the identity of $G$, together with the \emph{algebra structure}
    \[ [\bm{a},\bm{b}] \equiv [\bm{\ell_{a}},\bm{\ell_{b}}]|_{e} \quad\text{for any } \bm{a},\bm{b} \in \g, \]
    induced from the \emph{Lie bracket} of left invariant vector fields on $G$.
\end{definition}
\begin{remark}[Jacobi identity] \label{rem:jacobiidentity}
    We automatically obtain the \emph{Jacobi identity}
    \[ [\bm{a},[\bm{b},\bm{c}]] + [\bm{c},[\bm{a},\bm{b}]] + [\bm{b},[\bm{c},\bm{a}]] = \bm{0}, \quad\text{for any } \bm{a},\bm{b},\bm{c} \in \g, \]
    since it holds for the Lie bracket of vector fields.
\end{remark}
\begin{remark} \label{rem:linearizedmultiplicationliealgebra}
    Consider the behavior of group multiplication $m\colon G\times G\to G$ near the identity. The corresponding pushforward at the point $(e;e)$ is given by
    \[ m_{*}\big|_{(e;e)}\colon \g \oplus \g \to \g, \quad \bm{a}\oplus \bm{b} \mapsto \bm{a} + \bm{b}. \]
    This can be seen by considering curve $\gamma^{\bm{a}}_{\bullet}\colon \R \to G$ with the tangent vector $\bm{a}$ at the identity $e\equiv \gamma_{\tau=0}$. We denote this by $\bm{a} \equiv \frac{\bm{D}}{d\tau}\gamma_{\tau}\big|_{\tau=0}$. Then, specially choosing combination $\bm{a}\oplus \bm{0} \leftrightarrow (\gamma^{\bm{a}}_{\bullet};e)$, we have
    \[ m_{*}\big|_{(e;e)}(\bm{a}\oplus \bm{0}) \equiv m_{*}\big|_{(e;e)} \frac{\bm{D}}{d\tau} (\gamma^{\bm{a}}_{\tau};e) \big|_{\tau=0} = \frac{\bm{D}}{d\tau}m(\gamma^{\bm{a}}_{\tau};e)\big|_{\tau=0} = \frac{\bm{D}}{d\tau} \gamma^{\bm{a}}_{\tau} \big|_{\tau=0} \equiv \bm{a}. \]
    The result follows from an analogous argument for $\bm{b}$, using linearity of pushforward (a general property of pushforwards), and finally taking $\bm{a}\oplus \bm{b}=\bm{a}\oplus \bm{0} + \bm{0}\oplus \bm{b}$.
\end{remark}

Intuitively, near the identity the group multiplication of $G$ is reflected in the linear structure of vector addition in $\g$. This is very useful for the study of Lie groups.

\begin{figure}[!ht]
    \centering
    \includegraphics[width=1.0\linewidth]{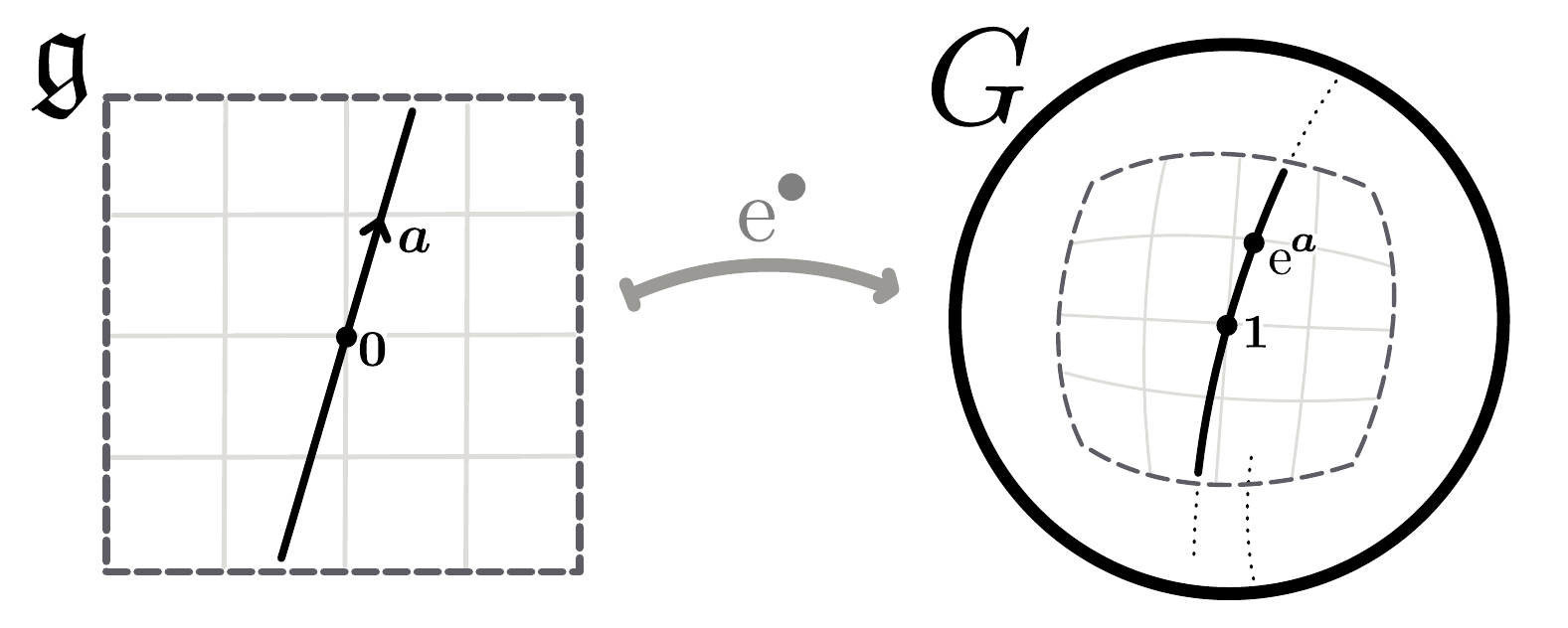}
    \caption{Illustration of the exponential map $\eunit^{\bullet}$ mapping from the Lie algebra $\g$ to the corresponding Lie group $G$. Locally around $\bm{0} \in \g$ it is a diffeomorphism.}
    \label{fig:exponentialmap}
\end{figure}

It is easy to go from $G$ to $\g$, we just look at the tangent vectors. After understanding the ``small'' transformations of $G$ near the identity, it is also desirable to go in the opposite way. There indeed exists the natural \emph{exponential map}.
\begin{definition}[Exponential map] \label{def:exponentialmap}
    Let $\g$ be a Lie algebra of a Lie group $G$.
    The \emph{exponential map} is defined by
    \[ \eunit^{\bullet} \equiv \exp(\bullet) \colon\g \to G, \quad \bm{a} \mapsto \gamma^{\bm{a}}_{1} \equiv \gamma^{\bm{a}}(1), \]
    where $\gamma^{\bm{a}}_{\bullet}:\R\to G$ is the integral curve of the left-invariant vector field $\bm{\ell}_{\bm{a}}$ (corresponding to $\bm{a}$ by \Cref{prop:lefinvariantlift}) going through the identity $\bm{1} \equiv \gamma^{\bm{a}}_{\tau=0}$.

    Since the curves $\sigma \mapsto \gamma^{\tau \bm{a}}(\sigma)$ and $\sigma \mapsto \gamma^{\bm{a}}(\tau \sigma)$ both correspond to the same vector $\tau \bm{a} \in \g$, they are equal, and we obtain \vspace{-0.3em}
    \[ \eunit^{\tau \bm{a}} \equiv  \gamma^{\tau\bm{a}}(1) = \gamma^{\bm{a}}(\tau)\equiv \gamma^{\bm{a}}_{\tau} \ \ \Longrightarrow\ \ \bm{a}\equiv \frac{\bm{D}}{d\tau} \eunit^{\tau \bm{a}}\Big|_{\tau=0}. \]
    From the properties of integral curves of the left-invariant vector fields we can also show that restrictions of \(\eunit^{\bullet}\) to the lines through the origin in \(\g\) are actually one-parameter subgroups of \(G\), that is for general \(\bm{a} \in \g\) we have
    \[ \eunit^{( \tau+\sigma )\bm{a}} = \gamma^{\bm{a}}_{\tau+\sigma} = \gamma^{\bm{a}}_{\tau}\gamma^{\bm{a}}_{\sigma} =  \eunit^{\tau \bm{a}} \eunit^{\sigma \bm{a}},\; \eunit^{0 \bm{a}} = \eunit^{\bm{0}} = \bm{1} \textcolor{gray}{\;\equiv e} \ \ \Longrightarrow\ \ \left( \eunit^{\tau \bm{a}} \right) ^{-1} = \eunit^{-\tau \bm{a}}. \]
\end{definition} \vspace{-0.4em}
\begin{remark}
    It can be easily seen that the differential (or pushforward) of the exponential map at the origin $\bm{0} \in \g$ is the identity, thus by the \emph{inverse function theorem} it is locally invertible at $\bm{0} \mapsto \bm{1}$. For an illustration see \Cref{fig:exponentialmap}. \vspace{-0.5em}
\end{remark}

Similar to groups, we can also represent Lie algebras on other vector spaces. \vspace{-0.4em}
\begin{definition}[Lie algebra representation] \label{def:liealgebrarepre}
    A representation of a Lie algebra~$\g$ on a finite-dimensional vector space $V$ is a homomorphism of Lie algebras $\xi_{\bullet} \colon \g\to \End(V)$, meaning it preserves the Lie algebra structure, that is
    \[ \xi_{[\bm{a},\bm{b}]}=[\xi_{\bm{a}},\xi_{\bm{b}}] \equiv \xi_{\bm{a}} \cdot \xi_{\bm{b}} - \xi_{\bm{b}} \cdot \xi_{\bm{a}} \quad\text{for any } \bm{a},\bm{b} \in \g,\]
    where the right side can be understood as the matrix commutator.
\end{definition} \vspace{-0.4em}
\begin{remark}
    Note that $\End(V)$ (with the Lie bracket given by the commutator) is the Lie algebra of $\GL(V)$.
\end{remark}

In the following definition we introduce the notion of \emph{inner automorphism} or \emph{conjugation}, and the related \emph{Adjoint} and \emph{adjoint} representations. A great deal of information about a Lie group can be extracted by the detailed study of their Lie algebra structure.
\begin{definition}[Conjugation, Adjoint and adjoint representations] \label{def:conjugationadjointrepre}
    Let $G$ be a Lie group. For any element $g \in G$ we have a \emph{conjugation} automorphism $\AD_{g}\colon G\to G$ given by $\AD_{g}\equiv L_{g}R_{g^{-1}}$, that is for any $h \in G$ we have
    \[ \AD_g h \equiv g h g^{-1}, \text{ also } \AD_{g}\AD_{h}=\AD_{gh}. \]
    In the following let $\gamma_{\bullet}\colon \R \to G$ be a curve with the tangent vector $\bm{a}$ at the identity $e\equiv \gamma_{\tau=0}$. For example, we can take $\gamma_{\tau}\equiv \eunit^{\tau \bm{a}}$.

    The corresponding induced map on the vectors at the identity gives rise to the \emph{Adjoint} representation $\Ad_{\bullet}\colon G\to \GL(\g)$ of $G$ on $\g$, that is $\Ad_g \equiv \AD_{g*}|_{e}$. Equivalently, using the previously defined curve we have
    \[ \Ad_{g}\bm{a} \equiv \frac{\bm{D}}{d\tau} \AD_{g}\gamma_{\tau}\Big|_{\tau=0}. \]
    Once more, the Adjoint representation induces the \emph{adjoint} representation $\ad_{\bullet}\colon \g\to \End(\g)$ of $\g$ on $\g$ by $\ad_{\bullet} \equiv \Ad_{\bullet*}|_{e}$, that is for any $\bm{b} \in \g$ we have
    \[ \ad_{\bm{a}}\bm{\bm{b}} \equiv \dv{\tau}\Ad_{\gamma_{\tau}}\bm{b}\Big|_{\tau=0}, \]
    where we write ``normal'' derivative, because if $V$ is a finite-dimensional vector space, then $\bm{T}_{p}V$ is canonically isomorphic to $V$ for any $p\in V$.
\end{definition}
\begin{remark}[Adjoint representation $\Ad_{\bullet}$ is orthogonal] \label{rem:adjointisorthogonal}
    By construction $\g$ is a real vector space. Hence, the Adjoint map $\Ad_{\bullet}$ is a \emph{real} representation. Indeed, induced maps are in general linear, and $\Ad_{g}$ is invertible with inverse $\Ad_{g^{-1}}$, since
    \[ \Ad_{g}\Ad_{g^{-1}}=\AD_{g*}\AD_{g^{-1}*}\big|_{e}=\left( \AD_{g}\AD_{g^{-1}} \right)_{*} \big|_{e} = \left( \AD_{gg^{-1}} \right)_{*} \big|_{e} = \AD_{e*}\big|_{e} = \bm{1}, \]
    and also satisfies the property $\Ad_{g}\Ad_{h}=\Ad_{gh}$, because $\AD_{g}\AD_{h}=\AD_{gh}$.

    Additionally, we can construct a real scalar product on $\g$ that is $\Ad$-invariant, by similar arguments as in \Cref{th:compactisunitary}. We just take any inner product $\left\langle \bullet,\bullet \right\rangle $ on $\g$ and average it over $G$ using the invariant Haar measure $\dd{g}$ as
    \[ \left( \bullet,\bullet \right) \equiv \int_{G} \left< \Ad_{g}\bullet,\Ad_{g}\bullet \right> \dd{g} \quad\Longrightarrow\quad \left( \Ad_{h}\bullet,\Ad_{h}\bullet \right)=\left( \bullet,\bullet \right) \quad\text{for any } h \in G.\]
    Thus, the Adjoint representation $\Ad_{\bullet}$ is \emph{orthogonal} (see \Cref{rem:bilinearformonSOgroup}).
\end{remark}
\begin{remark}
    One can show that the action of adjoint representation is actually given by the Lie bracket, that is $\ad_{\bm{a}}\bm{\bm{b}} = [\bm{a},\bm{b}]$. Hence, we can rewrite the Jacobi identity in \Cref{rem:jacobiidentity} as $\ad_{[\bm{a},\bm{b}]}=[\ad_{\bm{a}},\ad_{\bm{b}}]\equiv \ad_{\bm{a}}\cdot\ad_{\bm{b}}-\ad_{\bm{b}}\cdot\ad_{\bm{a}}$, which confirms that $\ad_{\bullet}$ is really a representation in the sense of \Cref{def:liealgebrarepre}.
\end{remark}
We also have very useful formulas relating the exponential and adjoint maps. They can be recognized as common identities valid for matrices in linear algebra, but they also hold generally in the abstract setting of Lie groups and algebras.
\vspace{-0.5em}
\begin{prop}[{Formulas for $\eunit^{\bullet}$, $\AD_{\bullet}$, $\Ad_{\bullet}$, and $\ad_\bullet$}] \label{prop:formulasexpadjoint}
    Let $\g$ be a Lie algebra of a Lie group $G$. Then for any $\bm{a} \in \g$ and $g \in G$ we have
    \[ \AD_{g}\!\eunit^{\bm{a}} = \eunit^{\Ad_{g}\bm{a}} \quad\text{and}\quad  \Ad_{\eunit^{\bm{a}}} = \eunit^{\ad_{\bm{a}}}, \]
    where $\eunit^{\ad_{\bm{a}}}$ is understood as a matrix exponential of $\ad_{\bm{a}} \in \End(\g)$.
\end{prop}
\vspace{-0.6em}
\begin{myproof}
    It is enough to check the following properties of the exponential:
    \vspace{-0.3em}
    \[ \AD_{g}\!\eunit^{(\tau+\sigma)\bm{a}} =\AD_{g}\left( \eunit^{\tau\bm{a}}\eunit^{\sigma\bm{a}} \right)  =  g \eunit^{\tau\bm{a}} \textcolor{gray}{g^{-1}g} \eunit^{\sigma\bm{a}}g^{-1} = \left( \AD_{g}\!\eunit^{\tau\bm{a}} \right)\left( \AD_{g}\!\eunit^{\sigma\bm{a}} \right), \vspace{-0.3em} \]
    \[ \AD_{g}\!\eunit^{0\bm{a}}= \AD_{g}\!\eunit^{\bm{0}}= \AD_{g}\bm{1} = \bm{1}, \quad \frac{\bm{D}}{d\tau} \AD_{g}\!\eunit^{\tau\bm{a}} \Big|_{\tau=0} = \Ad_{g} \frac{\bm{D}}{d\tau} \eunit^{\tau\bm{a}} \Big|_{\tau=0} = \Ad_{g}\bm{a}. \]
    Similarly for the second identity, where depending on the context \(\bm{1}\) denotes either the identity element in \(G\) or the identity element in \(\GL(\g)\): \vspace{-0.4em}
    \[ \Ad_{\eunit^{(\tau+\sigma)\bm{a}}}=\Ad_{\eunit^{\tau\bm{a}}\eunit^{\sigma\bm{a}}}= \Ad_{\eunit^{\tau\bm{a}}}\Ad_{\eunit^{\sigma\bm{a}}}, \vspace{-0.3em} \]
    \[ \Ad_{\eunit^{0\bm{a}}}=\Ad_{\eunit^{\bm{0}}} = \Ad_{\bm{1}} = \bm{1}, \quad  \frac{\bm{D}}{d\tau} \Ad_{\eunit^{\tau\bm{a}}} \big|_{\tau=0} = \ad_{\frac{\bm{D}}{d\tau} \eunit^{\tau\bm{a}} \raisebox{-0.1em}{$\scriptstyle\big|_{\tau=0}$}} = \ad_{\bm{a}}. \qedhere \]
\end{myproof}
In \Cref{ex:torus} we presented a basic example of a Lie group, namely a torus $\mathbb{T}^{n}$. Its simplicity stems from the commutative multiplication of complex numbers. Since any neighborhood of identity already generates a compact connected Lie group, this implies that the exponential map is surjective on the torus \cite{BrockerDieck1985}.

Perhaps surprisingly, we will be always able to restrict the integration of class functions on a compact connected Lie group to its \emph{maximal torus}. \vspace{-0.4em}
\begin{definition}[Maximal torus, rank, Cartan subalgebra] \label{def:torus}
    Let $G$ be a compact connected Lie group with the Lie algebra $\g$. Then we define:
    \begin{enumerate}
        \item A subgroup $T \subset G$ is a \emph{maximal torus} of $G$ if it is a torus in the sense of \Cref{ex:torus}, and there is no other torus $T'$ with $T \subsetneq T' \subset G$. Equivalently, $T$ is a maximal connected abelian subgroup of $G$, that is
              \[ g h = h g \quad\text{for any } g,h \in T. \]
              The dimension of the maximal torus $T$ in $G$ is called the \emph{rank} of $G$.
        \item A subalgebra $\mathfrak{t} \subset \g$ is a \emph{Cartan subalgebra} of $\g$ if it is a Lie algebra of some maximal torus $T$ of $G$, see \textbf{(1)}. Equivalently, $\mathfrak{t}$ is a maximal abelian subalgebra of $\g$, that is
              \[ [\bm{a},\bm{b}]=\bm{0}, \quad\text{for any } \bm{a},\bm{b} \in \mathfrak{t}. \]
    \end{enumerate}
\end{definition}
\vspace{-0.5em}
\begin{remark}
    Various alternative definitions must be proved to be equivalent. An interested reader can find the proofs for example in \cite{Sepanski2007, BrockerDieck1985}. They fundamentally exploit properties of the exponential map (see \Cref{def:exponentialmap} and \Cref{prop:formulasexpadjoint}).
\end{remark}

\begin{example}[Maximal torus of $\SO(\d)$] \label{ex:SOtorus}
    The maximal torus of $\SO(\d)$ is
    \[ T_{\SO(\d)}= \underbrace{\SO(2)\times \cdots \times \SO(2)}_{\r} \cong (S^{1})^{\r},\]
    where $\r\equiv \left\lfloor \d/2 \right\rfloor$ is the rank of $\SO(\d)\equiv \SO(2\r \textcolor{gray}{\,+\,1})$. The torus $T_{\SO(\d)}$ can be parametrized in the standard representation $\,\square\mspace{1mu}\equiv \C^{\d}\equiv \C^{2\r \textcolor{gray}{\scriptstyle+1}}$ as
    \begin{equation*}
        T =
        \left\{
            \begin{pmatrix}
                \begin{matrix}
                    \cos\theta_1 & -\sin\theta_1 \\
                    \sin\theta_1 & \mkern+15mu\cos\theta_1 \\
                \end{matrix} \\
                 & \ddots \\
                 &        &
                \begin{matrix}
                    \cos\theta_{\r} & -\sin\theta_{\r} \\
                    \sin\theta_{\r} & \mkern+15mu\cos\theta_{\r} \\
                \end{matrix} \\
                 &        &   & \textcolor{gray}{1\hspace{5pt}}
            \end{pmatrix}
        \middle|
            \begin{aligned}
                 & \theta_{j} \in [0,2\pi) \\
                 & j \in 1, \ldots, \r
            \end{aligned}
        \right\}
    \end{equation*}
    Alternatively, we can parametrize by $\r$ complex variables on the unit circle, namely by $\bm{x}\equiv (x_1, \ldots, x_{\r}) \equiv (\eunit^{\iunit\theta_{1}}, \ldots, \eunit^{\iunit\theta_{\r}})$.
\end{example}
\begin{definition}[Weyl group] \label{def:weylgroup}
    Let $T$ be a maximal torus of a connected compact Lie group $G$, and $N$ be the normalizer of $T$ in $G$, that is
    \[ N=\left\{ g \in G \middle| g T g^{-1} = T \right\}.  \]
    Then the quotient group $\mathfrak{W}\equiv N/T$ is called the \emph{Weyl group} of G.
\end{definition}
\begin{remark}
    It can be shown that the Weyl group $\mathfrak{W}$ is always finite. Intuitively, it is a group that permutes the factors in the torus.
\end{remark}

We are now ready to state one of the main theorems of this section. It is a generalization of the well-known fact that any $\SO(\d)$ matrix can be conjugated by a change of basis to the corresponding torus.
\begin{theorem}[Maximal torus theorem] \label{th:torustheorem}
    Let $T$ be a maximal torus of a connected compact Lie group $G$. Then every element $g\in G$ is \emph{conjugate} to an element of $T$, that is
    \[ \exists\,h\in G,t \in T\colon\ g = \AD_{h}t \equiv h t h^{-1}. \]
    Moreover, a general element of $G$ is conjugate to $\abs{\mathfrak{W}}$ such elements of $T$.
\end{theorem}
\begin{remark}
    One of the approaches (see \cite{BrockerDieck1985}) to proving this important theorem studies the map $\Psi$ defined in the ``proof'' of \Cref{th:weylint}.
\end{remark}
\begin{example}[Character of the standard representation $\,\square\,$ of $\SO(\d)$] \label{ex:charSO}
    \phantom{par} \\
    Since characters are \emph{class functions}, and by \Cref{th:torustheorem} every element can be conjugated to the maximal torus, to evaluate $\chi(g)$ it is enough to specify $\chi(t)$ for any corresponding torus element $t\leftrightarrow g$.
    Using the parametrization introduced in \Cref{ex:SOtorus}, we obtain
    \vspace{-0.4em}
    \begin{equation*}
        \chi_{\smallsquare}(\bm{x}\leftrightarrow g) =
        \begin{dcases}
            \mkern+31mu 2\sum_{i=1}^{\r} \cos(\theta_{i}) = \mkern+30mu \sum_{i=1}^{\r} \left( x_i + \frac{1}{x_i} \right) & \text{for } \d=2\r,   \\
            1+ 2\sum_{i=1}^{\r} \cos(\theta_{i}) = 1+\sum_{i=1}^{\r} \left( x_i + \frac{1}{x_i} \right)                    & \text{for } \d=2\r+1,
        \end{dcases}
    \end{equation*}
    \vspace{-0.4em}
    where $\,\square\mspace{1mu}\equiv \C^{\d}$ is the standard representation of $\SO(\d)$.
\end{example}
\begin{example}[Graded character of $\Sym(\square)$ for $\SO(\d)$] \label{ex:chargenSO}
    Similarly as in the previous example, using \Cref{prop:gradedchars}, \Cref{ex:SOtorus}, and realizing that eigenvalues of the torus element parametrized by $\bm{x} \leftrightarrow g$ are exactly $\{x_{i},1/x_{i}\textcolor{gray}{,1}\}_{i=1}^{\r}$, we obtain \vspace{-0.3em}
    \begin{equation*}
        P(t;\bm{x}\leftrightarrow g)\equiv \chi_{\Sym(\smallsquare)}(t;\bm{x} \leftrightarrow g) =
        \begin{dcases}
            \mkern+44mu \prod_{i=1}^{\r} \frac{1}{(1-t x_{i})(1-t/x_{i})}   & \text{for } \d=2\r,   \\
            \frac{1}{1-t} \prod_{i=1}^{\r} \frac{1}{(1-t x_{i})(1-t/x_{i})} & \text{for } \d=2\r+1.
        \end{dcases}
    \end{equation*}
    \vspace{-0.4em}
    We denoted $P\equiv \chi_{\Sym(\smallsquare)}$, since it will later turn out to be the \emph{``projection''} factor.
\end{example}

The maximal torus theorem shows the significance of the conjugation map $\AD_{\bullet}$. Naturally, it is easier to study its ``linearized'' version, the Adjoint representation $\Ad_{\bullet}$. If we consider only operators on $\g$ corresponding to the torus elements, that is \(\left\{\Ad_{t}\middle| t \in T\right\}\), and recalling that $\Ad_{\bullet}$ is orthogonal (see \Cref{rem:adjointisorthogonal}), we obtain a commuting set of normal operators. Therefore, it will be possible to perform \emph{simultaneous diagonalization}, but first we must \emph{complexify} $\g$. \vspace{-0.4em}
\begin{definition}[Complexification of Lie algebra]
    Let $\g$ be the Lie algebra of a Lie group $G$. The \emph{complexification} of $\g$ is defined by $\g_{\C} \equiv \g \otimes_{\R}\mkern+1mu\C \cong \g \oplus \mspace{2mu}\iunit\!\g$, where the algebra structure $[\bullet,\bullet]$ of $\g$ is extended to $\g_{\C}$ by $\C$-linearity. Similarly, we extend representations $(\Ad,\g)$ and $(\ad,\g)$ to $\g_{\C}$ by $\C$-linearity.
\end{definition}
\vspace{-0.2em}

\begin{definition}[Roots of Lie algebra] \label{def:rootsalgebra}
    Let $\g$ be the Lie algebra of a compact Lie group $G$ of rank $\r$, $T$ be a maximal torus of $G$, and $\mathfrak{t}$ be the corresponding Cartan subalgebra of $\g$. Then there is a set of nonzero elements $\rt(G) \subset \mathfrak{t}_{\C}^{*}$ (linear functionals on $\mathfrak{t}_{\C}$), called the \emph{\textcolor{gray}{real} roots} of $\g_{\C}$, such that we obtain the decomposition \vspace{-0.1em}
    \[ \g_{\C} = \mathfrak{t}_{\C} \oplus \bigoplus_{\alpha \smallin \rt(G)} \g_{\alpha},  \vspace{-0.5em} \]
    where $\g_{\alpha}$ is the nonzero space (called the \emph{root space}) corresponding to the \textcolor{gray}{real} root $\alpha$, which is defined by \vspace{-0.3em}
    \begin{align*}
        \g_{\alpha} & \equiv \left\{ \bm{a} \in \g_{\C} \middle| \ad_{\bm{m}}\bm{a} = [\bm{m},\bm{a}] = \iunit\alpha(\bm{m})\bm{a} \ \text{ for } \bm{m} \in \mathfrak{t} \right\} \\
                    & = \left\{ \bm{a} \in \g_{\C} \middle| \Ad_{t}\bm{a} = \Ad_{\eunit^{\bm{m}}}\bm{a} = \eunit^{\iunit\alpha(\bm{m})}\bm{a} \ \text{ for } t\equiv \eunit^{\bm{m}} \in T \right\},
    \end{align*}
    where we used \Cref{prop:formulasexpadjoint}. Choosing an orthonormal basis for $\mathfrak{t}$ and representing $\alpha \leftrightarrow \bm{\alpha}\equiv (\alpha_{1}, \ldots, \alpha_{\r})$ and $t\leftrightarrow \bm{x}\equiv (x_1, \ldots, x_{\r})\equiv (\eunit^{\iunit m_{1}}, \ldots, \eunit^{\iunit m_{\r}})$ in the spirit of \Cref{ex:SOtorus}, we also obtain
    \vspace{-0.3em}
    \[ \eunit^{\iunit\alpha(\bm{m})}=\eunit^{\iunit\sum_{i=1}^{\r}\alpha_{i} m_{i} } = \prod_{i=1}^{\r} (\eunit^{\iunit m_{i}})^{\alpha_{i}} \equiv \prod_{i=1}^{\r} x_{i}^{\alpha_{i}} \equiv \bm{x}^{\bm{\alpha}}. \]
\end{definition}
\begin{remark} \label{rem:evennumberofroots}
    We will use the standard fact that $R(G)$ is symmetric about the origin, that is if $\bm{\alpha} \in R(G)$ is a root, then $-\bm{\alpha}\in R(G)$ is a root as well. Since roots are nonzero, there is always an even number of roots.
\end{remark}

Finally, we are prepared to state and partially prove the main theorem of this section --- the \emph{Weyl integration formula}. Together with the \emph{maximal torus theorem} they open the gates to the deeper representation theory of compact Lie groups.
\begin{theorem}[Weyl integration formula] \label{th:weylint}
    Let $f$ be a class function on a connected compact Lie group $G$ of rank $\r$, $T$ be a maximal torus of $G$ parametrized by $\bm{x}\equiv (x_1, \ldots, x_{\r})$, and $\mathfrak{W}$ be the corresponding Weyl group. Then we have
    \[ \int_{G} f(g) \dd{g} = \frac{1}{\abs{\mathfrak{W}}} \varoiint\limits_{\abs{x_{i}}=1} f(\bm{x}) \left[ \prod_{\bm{\alpha} \smallin \rt(G)}(1-\bm{x^{\alpha}}) \right] \left[ \,\prod_{i=1}^{\r} \frac{\dd{x_{i}}}{2\pi i x_{i}} \right]  . \]
\end{theorem}

\begin{remark}
    We use the same notation $f$ for a class function on $G$, for its restriction to $T$, and also for its coordinate expression.
\end{remark}

\begin{myproof}[Sketch of the proof] \label{proof:weylintegrationformula}
    The idea is to restrict the integral of $f$ over the whole $G$ only to the suitable integral over $T$. Consider the map
    \[ \Psi\colon G/T \times T\to G, \quad  (gT;t) \mapsto \Psi_{gT}(t) \equiv \Psi(gT;t) \equiv g t g^{-1}. \]
    From \Cref{th:torustheorem} we know, that $\Psi$ is a finite-sheeted covering of $G$ with $\abs{\mathfrak{W}}$ sheets, where $\mathfrak{W}$ is the Weyl group of $G$. Hence, it follows that
    \[ \int_{G} f \dd{g} \equiv \int_{G} f \bm{\omega}_{G} = \int_{\Psi(G/T \times T)} f \bm{\omega}_{G} = \frac{1}{\abs{\mathfrak{W}}} \int_{G/T \times T} \Psi^{*}(f \bm{\omega}_{G}). \]
    Since $f$ is a class function, from definition of $\Psi$ we have
    \[ (\Psi^{*}f)|_{(gT;t)}=f\circ\Psi(gT,t)=f(gtg^{-1})=f(t), \]
    thus the only thing left to calculate is $\Psi^{*}\bm{\omega}_{G}$.

    To avoid dealing with many technicalities, we assume there exist left invariant normalized volume forms $\bm{\omega}_{G/T}$ on $G/T$ and $\bm{\omega}_{T}$ on $T$. In the following denote by $\pi_1$ and $\pi_2$ the natural projection maps from $G/T\times T$ to $G/T$ and $T$, respectively. We obtain a left-invariant volume form on $G/T\times T$ as
    \[ \bm{\omega}_{G/T\times T} \equiv (\pi_1^{*}\bm{\omega}_{G/T})\wedge(\pi_2^{*}\bm{\omega}_{T}). \]
    We want to find the factor $\mathfrak{D}$ relating the volume forms $\Psi^{*}\bm{\omega}_{G}$ and $\bm{\omega}_{G/T\times T}$, that is for any $(gT;t) \in G/T\times T$
    \[ \left( \Psi^{*}\bm{\omega}_{G} \right)\big|_{(gT;t)} =\mathfrak{D}(gT;t) \left[ (\pi_1^{*}\bm{\omega}_{G/T})\wedge(\pi_2^{*}\bm{\omega}_{T}) \right] \big|_{(gT;t)}. \]
    Using the left-invariance of $\bm{\omega}_{\bullet}$ we can translate the calculation to the identity by
    \begin{align*}
        \left( \Psi^{*}\bm{\omega}_{G} \right)\big|_{(gT;t)} & \equiv \Psi^{*}\left( \bm{\omega}_{G}|_{gtg^{-1}} \right)  = \Psi^{*} L^{*}_{gt^{-1}g^{-1}} \bm{\omega}_{G}|_{e} \\
                                                             & = \mathfrak{D}(gT;t)\; L^{*}_{(g^{-1};t^{-1})} \left( \left[ (\pi_1^{*}\bm{\omega}_{G/T})\wedge(\pi_2^{*}\bm{\omega}_{T}) \right] \big|_{(eT;e)} \right).
    \end{align*}
    Using linearity of $L^{*}_{(g^{-1};t^{-1})}$ and multiplying from the left by $L^{*}_{(g;t)}$, we obtain
    \begin{align*}
        \left( L_{gt^{-1}g^{-1}}\circ\Psi\circ L_{(g;t)} \right)^{*} \bm{\omega}_{G}|_{e} = \mathfrak{D}(gT;t)  \left[ (\pi_1^{*}\bm{\omega}_{G/T})\wedge(\pi_2^{*}\bm{\omega}_{T}) \right] \big|_{(eT;e)},
    \end{align*}
    so after identifying $\g$ with $\gt \oplus \mathfrak{t}$ we have \(\mathfrak{D}(gT;t) \equiv \det \left( L_{gt^{-1}g^{-1}}\circ\Psi\circ L_{(g;t)} \right)_{*}\).

    Taking any vector at identity $(\bm{a}\oplus \bm{b})\in \gt \oplus \mathfrak{t}=\g$ with the corresponding curves $x_{\bullet}\colon \R \to G/T$ and $y_{\bullet}\colon \R \to T$, that is $x_{\tau=0}\equiv eT$, $y_{\tau=0}\equiv e$, and
    \[ \bm{a} \equiv \frac{\bm{D}}{d\tau} x_{\tau}\big|_{\tau=0} \quad\text{and}\quad \bm{b} \equiv \frac{\bm{D}}{d\tau} y_{\tau}\big|_{\tau=0}, \]
    we can calculate
    \begin{align*}
        \left( L_{gt^{-1}g^{-1}}\circ\Psi\circ L_{(g;t)} \right)_{*}(\bm{a}\oplus \bm{b})                         & = \frac{\bm{D}}{d\tau}  L_{gt^{-1}g^{-1}}\circ\Psi\circ L_{(g;t)}(x_{\tau};y_{\tau}) \Big|_{\tau=0} \\
        \textcolor{gray}{\text{just apply } L_{(g;t)}} \quad                                                      & = \frac{\bm{D}}{d\tau}  L_{gt^{-1}g^{-1}}\circ\Psi(gx_{\tau};ty_{\tau}) \Big|_{\tau=0} \\
        \textcolor{gray}{\text{apply } \Psi(gT;t)\equiv gtg^{-1}} \quad                                           & = \frac{\bm{D}}{d\tau}  L_{gt^{-1}g^{-1}}(gx_{\tau})(ty_{\tau})(gx_{\tau})^{-1} \Big|_{\tau=0} \\
        \textcolor{gray}{\text{apply } L_{gt^{-1}g^{-1}} \text{ and cancel }gg^{-1}} \quad                        & = \frac{\bm{D}}{d\tau} \,g\,t^{-1}x^{\phantom{1}}_{\tau}\,t\,y^{\phantom{1}}_{\tau}\,x_{\tau}^{-1}g^{-1} \Big|_{\tau=0} \\
        \textcolor{gray}{\text{use } \AD_{g}h\equiv ghg^{-1}} \quad                                               & = \frac{\bm{D}}{d\tau} \AD_{g}\left( \AD_{t^{-1}}(x^{\phantom{1}}_{\tau})\,y^{\phantom{1}}_{\tau}\,x_{\tau}^{-1} \right)  \Big|_{\tau=0} \\
        \textcolor{gray}{\text{use } \Ad_{g} \equiv \AD_{g*}\big|_{e}\!}\quad                                     & = \Ad_{g}\frac{\bm{D}}{d\tau} \AD_{t^{-1}}(x^{\phantom{1}}_{\tau})\,y^{\phantom{1}}_{\tau}\,x_{\tau}^{-1} \Big|_{\tau=0} \\
        \phantom{\frac{D}{d\tau}}\textcolor{gray}{\text{use \Cref{rem:linearizedmultiplicationliealgebra}}} \quad & = \Ad_{g}\left( \Ad_{t^{-1}}\bm{a} + \bm{b} - \bm{a} \right) \\
        \textcolor{gray}{\text{matrix notation}} \quad                                                            & = \Ad_{g}
        \left(
            \begin{array}{c;{3pt/4pt}c}
                \Ad_{t^{-1}}\textcolor{gray}{|_{\!\gts}}\phantom{\big|}-\bm{1}_{\!\textcolor{gray}{\gts}} & \bm{0}_{\!\textcolor{gray}{\,\mathfrak{t} \to \gts}}\phantom{\big|}\mkern-7mu \\
                \hdashline[3.5pt/4pt]
                \textcolor{gray}{\cdots}                                                                  & \bm{1}_{\textcolor{gray}{\mathfrak{t}\phantom{\to \gts}}}\phantom{\big|}\mkern-7mu
            \end{array}
        \right) (\bm{a};\bm{b})
    \end{align*}

    Recalling \Cref{rem:adjointisorthogonal} for connected $G$ actually leads to the conclusion that $\Ad_{\bullet}$ is \emph{special orthogonal}, thus $\det \Ad_{g}=1$. Overall, we obtain
    \[ \mathfrak{D}(gT;t) = \det
        \left(
            \begin{array}{c;{3pt/4pt}c}
                \Ad_{t^{-1}}\textcolor{gray}{|_{\!\gts}}\phantom{\big|}-\bm{1}_{\!\textcolor{gray}{\gts}} & \bm{0}_{\!\textcolor{gray}{\,\mathfrak{t} \to \gts}}\phantom{\big|}\mkern-7mu \\
                \hdashline[3.5pt/4pt]
                \textcolor{gray}{\cdots}                                                                  & \bm{1}_{\textcolor{gray}{\mathfrak{t}\phantom{\to \gts}}}\phantom{\big|}\mkern-7mu
            \end{array}
        \right) = \det_{\gts}\left( \Ad_{t^{-1}}-\bm{1} \right). \]
    Since $\mathfrak{D}$ does not depend on $gT$, we can write $\mathfrak{D}(t)\equiv \mathfrak{D}(eT;t)$ and calculate \vspace{-0.5em}
    \[ \int_{G/T \times T} f\,\Psi^{*}(\bm{\omega}_{G}) = \int_{G/T \times T} f\,\mathfrak{D}\, \bm{\omega}_{G/T\times T} = \int_{T} f\,\mathfrak{D}\; \textcolor{gray}{\overbrace{\left( \int_{G/T} \bm{\omega}_{G/T} \right)}^{1} }  \bm{\omega}_{T}. \]
    The proof is thus finally concluded by utilizing \Cref{def:rootsalgebra} and \Cref{rem:evennumberofroots} to calculate
    \[ \mathfrak{D}(t) = \det_{\gts}\left( \Ad_{t^{-1}}-\bm{1} \right) = \prod_{\bm{\alpha}\in \rt(G)}\left( \bm{x}^{-\bm{\alpha}} - 1 \right) = \prod_{\bm{\alpha}\in \rt(G)}\left( 1- \bm{x}^{\bm{\alpha}} \right), \]
    because the invariant integration over the torus $T$ is given by
    \[ \int_{T}\textcolor{gray}{(\cdots\mkern-0.5mu)\;}\bm{\omega}_{T} = \idotsint_{0}^{2\pi} \textcolor{gray}{(\cdots\mkern-0.5mu)\;} \prod_{i=1}^{\r} \frac{\dd{\theta_{i}}}{2\pi} \equiv  \varoiint\limits_{\abs{x_{i}}=1} \textcolor{gray}{(\cdots\mkern-0.5mu)\;} \prod_{i=1}^{\r} \frac{\dd{x_{i}}}{2\pi i x_{i}}. \qedhere \]
\end{myproof}

\begin{remark}
    We chose the parametrization \(\bm{x}\equiv (x_1, \ldots, x_{\r})\equiv (\eunit^{\iunit \theta_{1}}, \ldots, \eunit^{\iunit \theta_{\r}})\) of the torus $T$ such that it is covered exactly once when \(\{\theta_{j}\}_{j}^{\r}\) run through the interval \([0,2\pi)\). This is important for the choice of normalization of the roots in \Cref{def:rootsalgebra}.
\end{remark}

\begin{remark}
    Every class function is invariant with respect to the Weyl group $\mathfrak{W}$, thus we also have the simplified form (for derivation see \cite[Appendix B]{Operator2017})
    \[ \int_{G} f(g) \dd{g} = \varoiint\limits_{\abs{x_{i}}=1} f(\bm{x}) \underbrace{\left[ \prod_{\bm{\alpha} \smallin \rt_{+}(G)}(1-\bm{x^{\alpha}}) \right]}_{\mathfrak{D}\mspace{-1mu}_{\mspace{-1mu}G}^{+}(\bm{x})}  \left[ \prod_{i} \frac{\dd{x_{i}}}{2\pi i x_{i}} \right], \]
    where $\rt_{+}(G)$ is the set of so-called \emph{positive roots}.
\end{remark}

\begin{example} \label{ex:restrictedmeasureSO}
    Specifically, for $G=\SO(\d)$ we have explicit forms \cite{Operator2017}
    \begin{equation*}
        \mathfrak{D}_{\mspace{-1mu}\SO(\d)}^{+}(\bm{x}) = \begin{dcases}
            \prod_{1\le i < j \le \r} ( 1-x_{i} x_{j} ) ( 1- x_{i}/x_{j} )                               & \text{for } \d=2\r,   \\
            \prod_{i=1}^{\r} ( 1-x_{i} )  \prod_{1\le i < j \le \r} ( 1-x_{i} x_{j} ) ( 1- x_{i}/x_{j} ) & \text{for } \d=2\r+1.
        \end{dcases}
    \end{equation*}
\end{example}

\chapter{The Master Formula} \label{ch:masterformuladerivation}

In this chapter we will derive the formula for the Hilbert series by utilizing everything we learned about the deep representation theory of compact Lie groups. At first, we will work in the special setting of a single real scalar (bosonic) field,
where the arguments are more transparent and thus easier to follow. Afterward, we will outline the generalization of the formalism to arbitrary field content.

\section{Strategy for scalar field} \label{sec:strategy}

The main strategy for a scalar field $\Phi$ follows as:
\begin{enumerate}
    \item Construct the \emph{Single Particle Graded Representation} $R_{\Phi}$ (SPGR) containing all possible operators consisting of only one $\Phi$ and any number of derivatives (of the type $\partial^{n}\Phi$), excluding ones that are trivially zero by use of EOM.
          \begin{flushright}
              --- addressing EOM relations
          \end{flushright}
    \item Combine multiple $R_{\Phi}$ (consistently with the statistics of $\Phi$) to create the \emph{Multi-Particle Graded Representation} $\J_{\Phi}$ (MPGR) capturing the highly redundant set of all possible non-trivial operators (of the type $\partial^{n}\Phi^{r}$).
          \begin{flushright}
              --- representation theory inherently\\
              addresses GDC relations
          \end{flushright}
    \item Project out the multiplicities of independent Lorentz and gauge invariant operators from $\J_{\Phi}$ by use of character orthogonality and formulas for graded characters, thus obtaining the Hilbert series for the operator basis $\B$, already viewed as a graded dimension of $\K\equiv \Span\B$.
          \begin{flushright}
              --- picking out only scalar operators\\
              and addressing IBP relations
          \end{flushright}
\end{enumerate}

This can be summarized in the diagram
\[ \begin{tikzcd}[column sep=10.5em]
        \Phi \arrow[r, mapsto, "\bigoplus_{n=0}^{\infty} \D^{n} \partial^{n}\bullet", "\text{and EOM}"' ] &[-5.2em]
        R_{\Phi} \arrow[r, mapsto, "\bigoplus_{r=0}^{\infty} \phi^{r}\Sym^{r}(\bullet )", "\text{since $\Phi$ is a boson}"' ] &[-3.5em]
        \J_{\Phi} \arrow[r, mapsto, "{\int_{\SO(\d)} \frac{1}{P(\D;g)} \chi_{\bullet }(\phi,\D; g) \dd{g}}", "\text{and Weyl integration formula}"' ] &
        H_{0}(\phi,\D),
    \end{tikzcd} \]
where $H_{0}(\phi,\D)$ is the ``main'' part (to be defined) of the full Hilbert series $H(\phi,\D)$, and the factor $1/P(\D;g)$ accounts for the IBP relations.


We will treat steps \textbf{(1)} and \textbf{(2)} in \Cref{sec:EOMredundancy}, afterward in \Cref{sec:IBPcohomology} we will derive the factor $1/P(\D;g)$ appearing in the step \textbf{(3)}, and we will finalize the derivation for the scalar field in \Cref{sec:masterformulaHilbertseries}, with a pretty straightforward generalization in \Cref{sec:generalizationarbitraryfieldcontent}.

\section{Equations of motion redundancy} \label{sec:EOMredundancy}

We would like to build $R_{\Phi}$ by repeatedly applying derivatives on $\Phi$, but we also need to continuously utilize all possible relations to avoid any redundancies. At this point only EOM relations can play a role, so we need to identify them.

The kinetic Lagrangian density of a single scalar field $\Phi$ has the form
\[ \mathcal{L}_{\text{kin}}(\Phi,\partial_{a}\Phi) \equiv \frac{1}{2} \partial_{a} \Phi \partial^{a} \Phi, \]
which leads to the free equation of motion (known as the \emph{wave equation})
\begin{equation*} \label{eq:scalarEOM}
    \wick{\c1\partial\c1\partial\Phi} \equiv  \partial^{a}\partial_{a}\Phi = 0.
\end{equation*}
Applying one derivative, we obtain $\partial_{a}\Phi$, which will be our prototype of an object transforming under the standard representation $\,\square\mspace{1mu}\equiv \C^{\d}$ of $\mspace{1mu}\SO(\d)$. Alternatively we could work with $\partial^{a}\Phi$, however for $\SO(\d)$ this leads to an equivalent representation, since we have the invariant metric tensor $\bm{\delta}_{ab}$ enabling us to lower indices.

When we apply multiple derivatives we obtain symmetric representations $\displaystyle\Sym^{n}(\square)$, since partial derivatives automatically commute. But additionally, we must not forget to apply EOM, thus we need to discard all \emph{trace} parts of any operator. Since for the fully symmetric tensor (representation) there is only one independent contraction, this motivates us to define the representations below.
\begin{definition}[Traceless symmetric representations of $\SO(\d)$] \label{def:tracelesssymmetricrepre}
    Let $V$ be a representation of $\SO(\d)$, and $Q(\bullet,\bullet)$ be a corresponding positive-definite bilinear form (see \Cref{rem:bilinearformonSOgroup}).
    We have a natural \emph{contraction} for $n\ge 2$
    \begin{align*}
        C\colon \Sym^{n}(V)      & \to \Sym^{n-2}(V), \\
        \bm{v}_1 \cdots \bm{v}_n & \mapsto \sum_{i,j} Q(\bm{v}_{i},\bm{v}_{j}) \bm{v}_{1} \cdots \widehat{\bm{v}}_{i} \cdots \widehat{\bm{v}}_{j}\cdots \bm{v}_{n},
    \end{align*}
    where hats indicate that corresponding vectors are to be omitted, and which can easily be checked to be a $G$-map using the $G$-invariance of $Q(\bullet,\bullet)$.
    Thus, $\Ker(C)$ is a subrepresentation of $\displaystyle\Sym^{n}(V)$, and it gives us a decomposition
    \[ \Sym^{n}(V) = \Ker(C) \oplus \frac{\Sym^{n}(V)}{\Ker(C)}\cong  \Ker(C) \oplus \Im(C) \equiv \Sym^{\{n\}}(V) \oplus \Sym^{n-2}(V), \]
    since $C$ is obviously surjective.

    Consequently, we have (by definition $\displaystyle\Sym^{\{0\}}(V)\equiv \C$ and $\displaystyle\Sym^{\{1\}}(V)\equiv V$)
    \[ \Sym^{n}(V) = \Sym^{\{n\}}(V) \oplus \Sym^{\{n-2\}}(V) \oplus \cdots \oplus \Sym^{\{n-2p\}}(V),\]
    where $p=\left\lfloor \frac{n}{2} \right\rfloor$. We call representations $\Sym^{\{n\}}(V)$ as \emph{traceless symmetric}.
\end{definition}
\begin{remark}
    A similar construction can be made for the tensor power $V^{\otimes n}$, only there we have more contractions, so decomposing it leads to the traceless part and multiple copies of $V^{\otimes n-2}$.
\end{remark}
\begin{remark}
    Suppose we have a representation $W$ and assume there exists some $G$-invariant contraction $C$ on $W$. Then if $W$ wants to be irreducible, it necessarily must be \emph{traceless}, otherwise $\Ker(C)$ would be its nonzero proper subrepresentation (nonzero since it maps to a lower-dimensional space).
\end{remark}
\begin{remark}
    It can be shown that the representations $\displaystyle\Sym^{\{n\}}(\square)$ for the standard representation $\,\square \mspace{1mu} \equiv \C^{\d}$ of $\SO(\d)$ are already irreducible \cite{Fulton1999}.
\end{remark}

We can create the \emph{traceless symmetric graded representation} $\displaystyle\Sym^{\{\bullet\}}(V)$ similarly to $\Sym(V)$.
It is in our interest to calculate the corresponding graded character.
\begin{prop}[Graded character of traceless symmetric graded representation] \label{prop:tracelesscharacter}
    Let $V$ be a representation of a compact Lie group $G$. We define the \emph{traceless symmetric graded representation} $\Sym^{\{\bullet\}}(V)$ of $V$ by
    \[ \Sym^{\{\bullet\}}(V) \equiv \bigoplus_{n=0}^{\infty} t^{n} \Sym^{\{n\}}(V). \]

    The corresponding graded character $\chi_{\Sym^{\{\bullet\}}(V)}(t;g)$ is given by
    \[ \chi_{\Sym^{\{\bullet\}}(V)}(t;g) \equiv \sum_{n=0}^{\infty} t^{n} \chi_{\Sym^{\{n\}}(V)}(g) =(1-t^2)\chi_{\Sym(V)}(t;g)=\frac{1-t^2}{\det_{V}(1-tg)}. \]
\end{prop}
\begin{myproof}
    Using \Cref{def:tracelesssymmetricrepre} we obtain for $n\ge 2$.
    \[ \Sym^{n}(V) \equiv \Sym^{\{n\}}(V) \oplus \Sym^{n-2}(V) \ \xRightarrow{\ref{prop:charactersproperies}} \ \chi_{\Sym^{n}(V)}(g)= \chi_{\Sym^{\{n\}}(V)}(g) + \chi_{\Sym^{n-2}(V)}(g). \]
    For convenience, we extend the definition of $\chi_{\Sym^{n}(V)}$ for negative $n$ to be zero. We can now simply express $\chi_{\Sym^{\{n\}}(V)}(g)$ and calculate
    \begin{align*}
        \sum_{n=0}^{\infty} t^{n} \chi_{\Sym^{\{n\}}(V)}(g) & = \sum_{n=0}^{\infty} t^{n} \left[\chi_{\Sym^{n}(V)}(g) - \chi_{\Sym^{n-2}(V)}(g)\right] \\
                                                            & = (1-t^{2})\sum_{n=0}^{\infty} t^{n} \chi_{\Sym^{n}(V)}(g) \equiv (1-t^{2}) \chi_{\Sym(V)}(t;g),
    \end{align*}
    thus the proof is finished by \Cref{prop:gradedchars}.
\end{myproof}

\begin{derivation}[Single Particle Graded Representation $R_{\Phi}$ for a single scalar field $\Phi$] \label{der:SPGRscalarfield}
    All operators composed of one $\Phi$ modulo EOM are therefore contained in
    \begin{equation*}
        R_{\Phi} = \Span
        \begin{pmatrix}
            \Phi \\
            \partial_{a}\Phi \\
            \partial_{\{a_{1}}\partial_{a_{2}\}}\Phi \\
            \textcolor{gray}{\vdots} \\
            \partial_{\{a_{1}}\cdots\partial_{a_{n}\}}\Phi \\
            \textcolor{gray}{\vdots}
        \end{pmatrix}
        \equiv \bigoplus_{n=0}^{\infty} \D^{n} \Sym^{\{n\}}(\square) \equiv \Sym^{\{\bullet \}}(\square),
    \end{equation*}
    where $_{\{\,\cdots\}}$ denotes the traceless symmetric part and $\,\square\mspace{1mu}\equiv \C^{\d}$ denotes the standard representation of $\SO(\d)$. We suggestively changed the grading labels $t \mapsto \D$.
\end{derivation}

\begin{derivation}[Multi-Particle Graded Representation $\J_{\Phi}$ for a single scalar field $\Phi$] \label{der:MPGRscalarfield}
    Since $\Phi$ is a boson, the corresponding operators must obey permutation symmetry. We can obtain all operators modulo EOM in the symmetric powers of $R_{\Phi}$, thus
    \[ \J_{\Phi} \equiv \bigoplus_{r=0}^{\infty} \phi^{r}\Sym^{r}(R_{\Phi}) \equiv \Sym(R_{\Phi}) \equiv  \Sym\left(\Sym^{\{\bullet \}}(\square)\right). \]

    The corresponding graded character can be calculated using \Cref{cor:plethysticexponentialgradedchars} and \Cref{prop:tracelesscharacter} as (where we used notation $P(\D;g)$ from \Cref{ex:chargenSO})
    \begin{equation*}
        \chi_{\J_{\Phi}}(\phi,\D;g) \equiv \chi_{\Sym(R_{\Phi})}(\phi,\D;g) = \PE\left[\phi\,\chi_{R_{\Phi}}(\D;g)\right] = \PE\left[\phi (1-\D^{2}) P(\D;g)\right].
    \end{equation*}
\end{derivation}
\begin{remark}[Gram determinant conditions]
    We are building $\J_{\Phi}$ in a certain sense constructively. Since the representation theory intimately understands that representations $\displaystyle\Ext^{n}(\square)$ for $n>\d$ are trivially zero, operators which generate GDC relations (see \Cref{sec:relations}) are automatically discarded during the process.
\end{remark}
\begin{remark}
    Alternatively, we can view $\J_{\Phi}$ as a graded differential polynomial quotient ring with the ideal generated by the equations of motion
    \begin{align*}
        \J_{\Phi} & =\bigslant{\C\left[\Phi;\partial_{a}\right]}{\left< \wick{\c1\partial\c1\partial\Phi} \right>} \\
                  & \equiv \C\left[\Phi,\partial_{a}\Phi,\partial_{\{a_1}\partial_{a_2\}}\Phi, \textcolor{gray}{\dots\,}, \partial_{\{a_1}\cdots\partial_{a_n\}}\Phi,\textcolor{gray}{\dots\,}\right].
    \end{align*}
\end{remark}

\section{Integration by parts redundancy} \label{sec:IBPcohomology}

This section will be already applicable in general, but we can always imagine we are working with a scalar field $\Phi$.

Suppose we have constructed the redundant representation $\J$ of all possible operators modulo EOM, together with its graded character. Our goal is to find the Hilbert series, which is the graded dimension of $\K\equiv \Span \B \subset \J$. We need to deal with IBP relations and also pick out only scalar (Lorentz and gauge invariant) operators. This will be addressed by a \emph{cohomology}-type calculation.

We would like to proceed similarly as in part \textbf{(3)} of \Cref{ex:dim1}, where we realized that any non-trivial operator generates exactly one IBP relation between operators containing one more derivative. Likewise for $\d\ge 2$, operators with one free index can generate IBP relations, but only those that are not trivially zero after applying the divergence. Prime example of operators which do not contribute to the IBP relations are operators of the form
\[ \partial^{a}\mathcal{O}_{ab} \;\text{ where }\; \mathcal{O}_{ab}\equiv \mathcal{O}_{[ab]} \quad\Longrightarrow\quad \partial^{b}\partial^{a}\mathcal{O}_{ab}=\partial^{(b}\partial^{a)}\mathcal{O}_{[ab]}=0, \]
that is so called co-exact 1-forms. Thus, we are naturally led to the following elementary concepts from homology/cohomology/Hodge theory.
\begin{definition}[Co-closed and co-exact forms] \label{def:coclosedcoexactforms}
    Let $\Sect \bm{\Lambda}^{\!k}\mathcal{M}$ be the space of $k$-forms on the space-time manifold $\mathcal{M}$. We define the following notation for a form $\bm{\omega} \in \Sect \bm{\Lambda}^{\!k}\mathcal{M}$ (thus $\bm{\sigma} \in \Sect \bm{\Lambda}^{\!k+1}\mathcal{M}$)
    \begin{alignat*}{4}
        \bm{\omega} & \text{ is \emph{co-closed}} & \hspace{1.3em} & \overset{\text{def}}{\Longleftrightarrow} & \hspace{1.3em} &  & \partial\cdot\bm{\omega}               & = 0, \\
        \bm{\omega} & \text{ is \emph{co-exact}}  &                & \overset{\text{def}}{\Longleftrightarrow} &                &  & \exists\,\bm{\sigma}\colon \bm{\omega} & =\partial\cdot\bm{\sigma}.
    \end{alignat*}
\end{definition}
\begin{remark} \label{rem:co-exactisco-closed}
    For forms we automatically have $\partial\cdot\partial\cdot\bullet = 0$ by symmetry of partial derivatives and asymmetry of forms, thus every co-exact form is also co-closed.
\end{remark}

In our case we work with the Euclidean space-time manifold $\mathcal{M}\equiv \R^{\d}$ together with the (Euclidean) Lorentz group $\SO(\d)$, so the $k$-forms transform under the representation $\displaystyle\Ext^{k}(\square)$, where $\,\square\mspace{1mu}\equiv \C^{\d}$ is the standard representation of $\SO(\d)$.

We would like to identify how many forms of a given degree are contained in $\J$, and use this information to count the operators in $\K$. This motivates us to study $\SO(\d)$-linear maps between $\displaystyle\Ext^{k}(\square)$ and $\J$, because images of such non-trivial maps identify subrepresentations of $\J$ with similar or same transformation properties as those of $k$-forms.

\begin{definition}[$k$-form graded representation]
    Let $\J$ be a multi-particle graded representation of the group $\SO(\d)$. The \emph{$k$-form graded representation} of $\J$ is defined by
    \[ \J_{[k]} \equiv \Hom_{\SO(\d)}\left(\Ext^{k}(\square),\J\right).\]
    We say that $\mathcal{O} \in \J$ is a \emph{$k$-form operator} (we denote it by $\mathcal{O}_{[k]}$) if $\mathcal{O}$ is in the image of some map from $\J_{[k]}$. Furthermore, by a straightforward extension of \Cref{def:coclosedcoexactforms} to $\J_{[k]}$ we can define its subspaces, for example
    \[ \J_{[k]\text{co-closed}} \equiv \Hom_{\SO(\d)}\left(\Ext^{k}(\square), \left\{ \mathcal{O} \in \J \middle| \partial\cdot\mathcal{O}=0 \right\} \right), \]
    \[ \J_{[k]\text{co-exact}} \equiv \Hom_{\SO(\d)}\left(\Ext^{k}(\square),\left\{ \mathcal{O} \in \J \middle| \exists\, \mathcal{O}'_{[k+1]}\colon \mathcal{O} = \partial\cdot\mathcal{O}'_{[k+1]} \right\} \right). \]
\end{definition}

To show this definition is indeed sensible, we need to make several comments about the exterior representations of $\SO(\d)$.
\begin{remark} \label{rem:hodgeduality}
    On any pseudo-Riemannian space, but we will specifically take $\R^{\d}$, we have the \emph{Hodge star operator} $\bm{*}\colon \bm{\Lambda}^{\!k}\R^{\d} \to \bm{\Lambda}^{\!\d-k}\R^{\d}$
    defined in components by
    \begin{equation*}
        \bm{\omega}_{a_1 \cdots a_{k}}\ \xmapsto{\ \bm{*}\ }\ (\bm{*}\bm{\omega})_{a_{k+1}\cdots a_{\d}}\equiv \frac{1}{k!}\bm{\omega}^{a_1 \cdots a_{k}} \bm{\epsilon}_{a_1 \cdots a_{k} a_{k+1} \cdots a_{\d}},
    \end{equation*}
    where $\bm{\epsilon}$ is the so-called ($\d$-dimensional) \emph{Levi-Civita} tensor.

    Since $\bm{\epsilon}$ is invariant with respect to $\SO(\d)$ transformations, after $\C$-linear extension of Hodge star $\bm{*}$ to $\,\square\mspace{1mu}\equiv \C^{\d}$ we get an isomorphism
    \[ \Ext^{k}(\square)\cong \Ext^{\d-k}(\square) \quad\Longrightarrow\quad \J_{[k]}\cong \J_{[\d-k]}. \]
    This just reflects that representation theory ``understands'' that any $k$-form can be transformed by contraction with $\bm{\epsilon}$ to a $(\d-k)$-form, or vice versa.

\end{remark}
\begin{remark}
    It is a standard result that representations $\displaystyle\Ext^{k}(\square)$ are irreducible for $k<\d/2$ \cite{BrockerDieck1985,Fulton1999} (thus also for $k>\d/2$, see \Cref{rem:hodgeduality}). This is very important, because for such $k$ we can use \Cref{cor:projectionformula}, or maybe even better \Cref{rem:gradedmultiplicity}.

    But then there is the case of $k=\d/2$ for even $\d$, which requires a little more work to be analyzed. The Hodge star operator satisfies the identity
    \[ \bm{*} \circ \bm{*} = (-1)^{k(\d-k)}\bm{1}, \]
    so for $\d=2k$ we obtain the automorphism of $k$th exterior representation of $\,\square$
    \[ \tau\equiv(-i)^{k}\bm{*}\colon \Ext^{k}(\square) \to \Ext^{k}(\square), \quad  \tau\circ\tau=\bm{1}, \]
    giving us the canonical decomposition of \(\displaystyle\Ext^{k}(\square)\) to two subrepresentations, namely to the \emph{self-dual} representation and the \emph{anti-self-dual} representation as
    \[ \Ext^{k}(\square) = \Ext\nolimits_{\,+}^{\!k}(\square) \oplus \Ext\nolimits_{\,-}^{\!k}(\square), \]
    which are eigenspaces of $\tau$ corresponding to the eigenvalues $1$ and $-1$, respectively.
    These already turn out to be irreducible (and mutually inequivalent), therefore
    \[ \dim \Hom_{\SO(\d)}\left(\displaystyle\Ext^{k}(\square),\displaystyle\Ext^{k}(\square) \right) = \sum_{\pm } \dim \Hom_{\SO(\d)}\left(\displaystyle\Ext^{k}(\square),\Ext\nolimits_{\,\pm }^{\!k}(\square) \right)=1+1=2. \]
    This is exactly what we want, because each of the $2$ independent $k=\d/2$ form representations plays their own role in the following derivation.
\end{remark}

\begin{derivation}[Addressing IBP relations by cohomology, splitting $H=H_0 + \Delta H$] \label{der:IBPrelationscohomology}
    Since IBP relations tell us that total divergence terms are equivalent to zero, the representation $\K\equiv \Span\B$ is composed of all $0$-forms contained in $\J$ modulo the co-exact ones. The Hilbert series is thus given by the graded multiplicities of trivial representations in $\J$ that are not co-exact, that is
    \[ H(\bm{\phi},\D) \equiv \dim_{(\bm{\phi},\D)}\K =\dim_{(\bm{\phi},\D)}\J_{[0]\text{not co-exact}}. \]
    For the sake of brevity, we will write just $\dim \bullet$ instead of $\dim_{(\bm{\phi},\D)}\bullet$. Since to every non-trivial co-exact $k$-form corresponds exactly one $(k+1)$-form that is not co-closed (see \Cref{def:coclosedcoexactforms}), by keeping track of the right grading we obtain
    \begin{align*}
        \dim \K
                                                             & = \overbrace{\dim \J_{[0]} - \phantom{\D}\ \dim \J_{[0]\text{co-exact}}}^{\dim \J_{[0]\text{not co-exact}} } \\
                                                             & = \dim \J_{[0]} - \D\ \dim \J_{[1]\text{not co-closed}} \\
                                                             & = \dim \J_{[0]} - \D\left( \dim \J_{[1]} - \dim \J_{[1]\text{co-closed}} \right) \\
                                                             & = \dim \J_{[0]} - \D\hspace{-0.05em}\bigg(\mkern-5mu \dim \J_{[1]} - \underbrace{\dim \J_{[1]\text{co-exact}}}_{\mathclap{\D\dim \J_{[2]\text{not co-closed}}}}  - \dim \J_{[1]\substack{\text{co-closed} \\ \text{not co-exact}}} \bigg)  \\[-1.8em]
        \raisebox{-0.27em}{\text{\scriptsize iteratively}\!} & \vdotswithin{=} \\[-1.04em]
                                                             & \vdotswithin{=} \\[-1em]
                                                             & = \underbrace{\sum_{k=0}^{\d} (-\D)^{k}  \dim \J_{[k]}}_{H_{0}} \,+\, \underbrace{\sum_{k=1}^{\d} (-1)^{k+1}\D^{k} \dim \J_{[k]\substack{\text{co-closed} \\ \text{not co-exact}}}}_{\Delta H},
    \end{align*}
    where we used the splitting of co-closed forms to the co-exact ones and to the rest (see \Cref{rem:co-exactisco-closed}). Iteration terminates because there are no $k$-forms with $k>\d$.
\end{derivation}

Hence, we found that the Hilbert series naturally splits into two pieces, where we will intuitively infer in \Cref{rem:contributionsDeltaH} that $\Delta H$ is a small correction to the ``main'' part $H_{0}$. Since $H_{0}$ usually contains almost the whole information of $H$, we sometimes refer to it also as the Hilbert series. We will obtain an exact formula for $H_0$, but calculation of $\Delta H$ is in some sense less straightforward.

\begin{remark}[Contributions to $\Delta H$] \label{rem:contributionsDeltaH}
    One such co-closed but not co-exact form is the Levi-Civita tensor $\bm{\epsilon}_{\bullet \cdots \bullet } \equiv  *1$ (where $*$ denotes the Hodge dual). This follows because it is constant, thus co-closed, and also a top-form, so there is no form of higher degree. We therefore always have a contribution of $-(-\D)^{\d}$ to $\Delta H$, because we overcount $\bm{\epsilon}_{\bullet \cdots \bullet }$ in the $H_{0}$ part.

    In the case of a single scalar field $\Phi$, the one-form $\partial_{a}\Phi$ also falls down to that category. Co-closedness results from EOM, as $\partial\cdot\partial\Phi\equiv \partial^{a}\partial_{a}\Phi =0$, and it is obvious that it is not co-exact. Since it contains one derivative, and we need another one to bring it down to a $0$-form, we obtain a contribution of $\phi \D^{2}$ to $\Delta H$.

    One could try to find other examples in $\J_{[k]\substack{\text{co-closed}\\ \text{not co-exact}}}$, but in the case of a scalar field there are none. That such terms are pretty rare sounds plausible. If the divergence of a form is not automatically zero by its components' symmetry, only other possibilities are that it is either constant (only $\bm{\epsilon}_{\bullet \cdots \bullet }$) or it vanishes by EOM. This typically leads to only a couple of instances, since the divergence of a form containing numerous particle fields is unlikely to be identically zero (for example $\partial^{a}\left( \partial_{a}\Phi \Phi \right) = \cancel{\partial^{a}\partial_{a}\Phi \Phi}{} + \partial_{a}\Phi \partial^{a} \Phi \neq 0$), and if we have to use only one particle field, there are not many such forms because $\partial_{[\bullet}\partial_{\bullet]}\equiv 0$.

    We conclude that for a single scalar field we have
    \[ \Delta H(\phi,\D) = (-1)^{\d+1}\D^{\d} + \phi\D^{2}. \]
    This agrees with an explicit formula for $\Delta H$, which is specific to the case where SPGR corresponds to a \emph{conformal} representation (in particular for a scalar field), derived by \textcite{Operator2017}.
\end{remark}

\section{Integral formula for the Hilbert series} \label{sec:masterformulaHilbertseries}

In this section we finalize the derivation of the formula for the Hilbert series, namely for the ``main'' part $H_0$. We essentially have done the hard work in \Cref{sec:IBPcohomology}, now we just (in the same spirit as in \Cref{cor:projectionformula}) express the graded dimension occurring in $H_0$ through integration of the graded characters.

\begin{derivation}[Projection factor $1/P(\D;g)$ addressing IBP relations] \label{der:projectionfactorIBP}
    First, we can perform the manipulations (just to obtain a nice alternative expression)
    \begin{multline*}
        H_{0}(\bm{\phi},\D) \equiv \sum_{k=0}^{\d} (-\D)^{k}  \dim_{(\bm{\phi},\D)} \J_{[k]} \equiv \sum_{k=0}^{\d} (-\D)^{k}  \dim_{(\bm{\phi},\D)} \Hom_{\SO(\d)}\left(\Ext^{k}(\square),\J\right) \\
        = \dim_{(\bm{\phi},\D)} \Hom_{\SO(\d)}\left(\bigoplus_{k=0}^{\d} (-\D)^{k}\!\Ext^{k}(\square),\J\right) \equiv \dim_{(\bm{\phi},\D)} \Hom_{\SO(\d)}\left(\Ext^{-}(\square),\J\right),
    \end{multline*}
    where $\displaystyle\Ext^{-}(\square)$ is the exterior graded representation of $\,\square\mspace{1mu}$, but with alternating signs in the grading. Using \Cref{prop:gradedchars} we can calculate its graded character as
    \[ \chi_{\Ext^{\raisebox{0.10em}{$\scriptstyle -$}}(\smallsquare)}(\D; g) \equiv \sum_{n=0}^{\infty} (-\D)^{n}\chi_{\Ext^{\raisebox{0.10em}{$\scriptstyle n$}}(\smallsquare)}(g) = \det_{\smallsquare}(1-\D g) = \frac{1}{\chi_{\Sym(\smallsquare)}(\D; g)} \equiv \frac{1}{P(\D;g)}. \]
\end{derivation}

\begin{derivation}[Master Formula for $H_0$] \label{der:masterformula}
    Since \Cref{der:IBPrelationscohomology} was not specific to a single scalar field, we quite generally obtain \emph{the Master Formula} for $H_{0}$ as
    \begin{align*}
        H_0(\bm{\phi},\D) & = \dim_{(\bm{\phi},\D)} \Hom_{\SO(\d)}\left(\Ext^{-}(\square),\J\right) = \int_{\SO(\d)} \chi_{\Ext^{\raisebox{0.10em}{$\scriptstyle -$}}(\smallsquare)}(\D;g^{-1})\,\chi_{\J}(\bm{\phi},\D;g) \dd{g} \\
                          & \!\!= \int_{\SO(\d)} \underbrace{\det_{\smallsquare}(1-\D g^{\intercal})}_{\det_{\smallsquare}(1-\D g)} \chi_{\J}(\bm{\phi},\D;g) \dd{g} \equiv \int_{\SO(\d)} \frac{1}{P(\D;g)}\chi_{\J}(\bm{\phi},\D;g) \dd{g},
    \end{align*}
    where we used equality $\bm{(2)}$ of \Cref{cor:projectionformula} extended to the case of graded representations, together with relation $g^{-1}=g^{\intercal}$ valid for $\SO(\d)$ (see \Cref{ex:compactliegroups}).
    Integration can be further simplified by restricting it to the torus $T$ of $\SO(\d)$ (using the Weyl integration formula discussed in \Cref{sec:weylintegrationformula}).
\end{derivation}

For the special case of a single scalar field we therefore obtain
\[ H(\phi,\D) = \underbrace{\int_{\SO(\d)} \frac{1}{P(\D;g)} \PE\left[ \phi (1-\D^{2}) P(\D;g) \right] \dd{g}}_{H_{0}(\phi,\D)}  \,+\, \underbrace{(-1)^{\d+1}\D^{\d} + \phi\D^{2}}_{\Delta H(\phi,\D)}.  \]
We will continue the calculation in \Cref{sec:scalarfield}.

\section{Generalization to arbitrary field content} \label{sec:generalizationarbitraryfieldcontent}

We are \emph{almost} ready to generalize the derivation to arbitrary field content. To say \emph{almost} is actually a big overstatement, since there are numerous places where it is necessary to use sophisticated tools far beyond the scope of this thesis. Nonetheless, the big picture can be understood very well after the worked out case of a single scalar field.

Our starting data for any EFT are:
\begin{itemize}
    \item Particle fields $\{\Phi_{i}\}$ together with a specification of their representations under the Lorentz group $\SO(\d)$ and a possible internal group $G$ composed of gauge groups and other (for example global) symmetry groups. We always assume particle fields transform linearly under $G$.
    \item EOM generated from the kinetic Lagrangian density $\mathcal{L}_{\text{kin}}$.
    \item Possibly some other constraints, an example being the Bianchi identities for the Maxwell tensor 
          \[ 3 \partial_{[a}F_{bc]} = \mathrm{d}_{a}F_{bc} = 0. \]
\end{itemize}

We follow the strategy in \Cref{sec:strategy} with slight modifications, see the diagram below. An additional internal group $G$ is accounted for simply by assigning to every operator its corresponding character with respect to $G$. Their only role will be played out at the end, where we just project out the scalar operators by integration over~$G$.

\( \begin{tikzcd}[column sep=10.0em]
    \{\Phi_{i}\}
    \arrow[r, mapsto, "\bigoplus_{n=0}^{\infty} \D^{n}\partial^{n}\bullet", "\text{and EOM}"' ]
    &[-5.1em] \{R_{\Phi_{i}}\}
    \arrow[r, mapsto, "{\bigotimes_{i} \left( \bigoplus\limits_{r=0}^{\infty} \phi_{i}^{r}\;\substack{\raisebox{-0.10em}{\scalebox{0.85}{$\textstyle\Sym$}}^{r}\mspace{-2mu}(\bullet_{i} ) \\ \textcolor{gray}{\Ext^{\raisebox{0.10em}{$\scriptstyle r$}}(\bullet_{i} ) }} \right)}", "\Phi_{i} \text{ is a }\, \substack{\text{boson}\\ \text{\textcolor{gray}{fermion}}}"' ]
    &[-3.3em] \J
    \arrow[r, mapsto, "{\int_{\SO(\d) \times G} \frac{1}{P(\D;g)} \chi_{\bullet }(\bm{\phi},\D; g) \dd{g}}", "\text{and Weyl integration formula}"' ]
    & H_{0}(\bm{\phi},\D)
\end{tikzcd} \)
\begin{remark}
    Now that we work with the direct product of groups $\SO(\d)\times G$ (see \Cref{def:groupsdirectproduct}), we will understand group elements as $g\equiv (g_{\SO(\d)};g_{G})$. In the following we will just write the dependence on $g$, but it is easy to recognize that some parts of expressions depend only on $g_{\SO(\d)}$, while others only on $g_{G}$.
\end{remark}
\begin{remark}[Fermions]
    We need to proceed more carefully with particle fields that are fermions (spinors), since objects transforming spinorially must be integrated over the covering group $\Spin(\d)$ of $\SO(\d)$. This leads to some technicalities, so we will focus mainly on bosons. We refer the interested reader to \cite{Operator2017,OperatorsSMEFT2015}.
\end{remark}

\begin{derivation}[{Single Particle Graded Representations $\{R_{\Phi_{i}}\}$ for general fields}] \label{der:SPGRgeneral}
    Once again, we start by building the single particle graded representations $\{R_{\Phi_{i}}\}$.
    In the case of a scalar field it was useful to decompose representations obtained just by applying derivatives to smaller pieces, since some of them were trivially zero by use of EOM. \emph{Decomposition to irreducible representations} under the action of $\SO(\d)$ is effective in general, because any constraint (EOM or other) always discards one or several whole irreducible representations. The more complicated example of electromagnetic field (photons) will be shown in \Cref{sec:photonselectomagneticfield}.

    If we successfully apply all EOM and other relations, we will be left with $\{R_{\Phi_{i}}\}$, which decompose to the space-time $\SO(\d)$ part and the internal $G$ part as
    \[ R_{\Phi_{i}}=R_{\SO(\d),\Phi_{i}}\otimes R_{G,\Phi_{i}}. \]
    The corresponding graded characters are then
    \[ \chi_{R_{\Phi_{i}}}\!(\D;g) \equiv \chi_{R_{\SO(\d),\Phi_{i}}\otimes R_{G,\Phi_{i}}}\!(\D;g) = \chi_{R_{\SO(\d),\Phi_{i}}}\!(\D;g)\, \chi_{G,\Phi_{i}}\mkern-2mu(g), \]
    where $\chi_{G,\Phi_{i}}$ is just the character of the internal group $G$ representation of the particle field $\Phi_{i}$, since all operators in $R_{\Phi_{i}}$ contain only one $\Phi_{i}$.
\end{derivation}
\begin{remark}[Young diagrams] \label{rem:youngdiagrams}
    Tensor powers $\,\square^{\otimes n}$ of the standard representation of $\SO(\d)$ are typically decomposed by usage of so-called \emph{Young symmetrizers} \cite{GoodmanWallach2009}. The idea it to exploit mutually commuting actions of $\SO(\d)$ and the \emph{symmetric} group $\mathfrak{S}_{n}$ (group of permutations on $n$ symbols), which acts by permuting the indices of the tensor. We can thus decompose $\,\square^{\otimes n}$ to subrepresentations already irreducible under $\mathfrak{S}_{n}$, particular examples already showed in \Cref{def:symmetricexterior}, where $\displaystyle\Sym^{n}(\square)$ corresponds to the trivial representation under permutations, and $\displaystyle\Ext^{n}(\square)$ corresponds to the \emph{sign} representation. Other irreducible representations have ``mixed'' symmetries. Any irreducible representation of $\mathfrak{S}_{n}$ is uniquely described by a \emph{Young diagram} with $n$ boxes, that will be intuitively used in \Cref{sec:photonselectomagneticfield}.
\end{remark}

\begin{derivation}[Multi-Particle Graded Representation $\J$ for general fields] \label{der:MPGRgeneral}
    \phantom{For}
    For $\Phi_{i}$ boson/\textcolor{gray}{fermion}, we obtain all operators containing only this particle field in symmetric/\textcolor{gray}{exterior} powers of $R_{\Phi}$, thus $\J_{\Phi_{i}}$ is
    \[ \J_{\Phi_{i}} \equiv  \bigoplus\limits_{r=0}^{\infty} \phi_{i}^{r}\;\substack{\raisebox{-0.10em}{\scalebox{0.9}{$\Sym$}}^{r}\mspace{-2mu}\left(R_{\Phi_i} \right) \\ \textcolor{gray}{\Ext^{\raisebox{0.10em}{$\scriptstyle r$}}\left(R_{\Phi_{i}} \right) }}
        = \substack{\raisebox{-0.10em}{\scalebox{0.9}{$\Sym$}}\mspace{1mu}\left( R_{\Phi_{i}} \right) \\ \textcolor{gray}{\Ext\left(R_{\Phi_{i}}\right)}}. \]
    The full \emph{Multi-Particle Graded Representation} $\J$ containing all possible operators (modulo EOM and other relations) is obtained just by taking the tensor product
    \[ \J \equiv \bigotimes_{i} \J_{\Phi_{i}} = \bigotimes_{i}
        \substack{\raisebox{-0.10em}{\scalebox{0.9}{$\Sym$}}\mspace{1mu}\left( R_{\Phi_{i}} \right) \\ \textcolor{gray}{\Ext\left(R_{\Phi_{i}}\right)}}. \]
    The corresponding graded character is given by \Cref{cor:plethysticexponentialgradedchars} as
    \[ \chi_{\J}(\bm{\phi},\D;g) = \prod_{i} \chi_{\J_{\Phi_{i}}}\!(\phi_{i},\D; g) = \prod_{i}\PE\left[ \phi_{i}\, \chi_{R_{\Phi_{i}}}\!(\D;g) \right] = \PE\left[ \sum_{i} \phi_{i}\, \chi_{R_{\Phi_{i}}}\!(\D;g) \right],  \]
    with the proper statistics understood in $\PE$ (see \Cref{rem:plethysticexponential}).
\end{derivation}

\begin{derivation}[General Master Formula for the Hilbert series] \label{der:MasterFormulageneral}
    Finally, by the same arguments as in \Cref{sec:IBPcohomology}, only with additional integration over the internal group $G$, we state \emph{the Master Formula}
    \begin{equation*}
        H(\bm{\phi},\D) = \underbrace{\int_{\SO(\d) \times G} \frac{1}{P(\D;g)}\chi_{\J}(\bm{\phi},\D;g) \dd{g}}_{H_0(\bm{\phi},\D)} \,+\, \Delta H(\bm{\phi},\D),
    \end{equation*}
    where $\Delta H$ corrects for miscalculations in $H_0$ caused by existence of co-closed but not co-exact forms in $\J$, namely
    \begin{align*}
        \Delta H(\bm{\phi},\D) & = \sum_{k=1}^{\d} (-1)^{k+1}\D^{k} \dim_{(\bm{\phi},\D)} \J_{[k]\substack{\text{co-closed} \\ \text{not co-exact}}} \\
                               & \equiv \sum_{k=1}^{\d} (-1)^{k+1}\D^{k} \dim_{(\bm{\phi},\D)} \Hom_{\SO(\d)\times G}\left( \Ext^{k}(\square), \J_{\substack{\hspace{-0.7em}\text{co-closed} \\ \text{not co-exact}}} \right).
    \end{align*}
\end{derivation}

In the next chapter we will apply the formalism to some simple examples. Hopefully, they will suffice to unveil the main techniques and solve some ambiguities of our exposition up to this point. Of course, there is much more to learn, and we will try to refer the interested reader to the relevant sources.

\chapter{Applications of the Formalism} \label{ch:applications}

We are now ready to employ the formalism developed in the preceding chapters.
We will not be able to obtain the full \emph{Hilbert series}, but by expanding the Plethystic Exponentials we will focus on parts with fixed field content (see \Cref{rem:partialhilbertseries}).

In general, the final integrals we will obtain are just, to put it simply, \emph{gigantic}.
Nevertheless, the results can be obtained by ``just'' calculating a finite number of residues, since the integrands are rational functions. This can be automated by use of \emph{computer algebra systems}, such as \texttt{Mathematica}. Since it is not possible to show all the details of the following computations, we encourage the reader to go through the accompanying \texttt{Mathematica notebook} \cite{Dujava2022}.

First, we will finish the calculation for a single scalar field. Analysis of the results will give us numerous enlightening inputs on the structure of the operator basis. We will give a brief discussion regarding the dependence of the Hilbert series on the dimension of the space-time $\d$. After that, we will show on a slightly more complicated example of the electromagnetic field (photons) how to proceed when the building block of the Lagrangian has non-trivial transformation properties. \vspace{-0.5em}

\section{Single scalar field} \label{sec:scalarfield}\vspace{-0.5em}
For concreteness, we will write out equations only for $\d=4$. Picking up where we left off the case of a single scalar field in \Cref{sec:masterformulaHilbertseries}, the usage of the Weyl integration formula  for $\SO(4)$ (see \Cref{th:weylint} and \Cref{ex:restrictedmeasureSO}) leads to
\begin{align*}
    H_{0}(\phi,\D) & = \int_{\SO(\mathsf{4})} \frac{1}{P(\D;g)} \overbrace{\PE\left[ \phi (1-\D^{2}) P(\D;g) \right]}^{\chi_{\J_{\Phi}}(\phi,\D;g)}  \dd{g} \\
                   & =\varoiint\limits_{\substack{\abs{x_{1}}=1 \\ \abs{x_{2}}=1 }}  (1-\D x_1)(1-\D/x_1)(1-\D x_2)(1-\D/x_2) \textcolor{gray}{\cross} \\[-2.2em]
                   & \hspace{5em}  \textcolor{gray}{\cross} \PE\left[ \frac{\phi (1-\D^{2})}{(1-\D x_1)(1-\D/x_1)(1-\D x_2)(1-\D/x_2)} \right] \textcolor{gray}{\cross} \\
                   & \hspace{15em} \textcolor{gray}{\cross} (1-x_1 x_2)(1-x_1/x_2)\frac{\dd{x_1}}{2\pi i x_1} \frac{\dd{x_2}}{2\pi i x_2},
\end{align*}
where we have $\abs{\D}<1$ (see \Cref{def:hilbertseries}), $\bm{x}=(x_1,x_2)$ parametrizes the torus $T$ of $\SO(4)$, and we used \Cref{ex:chargenSO} to obtain
\[ P(\D;\bm{x}) \equiv \chi_{\Sym(\smallsquare)}(\D;\bm{x}) = \frac{1}{(1-\D x_1)(1-\D/x_1)(1-\D x_2)(1-\D/x_2)}. \]

\begin{calculation}[Fixed field content $\Phi^{\bm{r}}, \bm{r}\le \bm{3}$]
    It is a straightforward exercise to show that the only non-trivial operators with the field content $\Phi^{\bm{r}}, \bm{r}=\bm{1},\bm{2},\bm{3}$, are exactly $\Phi^{\bm{1}}$, $\Phi^{\bm{2}}$, and $\Phi^{\bm{3}}$ (we use bold numbers only for visibility). This indeed checks out with the calculation through the Hilbert series formalism for any given $\d$, see the accompanying \texttt{Mathematica notebook}. We must not forget the term
    \[ \Delta H(\phi,\D) = (-1)^{\d+1}\D^{\d} + \phi\D^{2} \]
    for $\bm{r}=\bm{1}$, otherwise it does not contribute.
\end{calculation}

\begin{calculation}[Fixed field content $\Phi^{\bm{4}}$]
    The simplest non-trivial case are operators with the power $\bm{r}=\bm{4}$ of $\Phi$ (that is of the type $\partial^{n}\Phi^{\bm{4}}$). Just by expanding the Plethystic Exponential defined in \Cref{rem:plethysticexponential}, we get ($\bm{t}$ stands for any set of parameters, and the notation $|_{\phi=4}$ means the coefficient in front of $\phi^{4}$)
    \[ \PE\left[\phi f(\bm{t})\right]\big|_{\phi^{4}} = \frac{1}{24} \left(f(\bm{t})^4+6 f(\bm{t})^2 f(\bm{t}^2)+3 f(\bm{t}^2)^2+8 f(\bm{t}) f(\bm{t}^3)+6 f(\bm{t}^4)\right), \]
    where in the place of $f(\bm{t})$ we substitute $(1-\D^{2}) P(\D;\bm{x})$. For example, the term $f(\bm{t}^{4})$ corresponds to ($\bm{t}^{4}$ means we take the $4$th power of every argument)
    \[ (1-\D^{8}) P(\D^{4};\bm{x}^{4}) = \frac{(1-\D^{8})x_1^{4}x_2^{4}}{(1-\D^{4} x_1^{4})(x_1^{4}-\D^{4})(1-\D^{4} x_2^{4})(x_2^{4}-\D^{4})}. \]
    Just to carry out the calculation a little further, we multiply this term by the rest of the integrand, after which using the residue theorem twice and performing a considerable amount of algebra (not included here) results in
    \begin{align*}
         & \int_{\SO(4)} \frac{1}{P(\D;g)} \frac{1}{4} (1-\D^{8}) P(\D^{4};g^{4})  \dd{g} = \frac{1}{4}(1-\D^8) \varoiint\limits_{\abs{x_{i}}=1}  \frac{\dd{x_1}}{2\pi i} \frac{\dd{x_2}}{2\pi i} \textcolor{gray}{\cross} \\
         & \hspace{5em} \textcolor{gray}{\cross} \frac{(1-x_1 x_2)(x_2-x_1)(1-\D x_1)(x_1-\D)(1-\D x_2)(x_2-\D) x_1^2 x_2  }{(1-\D^4 x_1^4)(x_1^4-\D^4)(1-\D^4 x_2^4)(x_2^4-\D^4) } \\
         & \hspace{2em}=\frac{1}{4} \oint\limits_{\abs{x_{2}}=1} \frac{\dd{x_2}}{2\pi i} \frac{x_2 }{(1+\D^2 x_2^2)(x_2^2+\D^2)} = \frac{1}{4} \frac{1}{1-\D^{4}}
    \end{align*}
    Poles in the first integration are $(\cancel{\D}{removable},-\D,i\D,-i\D)$, and in the second $(i\D,-i\D)$.
\end{calculation}

If we wanted to continue the whole calculation by hand, we would probably lose that hand. Expanding the Plethystic Exponential gave us $5$ terms, each leading to a pretty large integrand. Poles are easy to find, but either we have poles of higher degree, so we must take derivatives to calculate the residues, or there are multiple (up to $3$) poles for a single term. And we must do that twice, for $x_1$ and $x_2$.

Thankfully, \texttt{Mathematica} comes to the rescue. \emph{Simply} by calculating the indicated residues and summing them all, we obtain (since $\Delta H|_{\phi^{4}}=0$)
\[ H_{\bm{4}}(\D) \equiv H(\phi,\D)\big|_{\phi^{4}} =\frac{1}{(1-\D^{4})(1-\D^{6})}=1+\D^{4}+\D^{6}+\D^{8}+\D^{10}+2\D^{12} + \cdots, \]
reproducing the result calculated in \cite{Operator2017}. This implies there is only one independent operator of the type $\partial^{0}\Phi^{\bm{4}}\equiv \Phi^{\bm{4}}$ (obviously), one of the type $\partial^{4}\Phi^{\bm{4}}$, one of the type $\partial^{6}\Phi^{\bm{4}}$, but two of the type $\partial^{12}\Phi^{\bm{4}}$, and so on.

\begin{remark}[Interpretation of the Hilbert series] \label{rem:interpretationhilbert}
    The interpretation of this Hilbert series is slightly different from the ones in \Cref{ex:dim1}, to which \Cref{rem:hilbertseriesform} is directly applicable. Since we have now fixed the operator field content, the form of $H_{\bm{4}}$ reflects independent possibilities of assigning a given number of derivatives to the $\bm{4}$ fields $\Phi$ and contracting them.

    For example, we can take the following assignments of $4$ and $6$ derivatives,
    \begin{equation*}
        \wick{ \c2\partial\c1\partial\Phi\c1\partial\Phi\c2\partial\Phi\Phi }
        \equiv  \partial^{b}\partial_{a}\Phi\partial^{a}\Phi\partial_{b}\Phi\Phi
        \qq{and}
        \wick{ \c2\partial\c1\partial\Phi\c1\partial\c1\partial\Phi\c1\partial\c2\partial\Phi\Phi }
        \equiv  \partial^{c}\partial_{a}\Phi\partial^{a}\partial_{b}\Phi\partial^{b}\partial_{c}\Phi\Phi,
    \end{equation*}
    to be \emph{``generators''}, and any operator of the type $\partial^{n}\Phi^{\bm{4}}$ can be already constructed by repeatedly assigning derivatives and contracting them in the same way.
\end{remark}

\begin{remark}[Challenge]
    It is pretty easy to transform any operator of the type $\partial^{4}\Phi^{\bm{4}}$ to the trivial zero operator or to the one stated above, for example \vspace{0.3em}
    \begin{align*}
        \wick{ \c1\partial\Phi\c1\partial\Phi\c1\partial\Phi\c1\partial\Phi }
        \equiv  \partial_{a}\Phi\partial^{a}\Phi\partial_{b}\Phi\partial^{b}\Phi
        =                                                                                            & \;\cancel{\partial^{b}( \partial_{a}\Phi\partial^{a}\Phi\partial_{b}\Phi\Phi )}{$\sim 0$ by IBP}   - \cancel{\partial_{a}\Phi\partial^{a}\Phi\left( \partial^{b}\partial_{b}\Phi \right) \Phi}{$\sim 0$ by EOM} \\
        \textcolor{gray}{\text{it is understood }} \qquad\qquad                                      & - \partial^{b}\partial_{a}\Phi\partial^{a}\Phi\partial_{b}\Phi\Phi  -\partial_{a}\Phi\partial^{b}\partial^{a}\Phi\partial_{b}\Phi\Phi \\
        \textcolor{gray}{\text{that } \partial\Phi\Phi\equiv (\partial\Phi)\Phi} \mkern+8mu \qquad = & - 2\partial^{b}\partial_{a}\Phi\partial^{a}\Phi\partial_{b}\Phi\Phi
        \equiv - 2 \wick{ \c2\partial\c1\partial\Phi\c1\partial\Phi\c2\partial\Phi\Phi}.
    \end{align*}
    We encourage the reader to show that any (non-trivial) operator of the type $\partial^{6}\Phi^{\bm{4}}$ is equivalent to the one stated above. If it does not feel like a challenge, have a go at operators of the type $\partial^{8}\Phi^{\bm{4}}$ or $\partial^{10}\Phi^{\bm{4}}$. Hopefully, it will be sufficient evidence, that even in the simplest example of a single scalar field, the combination of EOM and IBP relations leads to complexity that should not be underrated.
\end{remark}

And everything gets exponentially more tangled for a higher number of fields, where GDC relations start to kick in. Still, the Hilbert series formalism can give us (if we are able to perform the integration) information to all orders in powers of derivatives for any fixed field content $\Phi^{\bm{r}}$.

The authors of \cite{Operator2017} explicitly calculated the Hilbert series of a single scalar field $H_{\bm{r}}(\D)$ for $\bm{r}=\bm{1},\bm{\ldots}, \bm{8}$, but only in the dimension $\d=4$. As is expected, our calculations agree with theirs. It is very hard to go beyond the field content of $\Phi^{\bm{8}}$ (integrands become so lengthy, that even \texttt{Mathematica} has a really rough time), so we will rather explore the dependency of the Hilbert series on the dimension $\d$.

\begin{figure}[!ht]
    \centering
    \includegraphics[width=1\linewidth]{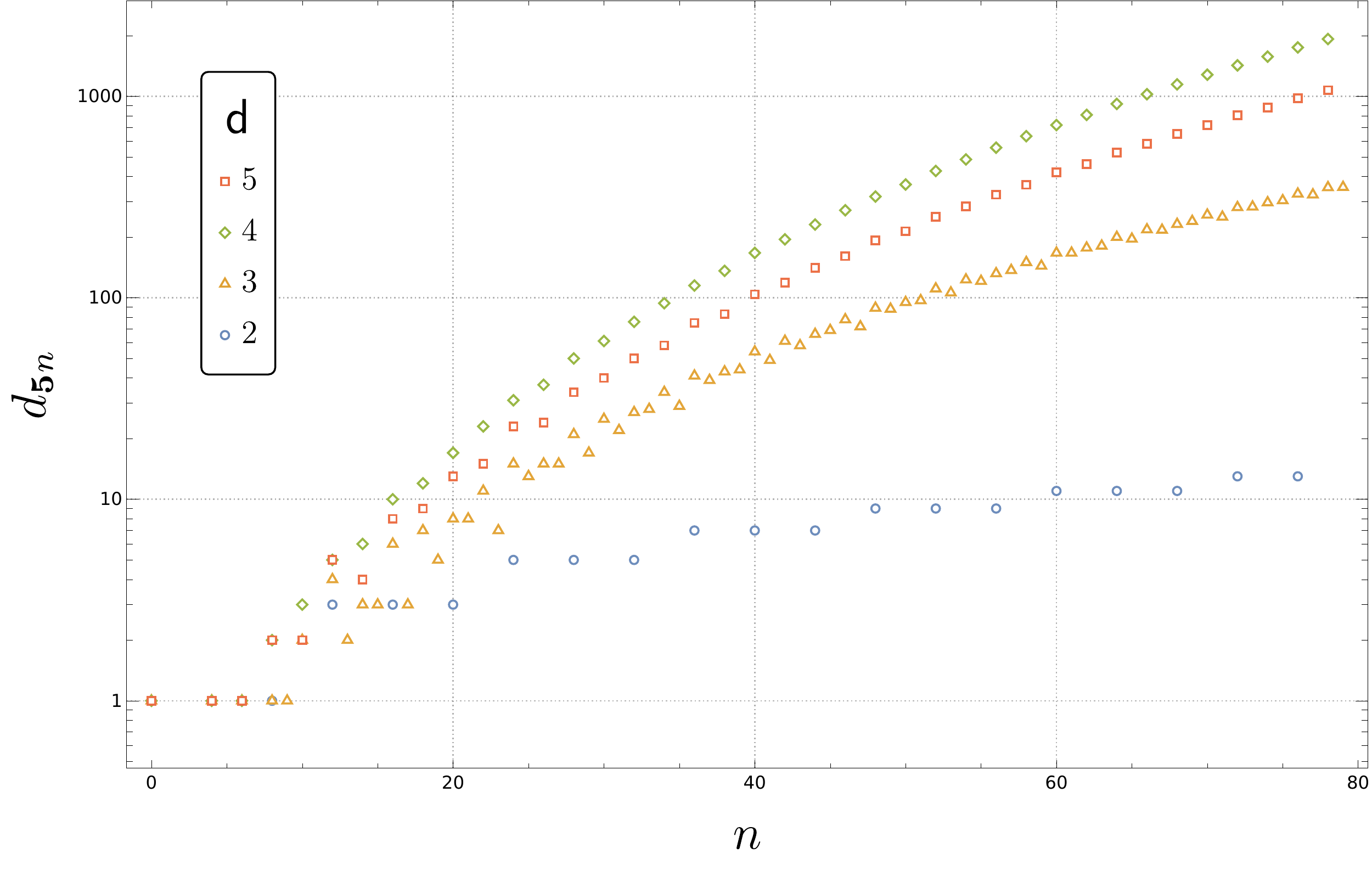}
    \caption{Semi-log plot of the coefficients $\opers{5}{n}$ in dimensions $\d=2, \ldots, 5$. In other words, the number of independent operators of the type $\partial^{n}\Phi^{\bm{5}}$.}
    \label{fig:singleboson5}
\end{figure}

\begin{figure}[!ht]
    \centering
    \includegraphics[width=1.0\linewidth]{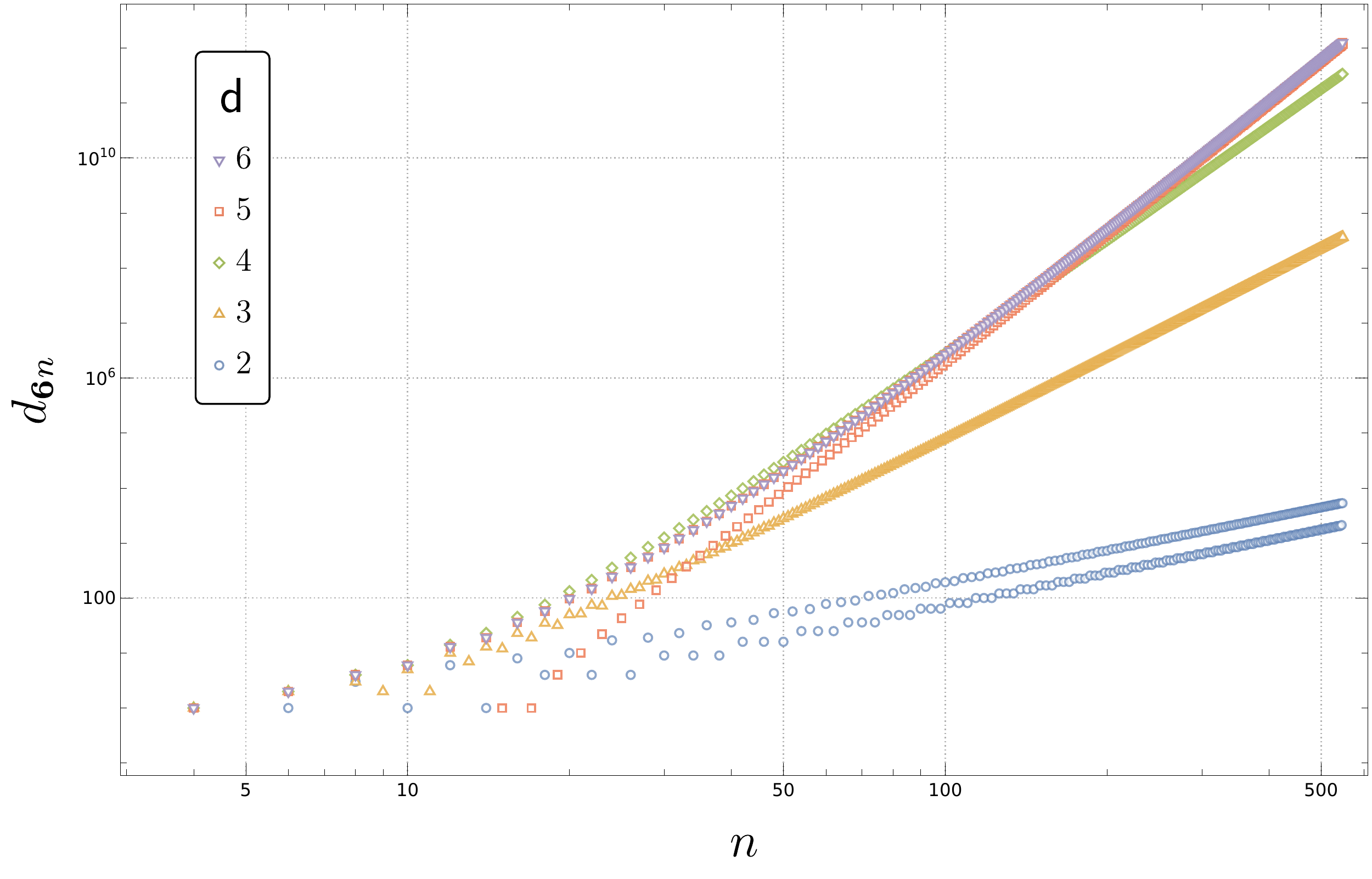}
    \caption{Log-log plot of the coefficients $\opers{6}{n}$ in dimensions $\d=2, \ldots, 6$. In other words, the number of independent operators of the type $\partial^{n}\Phi^{\bm{6}}$.}
    \label{fig:singleboson6}
\end{figure}

Below, in \Cref{tab:hilbert_series_for_a_single_scalar_particle_for_the_fixed_field_content_phi_4_5,tab:hilbert_series_for_a_single_scalar_particle_for_the_fixed_field_content_phi_6} we list the Hilbert series for $\bm{r}=\bm{4},\bm{5},\bm{6}$ and all dimensions $\d\ge 2$. For illustration, we list the corresponding expansions to the order of $\partial^{20}\Phi^{\bm{r}}$ in \Cref{tab:hilbert_series_coefs_4,tab:hilbert_series_coefs_5,tab:hilbert_series_coefs_6}. We will make a comment in \Cref{sec:sidenotegramdeterminantconditions} concerning the independence of $H_{\bm{r}}(\D)$ on $\d$ for $\d\ge \bm{r}$. For easier comparison of their dependence on $\d$ we also include \Cref{fig:singleboson5,fig:singleboson6}. To our knowledge, results of $H_{\bm{r}}(\D)$ for $\bm{r}=\bm{5},\bm{6}$ in dimensions other than $\d=4$ are \emph{novel}.

\begin{remark}[Analysis of $H_{\bm{4}}(\D)$]
    Comparing $H_{\bm{4}}(\D)$ for $\d=3$ and $\d=4$, we can see in \Cref{tab:hilbert_series_for_a_single_scalar_particle_for_the_fixed_field_content_phi_4_5} that for $\d=3$ we have one additional operator for every one in $\d=4$, but with $9$ more derivatives. This corresponds to the operator
    \[ \bm{\epsilon}^{abc}\wick{(\c3\partial\c2\partial\c1\partial\partial_{a}\Phi)(\c1\partial\c2\partial\partial_{b}\Phi)(\c3\partial\partial_{c}\Phi)(\Phi)}. \]
    Such an $\bm{\epsilon}$-assignment of derivatives ``can be used only once'', because by repeating it twice, we can deconstruct two $\bm{\epsilon}^{\bullet \cdots \bullet}$ as a combination of $\bm{\delta}^{\bullet\bullet}$, thus only contractions, which are already accounted for by the denominator.

    Why is there no such operator in $\d=4$, where we can contract derivatives with $\bm{\epsilon}^{abcd}$? Since derivatives commute, in order to obtain non-trivial operator, we must contract $\bm{\epsilon}^{abcd}$ with derivatives standing before different fields $\Phi$. But we are always able to move all derivatives in front of one $\Phi$ to the others by use of IBP relations, thus such an operator does not exist. This applies in general for scalar fields, so operators with $\bm{\epsilon}^{\bullet \cdots \bullet}$ exist only for $\d<r$. Note that the only source of operators with an odd number of derivatives are $\bm{\epsilon}^{\bullet \cdots \bullet}$ in odd dimensions.

    Somehow, by going down to $\d=2$, we completely lose the operator of the type $\partial^{6}\Phi^{\bm{4}}$ (and also its ``iterations''). We will study this in more detail in \Cref{sec:sidenotegramdeterminantconditions}.
\end{remark}

\begin{table}[!ht]
    \begin{adjustbox}{center}
        \begin{tabular}{rcccc}
            \toprule
            $\d$    &  & $H_{\bm{4}}(\D)$                                                      &  & $H_{\bm{5}}(\D)$                                                                                                                                                                                               \\
            \midrule\addlinespace[0.5em]
            $\ge 5$ &  &                                                                       &  & $\displaystyle\frac{1+\D^{12}+\D^{14}+\D^{16}+\D^{18}+\D^{30}}{\left(1-\D^4\right) \left(1-\D^6\right) \left(1-\D^8\right) \left(1-\D^{10}\right) \left(1-\D^{12}\right)}$                                     \\\addlinespace[1.0em]
            $4$     &  & $\displaystyle\frac{1}{\left(1-\D^4\right) \left(1-\D^6\right)}$      &  & $\displaystyle\frac{1+\D^{10}+\D^{12}+2 \D^{14}+2 \D^{16}+\D^{18}+\D^{22}+\D^{24}+\D^{28}+\D^{30}}{\left(1-\D^4\right) \left(1-\D^6\right) \left(1-\D^8\right) \left(1-\D^{10}\right) \left(1-\D^{12}\right)}$ \\\addlinespace[1.0em]
            $3$     &  & $\displaystyle\frac{1+\D^9}{\left(1-\D^4\right) \left(1-\D^6\right)}$ &  & $\displaystyle\frac{1+\D^9+\D^{12}+\D^{13}+\D^{14}+2 \D^{15}+\D^{16}+\D^{17}+\D^{18}+\D^{21}+\D^{30}}{\left(1-\D^4\right) \left(1-\D^6\right) \left(1-\D^{10}\right) \left(1-\D^{12}\right)}$                  \\\addlinespace[1.0em]
            $2$     &  & $\displaystyle\frac{1}{1-\D^4}$                                       &  & $\displaystyle\frac{1+\D^{12}}{\left(1-\D^4\right) \left(1-\D^{12}\right)}$                                                                                                                                    \\\addlinespace[0.5em]
            \bottomrule
        \end{tabular}
    \end{adjustbox}
    \caption{The Hilbert series for a single scalar field (fixed field content $\Phi^{\bm{4}}$ and $\Phi^{\bm{5}}$).}
    \label{tab:hilbert_series_for_a_single_scalar_particle_for_the_fixed_field_content_phi_4_5}
\end{table}

\begin{table}[!ht]
    \begin{adjustbox}{center}
        \begin{tabular}{rcc}
            \toprule
            $\d$    &  & $H_{\bm{6}}(\D)$                                                                                                                                                              \\
            \midrule\addlinespace[0.5em]
            $\ge 6$ &  & $\displaystyle\frac{\substack{1+2 \D^{10}+5 \D^{12}+7 \D^{14}+9 \D^{16}+11 \D^{18}+13 \D^{20}+14 \D^{22}+21 \D^{24}+24 \D^{26}                                                \\+28 \D^{28}+32 \D^{30}+26 \D^{32}+22 \D^{34}+13 \D^{36}+7 \D^{38}+3 \D^{40}+\D^{42}+\D^{44}}}{\left(1-\D^4\right) \left(1-\D^6\right)^2 \left(1-\D^8\right)^3 \left(1-\D^{10}\right)^2 \left(1-\D^{12}\right)}$                                                                                                                                                                    \\\addlinespace[1.0em]
            $5$     &  & $\displaystyle\frac{\substack{1+2 \D^{10}+5 \D^{12}+7 \D^{14}+\D^{15}+9 \D^{16}+\D^{17}+11 \D^{18}+3 \D^{19}+13 \D^{20}+7 \D^{21}+14 \D^{22}+13 \D^{23}+21 \D^{24}            \\ +22 \D^{25}+24 \D^{26}+26 \D^{27}+28 \D^{28}+32 \D^{29}+32 \D^{30}+28 \D^{31}+26 \D^{32}+24 \D^{33}+22 \D^{34}+21 \D^{35} \\ +13 \D^{36}+14 \D^{37}+7 \D^{38}+13 \D^{39}+3 \D^{40}+11 \D^{41}+\D^{42}+9 \D^{43}+\D^{44}+7 \D^{45}+5 \D^{47}+2 \D^{49}+\D^{59}}}{\left(1-\D^4\right) \left(1-\D^6\right)^2 \left(1-\D^8\right)^3 \left(1-\D^{10}\right)^2 \left(1-\D^{12}\right)}$ \\\addlinespace[1.0em]
            $4$     &  & $\displaystyle\frac{\substack{1+3 \D^{10}+6 \D^{12}+11 \D^{14}+17 \D^{16}+22 \D^{18}+31 \D^{20}+36 \D^{22}+48 \D^{24}+53 \D^{26}+58 \D^{28}                                   \\ +58 \D^{30}+48 \D^{32}+38 \D^{34}+23 \D^{36}+14 \D^{38}+6 \D^{40}+4 \D^{42}+2 \D^{44}+\D^{46}}}{\left(1-\D^4\right) \left(1-\D^6\right)^2 \left(1-\D^8\right)^3 \left(1-\D^{10}\right) \left(1-\D^{12}\right)}$                                                                                                                                                                   \\\addlinespace[1.0em]
            $3$     &  & $\displaystyle\frac{\substack{1+\D^8+2 \D^9+2 \D^{10}+2 \D^{11}+3 \D^{12}+5 \D^{13}+4 \D^{14}+6 \D^{15}+5 \D^{16}+6 \D^{17}+6 \D^{18}+6 \D^{19}+5 \D^{20}+6 \D^{21}+6 \D^{22} \\ +5 \D^{23}+6 \D^{24}+6 \D^{25}+6 \D^{26}+5 \D^{27}+6 \D^{28}+4 \D^{29}+5 \D^{30}+3 \D^{31}+2 \D^{32}+2 \D^{33}+2 \D^{34}+\D^{35}+\D^{43}}}{\left(1-\D^4\right) \left(1-\D^6\right)^2 \left(1-\D^8\right) \left(1-\D^{10}\right) \left(1-\D^{12}\right)}$                                                                                                                          \\\addlinespace[1.0em]
            $2$     &  & $\displaystyle\frac{1+\D^4+\D^6+2 \D^8+\D^{10}+3 \D^{12}+3 \D^{16}+\D^{18}+\D^{22}}{\left(1-\D^8\right) \left(1-\D^{12}\right)^2}$                                            \\\addlinespace[0.5em]
            \bottomrule
        \end{tabular}
    \end{adjustbox}
    \caption{The Hilbert series for a single scalar field (fixed field content $\Phi^{\bm{6}}$).}
    \label{tab:hilbert_series_for_a_single_scalar_particle_for_the_fixed_field_content_phi_6}
\end{table}

\begin{table}[!ht]
    \begin{adjustbox}{center}
        \begin{tabular}{rccccccccccccccccccccccc}
            \toprule
                    &  &     & \multicolumn{21}{c}{$\opers{4}{n}$}                                                          \\
            \cmidrule{3-24}
            $\d$    &  & $n$ & 0 & 1 & 2 & 3 & 4 & 5 & 6 & 7 & 8 & 9 & 10 & 11 & 12 & 13 & 14 & 15 & 16 & 17 & 18 & 19 & 20 \\
            \cmidrule{1-1} \cmidrule{3-24}\addlinespace[0.5em]
            $\ge 4$ &  &     & 1 & 0 & 0 & 0 & 1 & 0 & 1 & 0 & 1 & 0 & 1  & 0  & 2  & 0  & 1  & 0  & 2  & 0  & 2  & 0  & 2  \\
            3       &  &     & 1 & 0 & 0 & 0 & 1 & 0 & 1 & 0 & 1 & 1 & 1  & 0  & 2  & 1  & 1  & 1  & 2  & 1  & 2  & 1  & 2  \\
            2       &  &     & 1 & 0 & 0 & 0 & 1 & 0 & 0 & 0 & 1 & 0 & 0  & 0  & 1  & 0  & 0  & 0  & 1  & 0  & 0  & 0  & 1  \\
            \bottomrule
        \end{tabular}
    \end{adjustbox}
    \caption{Coefficients of the Hilbert series for a single scalar field (fixed field content $\Phi^{\bm{4}}$).}
    \label{tab:hilbert_series_coefs_4}
\end{table}

\begin{table}[!ht]
    \begin{adjustbox}{center}
        \begin{tabular}{rccccccccccccccccccccccccccc}
            \toprule
                    &  &     & \multicolumn{21}{c}{$\opers{5}{n}$}                                                          \\
            \cmidrule{3-24}
            $\d$    &  & $n$ & 0 & 1 & 2 & 3 & 4 & 5 & 6 & 7 & 8 & 9 & 10 & 11 & 12 & 13 & 14 & 15 & 16 & 17 & 18 & 19 & 20 \\
            \cmidrule{1-1} \cmidrule{3-24}\addlinespace[0.5em]
            $\ge 5$ &  &     & 1 & 0 & 0 & 0 & 1 & 0 & 1 & 0 & 2 & 0 & 2  & 0  & 5  & 0  & 4  & 0  & 8  & 0  & 9  & 0  & 13 \\
            4       &  &     & 1 & 0 & 0 & 0 & 1 & 0 & 1 & 0 & 2 & 0 & 3  & 0  & 5  & 0  & 6  & 0  & 10 & 0  & 12 & 0  & 17 \\
            3       &  &     & 1 & 0 & 0 & 0 & 1 & 0 & 1 & 0 & 1 & 1 & 2  & 0  & 4  & 2  & 3  & 3  & 6  & 3  & 7  & 5  & 8  \\
            2       &  &     & 1 & 0 & 0 & 0 & 1 & 0 & 0 & 0 & 1 & 0 & 0  & 0  & 3  & 0  & 0  & 0  & 3  & 0  & 0  & 0  & 3  \\
            \bottomrule
        \end{tabular}
    \end{adjustbox}
    \caption{Coefficients of the Hilbert series for a single scalar field (fixed field content $\Phi^{\bm{5}}$).}
    \label{tab:hilbert_series_coefs_5}
    \vspace{1em}
\end{table}

\begin{remark}[Analysis of $H_{\bm{5}}(\D)$] \label{rem:analysisH5scalar}
    Now turning our attention to $H_{\bm{5}}(\D)$, we see even higher level of complexity. Not only do we have more terms in the denominator, reflecting more \emph{building blocks} of how one can assign and contract derivatives, but also even in $\d\ge 5$ where we can not use $\bm{\epsilon}$, we have more \emph{starting blocks}. This simple interpretation of the numerator as extra starting operators is possible in general for a scalar field \cite{Operator2017}.

    However, in a more complicated EFT we can encounter a Hilbert series which can not be brought to the canonical form (see \Cref{rem:hilbertseriesform}) with only positive terms in the numerator. If this is the case, and we will encounter it in \Cref{sec:photonselectomagneticfield}, it is an indication of more complicated relations between operators.

    Comparing $H_{\bm{5}}(\D)$ for $\d=4$ and $\d=5$ we can see that in $\d=4$ we have several (similarly as in the discussion on $H_{\bm{4}}(\D)$) additional operators for every one in $\d=5$. For example, the term $\D^{10}$ corresponds to the operator
    \[ \bm{\epsilon}^{abcd}\wick{(\c3\partial\c2\partial\c1\partial\partial_{a}\Phi)(\c1\partial\c2\partial\partial_{b}\Phi)(\c3\partial\partial_{c}\Phi)(\partial_{d}\Phi)(\Phi)}. \]
    In $\d=3$ we have some similar ``starting'' operators (just check that some terms in the numerator are the same for $\d=3$ and $\d=4$, or even $\d=5$), but others dropped one $\D$, which corresponds to $\bm{\epsilon}$ with one index less.
\end{remark}

\begin{table}[!ht]
    \begin{adjustbox}{center}
        \begin{tabular}{rccccccccccccccccccccccc}
            \toprule
                    &  &     & \multicolumn{21}{c}{$\opers{6}{n}$}                                                           \\
            \cmidrule{3-24}
            $\d$    &  & $n$ & 0 & 1 & 2 & 3 & 4 & 5 & 6 & 7 & 8 & 9 & 10 & 11 & 12 & 13 & 14 & 15 & 16 & 17 & 18 & 19 & 20  \\
            \cmidrule{1-1} \cmidrule{3-24}\addlinespace[0.5em]
            $\ge 6$ &  &     & 1 & 0 & 0 & 0 & 1 & 0 & 2 & 0 & 4 & 0 & 6  & 0  & 13 & 0  & 19 & 0  & 36 & 0  & 58 & 0  & 97  \\
            5       &  &     & 1 & 0 & 0 & 0 & 1 & 0 & 2 & 0 & 4 & 0 & 6  & 0  & 13 & 0  & 19 & 1  & 36 & 1  & 58 & 4  & 97  \\
            4       &  &     & 1 & 0 & 0 & 0 & 1 & 0 & 2 & 0 & 4 & 0 & 6  & 0  & 14 & 0  & 23 & 0  & 45 & 0  & 75 & 0  & 132 \\
            3       &  &     & 1 & 0 & 0 & 0 & 1 & 0 & 2 & 0 & 3 & 2 & 5  & 2  & 10 & 7  & 13 & 12 & 23 & 19 & 35 & 32 & 50  \\
            2       &  &     & 1 & 0 & 0 & 0 & 1 & 0 & 1 & 0 & 3 & 0 & 1  & 0  & 6  & 0  & 1  & 0  & 8  & 0  & 4  & 0  & 10  \\
            \bottomrule
        \end{tabular}
    \end{adjustbox}
    \caption{Coefficients of the Hilbert series for a single scalar field (fixed field content $\Phi^{\bm{6}}$).}
    \label{tab:hilbert_series_coefs_6}
\end{table}

\begin{remark}[Analysis of $H_{\bm{6}}(\D)$]
    Although looking at \Cref{tab:hilbert_series_coefs_6} it seems there are more operators in $\d=4$, at around $n \approx 100$ the dimensions $\d=5,6$ overtake (see \Cref{fig:singleboson6}). For even $n$ we have $\opers{6}{n}^{\d=5}=\opers{6}{n}^{\d=6}$, but in $\d=5$ we additionally have odd operators containing $\bm{\epsilon}^{\bullet \cdots \bullet}$.

    Going down with $\d$, the number of terms in the denominator decreases due to GDC relations, leading to the \emph{exponential} suppression of the number of independent operators for a given power of $\D$. The asymptotics of the operator Hilbert series were studied by \textcite{EFTasymptotics}. Perhaps surprisingly, they seem to approximate pretty well even for small $n$.
\end{remark}

\section{Sidenote --- Gram determinant conditions} \label{sec:sidenotegramdeterminantconditions}

As we already mentioned several times, when lowering the dimension $\d$ some operators can vanish, as is evident from \Cref{tab:hilbert_series_for_a_single_scalar_particle_for_the_fixed_field_content_phi_4_5,tab:hilbert_series_for_a_single_scalar_particle_for_the_fixed_field_content_phi_6}. We will give an explicit example of such an operator and comment on this phenomenon.

The simplest example we already pointed out were operators of the type $\partial^{6}\Phi^{\bm{4}}$, which somehow vanish in $\d=2$. We will now show that EOM conditions are enough to eliminate one of the operators written out in \Cref{rem:interpretationhilbert}. Just expanding the contractions, using the EOM relation (in $\d=2$) for the \textcolor{green!60!black}{green terms}
\[ \wick{\c1\partial\c1\partial\Phi} = \partial^{a}\partial_{a}\Phi\equiv\partial^{1}\partial_{1}\Phi+\partial^{2}\partial_{2}\Phi\overset{!}{=}0 \quad\Longrightarrow\quad
    \textcolor{green!60!black}{\partial^{1}\partial_{1}\Phi}= \textcolor{green!60!black}{-\partial^{2}\partial_{2}\Phi}, \]
and commuting some of the second derivatives leads to (terms cancel pair by pair)
\begin{align*}
    \wick{
        \c2\partial\c1\partial\Phi\c1\partial\c1\partial\Phi\c1\partial\c2\partial\Phi\Phi =
    }
    \begin{subarray}{c}
        + \partial^{1}\partial_{1}\Phi\partial^{1}\partial_{1}\Phi\partial^{1}\partial_{1}\Phi\Phi                             \\
        + \partial^{2}\partial_{1}\Phi\partial^{1}\partial_{1}\Phi\partial^{1}\partial_{2}\Phi\Phi                             \\
        + \partial^{1}\partial_{2}\Phi\partial^{2}\partial_{1}\Phi\partial^{1}\partial_{1}\Phi\Phi                             \\
        + \partial^{2}\partial_{2}\Phi\partial^{2}\partial_{1}\Phi\partial^{1}\partial_{2}\Phi\Phi                             \\
        + \textcolor{green!60!black}{\partial^{1}\partial_{1}\Phi}\partial^{1}\partial_{2}\Phi\partial^{2}\partial_{1}\Phi\Phi \\
        + \partial^{2}\partial_{1}\Phi\partial^{1}\partial_{2}\Phi\textcolor{green!60!black}{\partial^{2}\partial_{2}\Phi}\Phi \\
        + \partial^{1}\partial_{2}\Phi\textcolor{green!60!black}{\partial^{2}\partial_{2}\Phi}\partial^{2}\partial_{1}\Phi\Phi \\
        + \textcolor{green!60!black}{\partial^{2}\partial_{2}\Phi\partial^{2}\partial_{2}\Phi\partial^{2}\partial_{2}\Phi}\Phi
    \end{subarray}
    =
    \begin{subarray}{c}
        \textcolor{black}{+ \partial^{1}\partial_{1}\Phi\partial^{1}\partial_{1}\Phi\partial^{1}\partial_{1}\Phi\Phi}         \\
        \textcolor{orange}{+ \partial^{2}\partial_{1}\Phi\partial^{1}\partial_{1}\Phi\partial^{1}\partial_{2}\Phi\Phi}        \\
        \textcolor{red!70!white}{+ \partial^{1}\partial_{2}\Phi\partial^{2}\partial_{1}\Phi\partial^{1}\partial_{1}\Phi\Phi}  \\
        \textcolor{blue!70!white}{+ \partial^{2}\partial_{2}\Phi\partial^{2}\partial_{1}\Phi\partial^{1}\partial_{2}\Phi\Phi} \\
        \textcolor{blue!70!white}{- \partial^{2}\partial_{2}\Phi\partial^{1}\partial_{2}\Phi\partial^{2}\partial_{1}\Phi\Phi} \\
        \textcolor{red!70!white}{- \partial^{2}\partial_{1}\Phi\partial^{1}\partial_{2}\Phi\partial^{1}\partial_{1}\Phi\Phi}  \\
        \textcolor{orange}{- \partial^{1}\partial_{2}\Phi\partial^{1}\partial_{1}\Phi\partial^{2}\partial_{1}\Phi\Phi}        \\
        \textcolor{black}{- \partial^{1}\partial_{1}\Phi\partial^{1}\partial_{1}\Phi\partial^{1}\partial_{1}\Phi\Phi}
    \end{subarray}
    = 0.
\end{align*}
But where are the celebrated GDC conditions? An attentive reader may recognize, that we encountered a very similar operator already in \Cref{sec:relations}. Just by appending one $\Phi$, by the same manipulations we obtain a Gram determinant condition
\[ 0 \overset{!}{=} 3!\,\partial^{\bm{[}a}\partial_{a}\Phi\partial^{b}\partial_{b}\Phi\partial^{c\bm{]}}\partial_{c}\Phi\Phi =  \wick{\c1\partial\c1\partial\Phi\c1\partial\c1\partial\Phi\c1\partial\c1\partial\Phi}\Phi + 2\wick{\c2\partial\c1\partial\Phi\c1\partial\c1\partial\Phi\c1\partial\c2\partial\Phi}\Phi -3\wick{\c2\partial\c1\partial\Phi\c1\partial\c2\partial\Phi\c1\partial\c1\partial\Phi}\Phi. \]
Since the first and the third term on the right side are automatically zero by EOM, the second term must necessarily vanish in $\d=2$ ``just'' by EOM.

Essentially the same thing happens for $H_{\bm{6}}(\D)$ (see \Cref{tab:hilbert_series_for_a_single_scalar_particle_for_the_fixed_field_content_phi_6}) when going down from $\d=5$ to $\d=4$, or from $\d=4$ to $\d=3$. It would be interesting to understand what happens from $\d=3$ to $\d=2$, where we lose $(1-\D^{4})(1-\D^{6})^2(1-\D^{10})$, but instead gain $(1-\D^{12})$ in the denominator, which is rather peculiar. As of now, we do not have any in-depth explanation.

On the other hand, it is interesting to study when GDC do not play a role, either when there are none or they are automatically satisfied. In the special case of scalar fields, GDC conditions can only exist for $\d<\bm{r}$, otherwise it is not possible to construct an operator that generates them. This is because we need to antisymmetrize in at least $\d+1$ indices, but since derivatives commute and $\Phi$ do not have any indices, the only possibility is to antisymmetrize indices of partial derivatives standing before different fields $\Phi$.

One would then expect some GDC relations manifesting already for $\d+1=\bm{r}$, but a quick examination of \Cref{tab:hilbert_series_for_a_single_scalar_particle_for_the_fixed_field_content_phi_4_5,tab:hilbert_series_for_a_single_scalar_particle_for_the_fixed_field_content_phi_6} shows otherwise. This is a consequence of IBP relations, which admit us to move all derivatives in front of one $\Phi$ away from it, thus we are able to find at most $\d$ indices in which we can antisymmetrize. Therefore, GDC conditions in the presence of IBP become effective for $\d+2 \le \bm{r}$.

The preceding discussion holds in general for scalar fields, but things can get more complicated very quick. For example see the next section.

\section{Photons, electromagnetic field} \label{sec:photonselectomagneticfield}

After obtaining some experience with the scalar field, we can try to apply our formalism to slightly more complicated case of the electromagnetic field. This section is mainly based on the work of \textcite{Chowdhury2019}.

It is advantageous to choose the Maxwell tensor $F_{ab}\equiv \mathrm{d}_{a}A_{b} = 2 \partial_{{\scriptscriptstyle\bm{[}}a}A_{b\scriptscriptstyle\bm{]}}$ as a building block of the Lagrangian density instead of the vector potential $A_{b}$, since it automatically satisfies the gauge invariance. The only essentially new thing will be the construction of the \emph{single particle graded representation} $R_{F}$, and the calculation of the corresponding graded character $\chi_{R_{F}}$.

\begin{calculation}[EOM and other relations]
    First, we must understand the structure of the building block $F_{\bullet\bullet}$. The kinetic Lagrangian density has the well-known form
    \[ \mathcal{L}_{\text{kin}}(A_{\bullet},\partial_{\bullet}A_{\bullet}) \equiv - \frac{1}{4} F_{ab}F^{ab}, \]
    which leads to the free equations of motion (the first set of \emph{Maxwell's equations})
    \begin{equation*} \label{eq:photonEOM}
        \partial^{a}F_{ab}=0. \tag{\textbf{EOM}}
    \end{equation*}
    Furthermore, $F_{\bullet\bullet}$ automatically satisfies the so-called \emph{Bianchi identities} due to its definition as the exterior derivative of $A_{\bullet}$ (the second set of \emph{Maxwell's equations})
    \begin{equation} \label{eq:photonBianchi}
        3 \partial_{[a}F_{bc]} = \mathrm{d}_{a}F_{bc} = \mathrm{d}_{a}\mathrm{d}_{b}A_{c} = 0.  \tag{\textbf{Bianchi}}
    \end{equation}

    One relation we will use is already a consequence of the previous two, because
    \[ \partial^{a}\partial_{a}F_{bc} = \partial^{a}\partial_{a}F_{bc}  + \cancel{\partial^{a}\partial_{c}F_{ab}}{by \eqref{eq:photonEOM}} + \cancel{\partial^{a}\partial_{b}F_{ca}}{by \eqref{eq:photonEOM}} = 3\partial^{a}\partial_{[a}F_{bc]} \overset{\text{\tiny\textcolor{gray}{by \eqref{eq:photonBianchi}}}}{=\mkern-2.7mu=\mkern-2.7mu=\mkern-2.7mu=\mkern-2.7mu=\mkern-2.7mu=}  0. \]
\end{calculation}
\begin{remark}
    The Hilbert series for the electromagnetic field will also include the kinetic term (because we work with $F_{\bullet\bullet}$ instead of $A_{\bullet}$), whereas in the case of a scalar field the kinetic term is always eliminated by IBP and EOM.

    Also note that we must make some corrections by hand, namely account for Chern-Simons topological terms of the type $A \wedge F \wedge \cdots \wedge F$, or overcounting of $F \wedge F \wedge \ldots \wedge F = \mathrm{d}\left( A \wedge \mathrm{d}A \wedge \ldots \wedge \mathrm{d}A \right)$. But more on that later.
\end{remark}

Now we would like to build $R_{F}$ by repeatedly applying derivatives on $F_{\bullet\bullet}$, but we also need to continuously utilize all possible relations to avoid any redundancies. It is therefore useful to decompose representations we obtain to smaller pieces, some of which will be zero by usage of the relations.

\begin{calculation}[Direct "manual" decomposition of $\partial_{a}F_{bc}$ using indices]
    Using antisymmetry of $F_{ab}$ we can perform the manipulations
    \begin{alignat*}{6}
                                                                      & \textcolor{gray}{+\;} \partial_{a}F_{bc} &  & +\textcolor{orange}{\overbrace{\partial_{c}F_{ab}}^{0} } &  & +\textcolor{red!70!white}{\overbrace{\partial_{b}F_{ca}}^{0} } &  & \mathrel{\textcolor{gray}{+}} 3\partial_{[a}F_{bc]}     &  & \mathrel{\textcolor{gray}{+}} 3\partial_{[a}F_{bc]} \\
        3\partial_{a}F_{\textcolor{gray}{[}bc\textcolor{gray}{]}} =\; & \textcolor{gray}{+\;} \partial_{a}F_{bc} &  &                                                          &  & +\textcolor{red!70!white}{\partial_{b}F_{ac}} =\;              &  & \mathrel{\textcolor{gray}{+}} 2\partial_{(a}F_{b)c} =\; &  & \mathrel{\textcolor{gray}{+}} 2\partial_{\{a}F_{b\}c} + \tfrac{2}{\d}\bm{\delta}_{ab}\partial^{\textcolor{gray}{d}}F_{\textcolor{gray}{d}c} \\
                                                                      & \textcolor{gray}{+\;} \partial_{a}F_{bc} &  & +\textcolor{orange}{\partial_{c}F_{ba}}                  &  &                                                                &  & -2\partial_{(a}F_{c)b}                                  &  & -2\partial_{\{a}F_{c\}b} - \tfrac{2}{\d}\bm{\delta}_{ac}\partial^{\textcolor{gray}{d}}F_{\textcolor{gray}{d}b},
    \end{alignat*}
    where ${\{\sbullet[0.7]\mkern+3mu\sbullet[0.7]\}}$ denotes the traceless symmetric part, thus obtaining the decomposition
    \[ \partial_{a}F_{\textcolor{gray}{[}bc\textcolor{gray}{]}} = \partial_{[a}F_{bc]} + \frac{4}{3}\partial_{\textcolor{gray}{\{}a}F_{[b\textcolor{gray}{\}}c]} - \frac{4}{3 \d}\bm{\delta}_{a[b}\partial^{\textcolor{gray}{d}}F_{c]\textcolor{gray}{d}} ,   \]
    where it is understood that first we perform the antisymmetrization $[bc]$, and only after that the traceless symmetrization $\textcolor{gray}{\{}\mkern-4mu\sbullet[0.7]\sbullet[0.7]\textcolor{gray}{\}}$ in the corresponding indices. The first fully antisymmetric term is zero by \eqref{eq:photonBianchi} and the contraction in the last term is zero by \eqref{eq:photonEOM}.
\end{calculation}

\begin{remark}[Young diagrams and symmetrizers]
    Recalling \Cref{rem:youngdiagrams}, we can efficiently describe this decomposition by usage of \emph{Young diagrams}. Since the theory is rather complicated, in the following we will work just intuitively. We assign to every index one box, and represent their symmetries by a certain left- and top-justified diagram. For example, to a fully symmetric traceless tensor and fully antisymmetric tensor we assign the diagrams
    \vspace{-1em}
    \begin{align*}
        \ytableausetup{smalltableaux,centertableaux}
        T_{\{a_1a_2\ldots a_{n}\}} \ \longleftrightarrow\
        \begin{ytableau}
            1 & 2 & \none[\scriptscriptstyle^{_{_{\mkern+2.5mu\cdots}}}] & n
        \end{ytableau}
        \qquad \text{and} \qquad
        T_{[a_1a_2\ldots a_{n}]} \ \longleftrightarrow\
        \begin{ytableau}
            1 \\
            2 \\
            \none[\raisebox{-.15ex}{\scalebox{.6}{\vdots}}] \\
            n
        \end{ytableau}\,.
    \end{align*}
    More complicated \emph{mixed} symmetries (irreducible under $\mathfrak{S}_{n}$) are represented with diagrams of multiple rows and columns, for example
    \begin{align*}
        \ytableausetup{smalltableaux,centertableaux}
        \partial_{\textcolor{gray}{\{}a}F_{[b\textcolor{gray}{\}}c]}
        \ \longleftrightarrow\
        \begin{ytableau}
            b & a \\
            c
        \end{ytableau},
    \end{align*}
    where the Young diagram $\lambda$ is always interpreted as follows:
    \begin{itemize}
        \item First perform all permutations preserving each column of $\lambda$, and sum over them multiplied with the corresponding sign of permutation.
        \item Then perform all permutations preserving each row of $\lambda$ and sum over them.
    \end{itemize}
    This is essentially the definition of a \emph{Young symmetrizer} corresponding to $\lambda$.
\end{remark}

\begin{calculation}[Decomposition using Young diagrams]
    To decompose a tensor of the form $\lambda \otimes \square\mkern+1mu$, we essentially make all available antisymmetric products and all available symmetric products, which we further decompose to a traceless symmetric product and the corresponding traces. Leaving out the traces, this can be represented by attaching the box $\,\square\,$ to all possible places of the Young diagram $\lambda$. Thus, the decomposition performed earlier directly using indices can be done diagrammatically as (gray boxes represent contraction)
    \[ \renewcommand{\bullet}{\textcolor{gray}{\oldbullet}}
        \ytableausetup{smalltableaux,boxsize=1.2em,centertableaux}
        \everymath={\scriptstyle}
        \begin{ytableau}
            F_{\bullet \sbullet} \\
            F_{\sbullet \bullet}
        \end{ytableau}
        \otimes
        \begin{ytableau}
            \partial_{\bullet}
        \end{ytableau}
        =
        \cancel{
            \begin{ytableau}
                F_{\bullet \sbullet} \\
                F_{\sbullet \bullet} \\
                \partial_{\bullet}
            \end{ytableau}
        }{Bianchi}
        \oplus
        \begin{ytableau}
            F_{\bullet \sbullet} & \partial_{\bullet} \\
            F_{\sbullet \bullet}
        \end{ytableau}
        \oplus
        \cancel{
            \begin{ytableau}
                \none & F_{\bullet \sbullet} \\
                *(lightgray) \partial_{\bullet} & *(lightgray) F_{\sbullet \bullet}
            \end{ytableau}
        }{EOM}
        \,.\]
    Following the prescription, we easily find the decomposition of $\partial_{a}\partial_{\textcolor{gray}{\{}b}F_{[c \textcolor{gray}{\}}d]}$ as
    \[ \renewcommand{\bullet}{\textcolor{gray}{\oldbullet}}
        \ytableausetup{smalltableaux,boxsize=1.2em,centertableaux}
        \everymath={\scriptstyle}
        \begin{ytableau}
            F_{\bullet \sbullet} & \partial_{\bullet} \\
            F_{\sbullet \bullet}
        \end{ytableau}
        \otimes
        \begin{ytableau}
            \partial_{\bullet}
        \end{ytableau}
        =
        \cancel{
            \begin{ytableau}
                F_{\bullet \sbullet} & \partial_{\bullet} \\
                F_{\sbullet \bullet} \\
                \partial_{\bullet}
            \end{ytableau}
        }{Bianchi}
        \oplus
        \cancel{
            \begin{ytableau}
                F_{\bullet \sbullet} & \partial_{\bullet} \\
                F_{\sbullet \bullet} & \partial_{\bullet}
            \end{ytableau}
        }{\normalfont$\partial_{[\bullet}\partial_{\bullet]}=0$}
        \oplus
        \begin{ytableau}
            F_{\bullet \sbullet} & \partial_{\bullet} & \partial_{\bullet} \\
            F_{\sbullet \bullet}
        \end{ytableau}
        \oplus
        \cancel{
            \begin{ytableau}
                \none & F_{\bullet \sbullet} & \partial_{\bullet} \\
                *(lightgray) \partial_{\bullet} & *(lightgray) F_{\sbullet \bullet}
            \end{ytableau}
        }{EOM}
        \oplus
        \cancel{
            \begin{ytableau}
                F_{\bullet \sbullet} & *(lightgray) \partial_{\bullet} & *(lightgray) \partial_{\bullet} \\
                F_{\sbullet \bullet}
            \end{ytableau}
        }{Bianchi + EOM}\,. \]
\end{calculation}

\begin{calculation}[Single Particle Graded Representation $R_{F}$ for electromagnetic field]
    Continuing further, same arguments give us that any trace is zero by \eqref{eq:photonEOM} and any antisymmetrization containing derivative is zero by \eqref{eq:photonBianchi} or symmetry of derivatives, therefore we obtain $R_{F}$ of the form
    \begin{equation*}
        \ytableausetup{smalltableaux,boxsize=0.7em,centertableaux}
        R_{F} =
        \Span\begin{pmatrix}
            F_{[ab]} \\
            \partial_{\textcolor{gray}{\{}a_1}F_{[a \textcolor{gray}{\}}b]} \\
            \partial_{\textcolor{gray}{\{}a_{1}}\partial_{a_{2}}F_{[a \textcolor{gray}{\}}b]} \\
            \textcolor{gray}{\vdots} \\
            \partial_{\textcolor{gray}{\{}a_{1}}\cdots\partial_{a_{n}}F_{[a \textcolor{gray}{\}}b]} \\
            \textcolor{gray}{\vdots}
        \end{pmatrix}
        \equiv \D\,\ydiagram{1,1} \oplus \D^2\,\ydiagram{2,1} \oplus \D^3\,\ydiagram{3,1} \oplus \textcolor{gray}{\cdots} \oplus
        \D^{n}\,
        \overbrace{\begin{ytableau}
                \phantom{1} & \phantom{2} & \none[\scriptscriptstyle^{_{\mkern+2.5mu\cdots}}] & \phantom{n}  \\
                \phantom{1}
            \end{ytableau}}^{n}
        \oplus\,\textcolor{gray}{\cdots}
    \end{equation*}
    with the graded character (because $F_{ab}= \mathrm{d}_{a}A_{b}$ already contains one derivative)
    \ytableausetup{boxsize=0.35em,centertableaux}
    \[ \chi_{R_{F}}\!(\D;\bm{x}) \equiv \D\chi_{\ydiagram{1,1}}(\bm{x}) + \D^2\chi_{\ydiagram{2,1}}(\bm{x}) + \D^{3}\chi_{\ydiagram{3,1}}(\bm{x}) + \D^{4}\chi_{\ydiagram{4,1}}(\bm{x}) + \cdots. \]
\end{calculation}

\begin{calculation}[Graded character of $R_{F}$]
    Casting also the SPGR $R_{\Phi}$ for a single scalar field (see \Cref{der:SPGRscalarfield}) in the language of Young diagrams, it takes the form (we denote the trivial representation by $\bigcirc$) \vspace{-1em}
    \begin{equation*}
        R_{\Phi} \equiv \Sym^{\{\bullet\}}(\square) =
        \ytableausetup{smalltableaux,boxsize=0.7em,centertableaux}
        \bigcirc \oplus \D\,\ydiagram{1} \oplus \D^2\,\ydiagram{2} \oplus \textcolor{gray}{\cdots} \oplus \D^{n}\,
        \overbrace{\begin{ytableau}
                \phantom{1} & \phantom{2} & \none[\scriptscriptstyle^{_{\mkern+2.5mu\cdots}}] & \phantom{n}
            \end{ytableau}}^{n}
        \oplus\,\textcolor{gray}{\cdots}
    \end{equation*}
    with the corresponding character
    \ytableausetup{boxsize=0.35em}
    \[ \chi_{R_{\Phi}}\!(\D;\bm{x}) \equiv \chi_{\Sym^{\{\bullet\}}(\smallsquare)}(\D;\bm{x}) = 1 + \D\chi_{\ydiagram{1}}(\bm{x}) + \D^2\chi_{\ydiagram{2}}(\bm{x}) + \D^{3}\chi_{\ydiagram{3}}(\bm{x}) + \cdots. \]

    Using \Cref{prop:charactersproperies} and some reordering, we can calculate
    \begin{alignat*}{5}
        \chi_{R_{\Phi}}\!(\D;\bm{x})\left( \D \chi_{\ydiagram{1}}(\bm{x}) \right) &= \lefteqn{\left( 1 + \D\chi_{\ydiagram{1}}(\bm{x}) + \D^2\chi_{\ydiagram{2}}(\bm{x}) + \D^{3}\chi_{\ydiagram{3}}(\bm{x}) + \cdots  \right) \left( \D \chi_{\ydiagram{1}}(\bm{x}) \right)} \\
        &=  \D \chi_{\ydiagram{1}}(\bm{x}) &&+ \D^2\chi_{\ydiagram{1}\smallotimes\ydiagram{1}}(\bm{x}) &&+ \D^3\chi_{\ydiagram{2}\smallotimes\ydiagram{1}}(\bm{x}) &&+ \D^{4}\chi_{\ydiagram{3}\smallotimes\ydiagram{1}}(\bm{x}) &&+ \cdots    \\
        &=  \D \chi_{\ydiagram{1}}(\bm{x}) &&+ \D^2\chi_{\ydiagram{2}}(\bm{x}) &&+ \D^3\chi_{\ydiagram{3}}(\bm{x}) &&+ \D^{4}\chi_{\ydiagram{4}}(\bm{x}) &&+ \cdots    \\
        &\phantom{ = \D \chi_{\ydiagram{1}}(\bm{x})\;} &&+ \D^2\chi_{\ydiagram{1,1}}(\bm{x}) &&+ \D^3\chi_{\ydiagram{2,1}}(\bm{x}) &&+ \D^{4}\chi_{\ydiagram{3,1}}(\bm{x}) &&+ \cdots \\
        &\phantom{ = \D \chi_{\ydiagram{1}}(\bm{x})\;} &&+ \D^2 \hspace{2.95em} &&+ \D^3\chi_{\ydiagram{1}}(\bm{x}) &&+ \D^{4}\chi_{\ydiagram{2}}(\bm{x}) &&+ \cdots    \\
        &= \lefteqn{\left( \chi_{R_{\Phi}}\!(\D;\bm{x}) - 1 \right) + \D \chi_{R_{F}}\!(\D;\bm{x}) + \D^2\chi_{R_{\Phi}}\!(\D;\bm{x}),}
    \end{alignat*}
    where in the third equality the first line is the traceless symmetrized product, the second line is the antisymmetric product, and the third line is the trace. We can thus express the graded character of $R_{F}$ as
    \begin{align*}
        \chi_{R_{F}}\!(\D;\bm{x}) & = \frac{\left( \D \chi_{\smallsquare}(\bm{x}) -(1+\D^2) \right) \chi_{R_{\Phi}}\!(\D;\bm{x}) + 1}{\D} \\
                                  & = \frac{\left( (\D -\D^{3}) \chi_{\smallsquare}(\bm{x}) -(1-\D^4) \right) P(\D;\bm{x}) + 1}{\D},
    \end{align*}
    where the character $\chi_{\smallsquare}$ is given in \Cref{ex:charSO}.
\end{calculation}

To calculate the Hilbert series $H_{F^{\bm{r}}}(\D)$ with a fixed field content $F^{\bm{r}}$, we proceed exactly as in \Cref{sec:scalarfield}, just with the substitution $\chi_{R_{\Phi}}\!(\D;\bm{x}) \mapsto \chi_{R_{F}}\!(\D;\bm{x})$. We list some of the results in \Cref{tab:hilbert_series_elmag_photon_4}, partly reproducing and partly extending calculations of \textcite{Operator2017} and \textcite{Chowdhury2019}.

\begin{table}[!ht]
    \begin{adjustbox}{center}
    \begin{tabular}{rcc@{\hskip -0.4em}c@{\hskip -2em}c}
        \toprule
        $\d$ && $\tfrac{1}{\D^{4}} H_{F^{\bm{4}}}(\D)$ & \textcolor{gray}{miscount} & $\tfrac{1}{\D^{5}} H_{0,F^{\bm{5}}}(\D)$  \\
        \midrule\addlinespace[0.5em]
        $\ge 10$  && $\displaystyle \frac{2+3 \D^2+2 \D^4}{\left(1-\D^4\right) \left(1-\D^6\right)}$ & & \textcolor{gray}{$\cdots$} \\\addlinespace[1.0em]
        $9$  && $\displaystyle \frac{ 2+3 \D^2+2 \D^4}{\left(1-\D^4\right) \left(1-\D^6\right)}$ & $\textcolor{gray}{-\,\D}$ & \textcolor{gray}{$\cdots$}  \\\addlinespace[1.0em]
        $8$  && $\displaystyle \frac{ 2+3 \D^2+2 \D^4}{\left(1-\D^4\right) \left(1-\D^6\right)}$ & $\textcolor{gray}{+\,1}$ & \textcolor{gray}{$\cdots$}  \\\addlinespace[1.0em]
        $7$  && $\displaystyle \frac{ \textcolor{orange}{\frac{1}{\D}}+2+3 \D^2+2 \D^4}{\left(1-\D^4\right) \left(1-\D^6\right)}$ & $\textcolor{gray}{-\,\frac{1}{\D}}$ & $\displaystyle\frac{\substack{\D^3+4 \D^4+4 \D^5+16 \D^6+10 \D^7+39 \D^8+17 \D^9+69 \D^{10}+28 \D^{11}\\ +99 \D^{12}+36 \D^{13}+125 \D^{14}+41 \D^{15}+135 \D^{16}+43 \D^{17}+126 \D^{18}\\+38 \D^{19}+105 \D^{20}+28 \D^{21}+73 \D^{22}+19 \D^{23}+41 \D^{24}+10 \D^{25}\\ +19 \D^{26}+2 \D^{27}+5 \D^{28}-2 \D^{30}-\D^{31}-2 \D^{32}-\D^{33}+\D^{37}}}{\left(1-\D^4\right) \left(1-\D^6\right) \left(1-\D^8\right) \left(1-\D^{10}\right) \left(1-\D^{12}\right)}$ \\\addlinespace[1.0em]
        $6$  && $\displaystyle \frac{ 2+3 \D^2+2 \D^4+\D^8}{\left(1-\D^4\right) \left(1-\D^6\right)}$ & & $\displaystyle\frac{\substack{1+4 \D^2+13 \D^4+34 \D^6+73 \D^8+121 \D^{10}+168 \D^{12}+210 \D^{14}+226 \D^{16}\\ +213 \D^{18}+182 \D^{20}+131 \D^{22}+79 \D^{24}+42 \D^{26}+16 \D^{28}+\D^{30}-\D^{32}-\D^{36}}}{\left(1-\D^4\right) \left(1-\D^6\right) \left(1-\D^8\right) \left(1-\D^{10}\right) \left(1-\D^{12}\right)}$ \\\addlinespace[1.0em]
        $5$  && $\displaystyle \frac{ 2+3 \D^2+2 \D^4}{\left(1-\D^4\right) \left(1-\D^6\right)}$ & & $\displaystyle\frac{\substack{\D+5 \D^3+4 \D^4+16 \D^5+16 \D^6+30 \D^7+36 \D^8+51 \D^9+63 \D^{10}+73 \D^{11}+89 \D^{12}+92 \D^{13}\\ +110 \D^{14}+103 \D^{15}+117 \D^{16}+103 \D^{17}+108 \D^{18}+91 \D^{19}+88 \D^{20}+71 \D^{21}+59 \D^{22}\\ +49 \D^{23}+32 \D^{24}+27 \D^{25}+13 \D^{26}+12 \D^{27}+2 \D^{28}+3 \D^{29}-3 \D^{30}-2 \D^{32}-\D^{33}}}{\left(1-\D^4\right) \left(1-\D^6\right) \left(1-\D^8\right) \left(1-\D^{10}\right) \left(1-\D^{12}\right)}$ \\\addlinespace[1.0em]
        $4$  && $\displaystyle \frac{ 3+5 \D^2+\D^4-2 \D^6}{\left(1-\D^4\right) \left(1-\D^6\right)}$ & & $\displaystyle\frac{2 \left(\substack{2 \D^4+7 \D^6+17 \D^8+28 \D^{10}+35 \D^{12}+42 \D^{14}+39 \D^{16}+28 \D^{18}\\ +18 \D^{20}+4 \D^{22}-7 \D^{24}-8 \D^{26}-7 \D^{28}-7 \D^{30}-\D^{32}+2 \D^{34}}\right)}{\left(1-\D^4\right) \left(1-\D^6\right) \left(1-\D^8\right) \left(1-\D^{10}\right) \left(1-\D^{12}\right)}$  \\\addlinespace[1.0em]
        $3$  && $\displaystyle \frac{ 1+\D^2+\D^5-\D^6}{\left(1-\D^4\right) \left(1-\D^6\right)}$ & & $\displaystyle\frac{\substack{\D^5+2 \D^7+\D^8+\D^9+2 \D^{10}+\D^{11}+\D^{12}\\+\D^{13}+\D^{14}+2 \D^{16}-\D^{17}+\D^{25}-\D^{26}}}{\left(1-\D^4\right) \left(1-\D^6\right) \left(1-\D^{10}\right) \left(1-\D^{12}\right)}$ \\\addlinespace[0.5em]
        \bottomrule
    \end{tabular}
    \end{adjustbox}
    \caption{The Hilbert series for the electromagnetic field (fixed field content $F^{\bm{4}}$ and $F^{\bm{5}}$).}
    \label{tab:hilbert_series_elmag_photon_4}
\end{table}

\begin{remark}[Analysis of the highlighted terms in $H_{F^{\bm{4}}}$]
    In the listing of the Hilbert series, we divided by $\D^{4}$ and $\D^{5}$, because such number of derivatives is carried automatically by $F^{\bm{4}}$ and $F^{\bm{5}}$, respectively. It is peculiar that for $H_{F^{\bm{4}}}$ in $\d=7$, it seems there is a \textcolor{orange}{``starting operator''} with only $3$ derivatives instead of $4$. This is exactly the already mentioned topological Chern-Simons term $\textcolor{orange}{A\wedge F\wedge F\wedge F}$, which is a top-form that can be integrated over the space-time.~The gauge transformation $A \mapsto A + \mathrm{d}u$ only contributes a total derivative term, because
    \[ \mathrm{d}u \wedge F\wedge F\wedge F = \mathrm{d}u \wedge \mathrm{d}A\wedge \mathrm{d}A\wedge \mathrm{d}A  =  \mathrm{d}\left( u \wedge \mathrm{d}A\wedge \mathrm{d}A\wedge \mathrm{d}A \right).  \]
    When calculating $H_{0,F^{\bm{4}}}$ in $\d=7$ we also obtain an additional term of $\textcolor{gray}{-\D^{3}}$, which is outside the fraction. It reflects the fact that the Hilbert series approach does not see $A\wedge F\wedge F\wedge F$ itself, it was able to find operators of the type $\partial^{n}A\wedge F\wedge F\wedge F$, that is with $n>1$ derivatives distributed in some way between $A$ and $F$s, because one derivative can always be borrowed to form $\partial^{n-1} F_{\bullet[\bullet}F_{\bullet\bullet}F_{\bullet\bullet}F_{\bullet\bullet]}$ up to total derivatives. For an example of this procedure see Subsection 5.4 of \cite{Chowdhury2019}.

    Similarly, the miscount $\textcolor{gray}{+\D^{4}}$ in $\d=8$ is caused by directly working with $F$ instead of $A$. The Hilbert series does not realize that the operator $F\wedge F\wedge F\wedge F$ is actually a total derivative, simply $F \wedge F\wedge F\wedge F =  \mathrm{d}\left( A \wedge \mathrm{d}A\wedge \mathrm{d}A\wedge \mathrm{d}A \right).$

    We can immediately see that the term $\textcolor{gray}{-\D^{5}}$ in $\d=9$ is a miscount, since it would imply $-1$ independent operators of the type $\partial F\wedge F\wedge F\wedge F$. It has to do with $\bm{*} (F \wedge F\wedge F\wedge F)$ being a co-closed one-form, because of Bianchi identities. It is also co-exact, since analogously as before we have \vspace{-0.8em}
    \[ \bm{*} (F \wedge F\wedge F\wedge F) = \bm{*}\mathrm{d}\left( A \wedge F\wedge F\wedge F \right) = \overbrace{\bm{*}\mathrm{d}\bm{*}^{-1}}^{\partial \cdot} \!\bm{*}\left( A \wedge F\wedge F\wedge F \right), \]
    but our approach does not ``know'' about the presence of $A$, so it miscalculates.

    After correcting for these miscalculations, it seems we obtain the full Hilbert series $H_{F^{\bm{4}}}$. For the construction of the operator basis in $\d=4$ see \cite{Chowdhury2019}.
\end{remark}
\begin{remark}
    The Hilbert series $H_{F^{\bm{4}}}$ in $\d=3,4$ can not be brought to the canonical form with only positive terms in the numerator (see \Cref{rem:analysisH5scalar}). For further discussion on the general structure of the Hilbert series see \cite{Operator2017}.
\end{remark}
\begin{remark}
    To our knowledge, results of $H_{0,F^{\bm{5}}}$ in $\d=3,6,7$ are \emph{novel}.
\end{remark}

\section{\texorpdfstring{What's}{What’s} next?}

We did not even come close to exhausting all interesting EFTs, to which our formalism could be applied. We will just list some of the possibilities:
\begin{itemize}
    \item We only worked with one particle field at a time, but it is straightforward to consider \emph{multiple particle fields} at once (see \Cref{der:MPGRgeneral}).
    \item EFTs of \emph{gravity} are pretty similar to the case of the electromagnetic field, see for example \cite{EFTGravity} or \cite{Chowdhury2019}.
    \item One of the first applications of the Hilbert series approach was done for the \emph{Standard Model EFT} (SMEFT) specially in $\d=4$ \cite{OperatorsSMEFT2015}. Gauge invariance was accounted for simply by integration over the gauge group (see \Cref{der:MasterFormulageneral}). Since SMEFT also contains fermions, integration over the covering group $\Spin(4)$ of $\SO(4)$ was necessary, but also easily realized by the isomorphism of groups $\SU(2)\times\SU(2) \cong \Spin(4)$.
    \item From the beginning we restricted ourselves to the case of \emph{linearly realized} symmetries, but for example \emph{Chiral Perturbation Theory} ($\chi$PT) does not fall into this category. For a pretty detailed treatment see \cite{Operator2017} and \cite{Operator2020}.
\end{itemize}

\chapter*{Conclusion}
\addcontentsline{toc}{chapter}{Conclusion}

In this thesis we studied operators within the context of Effective Field Theories. Our goal was to compute the Hilbert series that encodes information about the number of independent operators of a given type.

Recognizing possible relations between operators --- Equations Of Motion (EOM), Integration By Parts (IBP), and Gram Determinant Conditions (GDC) --- led us to the study of representation theory of compact Lie groups.

Along the way we learned a handful of things, which are not only interesting purely from a mathematical standpoint, but also equipped us with the necessary tools to derive the Master Formula for the Hilbert series. More precisely, we obtained the splitting $H=H_0+\Delta H$, where we found an explicit formula only for the main term $H_0$. Thankfully, the correction term $\Delta H$ typically contains only a couple of contributions, which can be calculated by hand.

Finally, we illustrated usage of the formalism in the case of a single scalar and electromagnetic field. Utilizing \texttt{Mathematica}, we successfully computed the Hilbert series for various fixed powers of a particle field in several dimensions \cite{Dujava2022}. To our knowledge, some of our results are novel, as we pointed out in the text.

It would be worthwhile to optimize the computations in \texttt{Mathematica}, in particular with regard to the computational complexity, since going to higher dimensions or higher number of particle fields results in enormous computation time.

The developed formalism is applicable in quite a general setting. However, we did not have the space to adequately cover certain aspects, such as the treatment of fermions (spinors) and the inclusion of parity and charge conjugation. While a brief comment on the former can be found in \cite{Operator2017} (which still warrants a more comprehensive discussion), the latter was thoroughly analyzed in \cite{Operator2020}.

Naturally, numerous questions remain unanswered: How should we proceed in general if the particle fields transform non-linearly under the action of the group? Can we obtain the Hilbert series in a fully closed form? Is it possible to automate the calculation of the correction term $\Delta H$? There is definitely a potential for further exploration and advancements.

\printbibliography[heading=bibintoc, title=References]

\end{document}